\documentclass[12pt,notitlepage,aps,preprintnumbers,superscriptaddress,nofootinbib,aps,prd,floatfix]{revtex4-2}
\usepackage{float}
\usepackage{physics,slashed}
\usepackage{amsfonts,amsmath,amssymb}
\usepackage{mathrsfs}
\usepackage{bm}
\usepackage{epsfig}
\usepackage{graphicx}   
\usepackage{subfigure}  
\usepackage{verbatim}   
\usepackage{color}      
\usepackage{hyperref}   
\usepackage{url}
\definecolor{nicered}{rgb}{0.5,0.,0.}
\definecolor{nicegreen}{rgb}{0.,0.5,0.}
\definecolor{niceblue}{rgb}{0.,0.,0.5}
\hypersetup{colorlinks,citecolor=nicegreen,linkcolor=nicered,urlcolor=niceblue}
\usepackage{multirow}

\newcommand{\GeV}{\textrm{GeV}}
\newcommand{\TeV}{\textrm{TeV}}

\begin{document}
\setlength{\abovedisplayskip}{3pt}
\setlength{\belowdisplayskip}{3pt}
\preprint{ANL-181430, MSUHEP-23-003, PITT-PACC-2302}

\title{High-energy neutrino deep inelastic scattering cross sections}

\author{Keping Xie}
\email{xiekeping@pitt.edu}
\affiliation{
Pittsburgh Particle Physics Astrophysics and Cosmology Center,
Department of Physics and Astronomy, University of Pittsburgh, Pittsburgh, PA 15260, USA\looseness=-1}
\affiliation{Department of Physics and Astronomy, Michigan State University, East Lansing, MI 48824, USA\looseness=-1}

\author{Jun Gao}
\email{jung49@sjtu.edu.cn}
\affiliation{NPAC, Shanghai Key Laboratory for Particle Physics and Cosmology,
School of Physics and Astronomy, Shanghai Jiao Tong University, Shanghai 200240, China}
\affiliation{Key Laboratory for Particle Astrophysics and Cosmology (MOE), Shanghai 200240, China}
\author{T.~J.~Hobbs}
\email{tim@anl.gov}
\affiliation{High Energy Physics Division, Argonne National Laboratory, Argonne, IL 60439, USA}

\author{Daniel R. Stump}
\email{stump@msu.edu}
\affiliation{Department of Physics and Astronomy, Michigan State University, East Lansing, MI 48824, USA\looseness=-1}

\author{C.-P. Yuan}
\email{yuanch@msu.edu}
\affiliation{Department of Physics and Astronomy, Michigan State University, East Lansing, MI 48824, USA\looseness=-1}

\collaboration{CTEQ-TEA Collaboration}
\date{\today}

\begin{abstract}
We present a state-of-the-art prediction for cross sections of neutrino deep inelastic scattering (DIS) from nucleon at high neutrino energies, $E_\nu$, up to 1000 EeV ($10^{12}~\GeV$).
Our calculations are based on the latest CT18 NNLO parton distribution functions (PDFs) and their associated uncertainties. To make predictions for the highest energies, we extrapolate the PDFs to small $x$ according to several procedures and assumptions, thus affecting the uncertainties at ultra-high $E_\nu$; we quantify the uncertainties corresponding to these choices. Similarly, we quantify the uncertainties introduced by the nuclear corrections that are required to evaluate neutrino-nuclear cross sections for the neutrino observatories.
These results can be applied to currently-running astrophysical neutrino observatories, such as IceCube and KM3NeT, as well as various future experiments that have been proposed.
\end{abstract}

\maketitle
\tableofcontents{}

\section{Introduction}
\label{sec:intro}

Empirical information furnished by the deeply inelastic scattering (DIS) of high-energy neutrinos from nucleons and nuclei played an 
an important role in establishing Quantum Chromodynamics (QCD) as the microscopic theory of strong interaction.
Within this context, charged-current (CC) neutrino DIS ($\nu$DIS) has the potential to be particularly enlightening in that it accesses unique combinations of quark-flavor currents inside QCD matter, having been measured in accelerator-based DIS experiments with neutrino energies up to $E_\nu\! \sim\! 300\, \mathrm{GeV}$ (see Ref.~\cite{Zyla:2020zbs} for an overview). 
In parallel with these accelerator-based experiments, which have played an invaluable role in understanding the hadronic and nuclear structure, considerable interest also attaches to
neutrino measurements recorded at energies several orders of magnitude beyond those accessed in
terrestrial experiments; conventionally, such measurements are designated as high-energy (HE, $10^3\!<\! E_\nu\! <\! 10^8$ GeV) and
ultra-high energy (UHE, $E_\nu\!>\! 10^8$ GeV). A significant share of the interest in HE and UHE neutrino processes derives from the fact that such measurements
may possess sensitivity to a variety of beyond-the-standard model (BSM) scenarios, including non-standard interactions, leptoquarks, and the possibility of hidden extra dimensions (see, {\it e.g.},
Ref.~\cite{Ackermann:2022rqc} for a recent review). In addition, (U)HE astrophysical neutrinos can provide information constraining at least 6 of the 9 available neutrino flavor-oscillation channels, heightening
their sensitivity to various BSM and non-standard interaction possibilities~\cite{Ackermann:2022rqc}.
It is notable that high-energy neutrino measurements have motivations extending beyond fundamental high-energy physics (HEP) to particle astrophysics; this is due to the fact that the weakness
of neutrinos' interactions with matter renders them ideal messengers for astronomy beyond the visible electromagnetic spectrum, possibly conveying information on the nature
of their astrophysical sources related to the unsolved question regarding the origin of high-energy cosmic rays. 

An array of experimental facilities, therefore, aims to measure neutrino cross sections at high energies. For example, the IceCube Neutrino Observatory is capable of detecting high-energy neutrinos with its unique instrumental volume of about 1 km$^3$ within the Antarctic ice sheet~\cite{IceCube:2003llu}. 
Other neutrino observatories under development with complementary reach include IceCube-Gen2~\cite{vanSanten:2017chb}, KM3NeT~\cite{KM3Net:2016zxf}, Baikal-GVD~\cite{Avrorin:2011zzc}, GRAND~\cite{GRAND:2018iaj}, POEMMA~\cite{Olinto:2017xbi}, P-ONE~\cite{P-ONE:2020ljt} and TRIDENT~\cite{Ye:2022vbk}.
As pointed by Ref.~\cite{Denton:2020jft}, the proposed and/or under-construction GRAND~\cite{GRAND:2018iaj} and POEMMA~\cite{Olinto:2017xbi} facilities have the opportunity to measure the $\nu_\tau$-nucleon cross section with neutrino energies up to $E_\nu\sim10^{9}~\GeV$. Meanwhile, the neutrino DIS cross sections are essential inputs to the modeling of the detection of high-energy neutrinos as well as their propagation through the earth, and directly impact the IceCube's extracted flux of astrophysical neutrinos with energies of about $10^{4}$ GeV to $10^7$ GeV~\cite{IceCube:2017roe,IceCube:2020rnc} as well as output of earth tomography~\cite{Donini:2018tsg}.
We may expect the observed neutrino energies extended by a few orders of magnitudes due to large exposure from the next generation of neutrino observatories.
More interestingly, at IceCube one can also independently extract the total cross sections using earth absorption~\cite{IceCube:2017roe} or even differential cross sections in elasticity~\cite{IceCube:2018pgc} in neutrino DIS though with large uncertainties.
See Refs.~\cite{Valera:2022ylt,Valera:2022wmu} for various systematic analyses, in particular the one from the astrophysical neutrino flux.
We also note the recently proposed FASER (ForwArd Search ExpeRiment) program~\cite{Feng:2017uoz,FASER:2018bac} as well as other experiments at the Forward Physics Facility (FPF)~\cite{Anchordoqui:2021ghd,Feng:2022inv} at the LHC, which can potentially fill the gap between neutrino energies measured at IceCube and fixed-target experiments. 

For all these reasons, a better understanding of $\nu$DIS is also crucial for programs centered on neutrino observatories and neutrino-based particle astrophysics.
Ultimately, the ability of UHE neutrino measurements to impose stringent constraints on such scenarios depends on the current theoretical accuracy
for predictions of the purely standard model neutrino-nucleon (-nucleus) interactions.
Theoretical predictions on neutrino DIS rely on the theorem of QCD factorization~\cite{Collins:1989gx} and thus the hard coefficient functions that can be calculated perturbatively and the parton distributions (PDFs) of nucleons and nuclei.
For neutrino energies greater than $10^7$ GeV, the corresponding DIS cross sections are potentially sensitive to QCD dynamics in the region of very small $x\! <\! 10^{-5}$.
The PDFs at small $x$ are only loosely constrained by experimental data from HERA~\cite{H1:2015ubc} and LHC measurements in the forward region~\cite{Belyaev:2021cyr}, which can lead to large uncertainties in theoretical predictions.
Besides, the astrophysical neutrino DIS cross sections also depend on the effects of nuclear modifications since the neutrinos are colliding with nuclei inside water or the earth's crust rather than free nucleons.

In the current work, we present state-of-the-art predictions on total cross sections for neutrino DIS with neutrino energies up to $10^{12}$ GeV for both the charged-current (CC) and neutral-current (NC) interactions.
We use the CT18 PDFs~\cite{Hou:2019efy} as our baseline for the nucleon PDFs; these were fitted at next-to-next-to-leading order (NNLO) to a broad selection of the world's high-energy data, including the most recent LHC precision experiments.
For the main NNLO calculations presented in this work, the hard coefficient functions are determined consistently in the S-ACOT-$\chi$ heavy-quark scheme~\cite{Gao:2021fle}, and are therefore based on a systematic evaluation of heavy-quark mass effects.
It will be referred to as the ACOT scheme in this work.
In addition, we explore approximate N3LO corrections using massless Wilson coefficients for both neutral- and charged-current DIS structure functions~\cite{Salam:2008qg,Vermaseren:2005qc,Vogt:2006bt,Moch:2007rq,Davies:2016ruz,Moch:2004xu}.
Going beyond evaluations of free-nucleon cross sections, we also estimate nuclear effects and incorporate these into our analysis by considering recent nuclear PDF studies, specifically, EPPS21~\cite{Eskola:2021nhw} and nCTEQ15WZ~\cite{Kusina:2020lyz}.
The impact of higher-order corrections including small-$x$ resummed corrections is also discussed. 

Our predictions show the combined PDF-driven and theoretical uncertainties for $\nu$DIS cross sections on nuclei to be several percent at high energies, increasing to a few tens of percent at ultra-high energies.
Controlling these uncertainties, especially those at the highest neutrino energies, will require improved
knowledge of free-nucleon PDFs in the extrapolated region of extremely small $x$ as well as of nuclear
corrections in the far-shadowing regime at similarly small $x$.
We compare our cross-section predictions to those of CSMS~\cite{Cooper-Sarkar:2011jtt}, which are frequently used in IceCube publications, and discuss the implications of these comparisons for IceCube. 
In parallel, we briefly discuss some of the other calculations in the literature, such as GQRS~\cite{Gandhi:1998ri}, CTW~\cite{Connolly:2011vc}, and BGR~\cite{Bertone:2018dse,Garcia:2020jwr}, as well as the updated ones to include the shallow inelastic scattering~\cite{Jeong:2023hwe} and nuclear structure functions~\cite{Candido:2023utz}.
An alternative approach based on the color-dipole model~\cite{Arguelles:2015wba} results in a growth of the neutrino cross section like $\ln^2s$, with $\sqrt{s}$ being the central-of-mass energy of the scattering process, in agreement with perturbative QCD predictions but with larger theoretical uncertainty.
Recently, Ref.~\cite{Candido:2023utz} proposed a neural network-based method to parametrize neutrino DIS purely in terms of structure functions, with extrapolations down to $Q\lesssim1~\GeV$; in this work, however, we concentrate primarily on theory predictions for high-energy neutrino cross sections with neutrino energy above TeV.

The organization of the paper is as follows. 
Sec.~\ref{sec:vN} discusses the neutrino-nucleon cross section, which we calculate and compare for several assumptions with respect to the small-$x$ PDF behavior.
In Sec.~\ref{sec:PDFs}, we discuss the impact on the neutrino-nucleon DIS cross section deriving from the behavior and uncertainty of small-$x$ PDFs, and the role of nuclear corrections in the corresponding cross sections on nuclei.
We compare our results with other calculations and with existing cosmic neutrino data in Sec.~\ref{sec:compare} before concluding in Sec.~\ref{sec:conc}.
In addition, we provide more detailed discussion and supplementary material in several appendices. These provide 
practical neutrino cross section tables (App.~\ref{app:num}); a
more detailed discussion of kinematics (App.~\ref{app:xQlimit});
estimates of neutrino Earth absorption (App.~\ref{app:abspt});
and comparative predictions of the neutrino-electron interaction and Glashow
resonance (App.~\ref{sec:Glashow}).

\section{The theoretical framework}
\label{sec:vN}
\label{sec:th}
\begin{figure}
    \centering
    \includegraphics{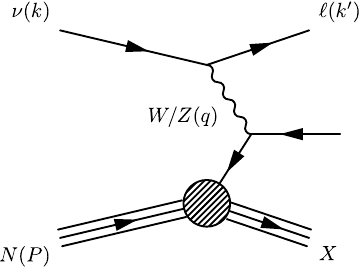}
    \caption{Feynman Diagram for neutrino-nucleon deep-inelastic scattering.}
    \label{fig:feynDIS}
\end{figure}

The charged-current (CC) or neutral-current (NC) process for the deep-inelastic scattering of a neutrino
of flavor $\ell$ can be written as
\begin{equation}\label{ccdis}
\nu_\ell(k) + N(P) \longrightarrow \ell(k^{\prime}) + X(P+q)\ ,
\end{equation}
which we illustrate in Fig.~\ref{fig:feynDIS}. Here, the final-state lepton $\ell$ can either be a neutrino, $\nu_\ell$ (corresponding to NC DIS), or a charged-lepton, $\ell^\pm$ (for CC scattering).
The 4-momentum transfer is $q = k - k^{\prime}$.
$N$ denotes a nucleon target, which can be a proton or neutron with mass $m_N$, while $X$ represents the inclusive hadronic final state. The 4-momenta $k,~k^{\prime}$ and $P$
are indicated in Fig.~\ref{fig:feynDIS}.
The familiar kinematic invariants are
\begin{equation}\begin{aligned}
\label{eq:invariants}
s  & =  (k+P)^{2} = m_N^{2} + 2 m_N E_{\nu}\ , \\
Q^{2} & =  - q^{2}\ ,  \\
x & =  \frac{Q^{2}}{2 P\cdot{q}} = \frac{Q^{2}}{2m_N(E_\nu-E_\nu')}\ ,\\
y & =  \frac{P\cdot{q}}{P\cdot{k}} =\frac{E_\nu-E_\nu'}{E_\nu}=\frac{Q^2}{2xm_NE_\nu}\ ,
\end{aligned}
\end{equation}
where $E_{\nu}$ is the (initial) neutrino energy in the nucleon rest frame.

The inclusive DIS (anti)neutrino-nucleon cross section can be written as~\cite{Formaggio:2012cpf}
\begin{equation}\label{eq:diffXS_mass}
\frac{\dd^2\sigma^{\nu(\bar{\nu})}}{\dd x\dd y}
=\frac{G_F^2m_NE_\nu}{\pi(1+Q^2/M_{W,Z}^2)^2}
\left[\frac{y^2}{2}2xF_1+
\left(1-y-\frac{m_Nxy}{2E_\nu}\right)F_2
\pm y\left(1-\frac{y}{2}\right)xF_3\right].
\end{equation}
Here $F_i~(i=1,2,3)$ are the corresponding structure functions in the case of charged or neutral current scattering. The positive (negative) sign in the last term related to $F_3$ corresponds to the neutrino (antineutrino) scattering, as a result of spin correlation.
We can substitute the longitudinal structure function
\begin{equation}
F_L=F_2(1+4x^2m_N^2/Q^2)-2xF_1,
\end{equation}
and obtain
\begin{equation}\label{eq:diffXS}
\frac{\dd^2\sigma^{\nu(\bar{\nu})}}{\dd x\dd Q^2}=
\frac{G_F^2}{4\pi x(1+Q^2/M_{W,Z}^2)^2}
\left[Y_{+}F_2-y^2 F_L \pm Y_- xF_3\right],
\end{equation}
where $Y_{\pm}=1\pm(1-y)^2$. 
At the leading order (LO), the neutrino CC structure functions can be written as
\begin{equation}\begin{aligned}\label{eq:F23CC}
&F_{2}^{\nu(W)}=2x\left(\sum_{i}d_i+\sum_{j}\bar{u}_j\right),~
&&xF_{3}^{\nu(W)}=2x\left(\sum_{i}d_i-\sum_{j}\bar{u}_j\right),\\
&F_{2}^{\bar{\nu}(W)}=2x\left(\sum_{j}u_j+\sum_{i}\bar{d}_i\right),~
&&xF_{3}^{\bar{\nu}(W)}=2x\left(\sum_{j}u_j-\sum_{i}\bar{d}_i\right),
\end{aligned}\end{equation}
where the index $i(j)$ runs over all the light $d(u)$-type quarks. 
Similarly, the LO neutral-current structure functions are
\begin{equation}\begin{aligned}\label{eq:F23NC}
F_{2}^{\nu,\bar{\nu}(Z)}=x\sum_{i}^{n_f}(a_i^2+v_i^2)(q_i+\bar{q}_i),~
xF_3^{\nu,\bar{\nu}(Z)}=x\sum_{i}^{n_f}a_iv_i(q_i-\bar{q}_i),
\end{aligned}\end{equation}
where $(a_{i})~v_{i}$ are the NC (axial) vector couplings for the quark $q_i$ and $i$ runs over all quarks in the $n_f$ flavor number scheme. At the LO, the longitudinal structure functions are zeros, $F_L^{i}=0$. In perturbative QCD, the structure functions in Eqs.~(\ref{eq:F23CC}) and (\ref{eq:F23NC}) as well as $F_L^i$ receive higher-order corrections, which is one main focus of this work.

In the isospin symmetric limit, the isoscalar $u(\bar{u})$ and $d(\bar{d})$ PDFs can be constructed in terms of the proton PDFs as
\begin{eqnarray}
f_{u/I}=f_{d/I}=(f_{u/p}+f_{d/p})/2,~ 
f_{\bar{u}/I}=f_{\bar{d}/I}=(f_{\bar{u}/p}+f_{\bar{d}/p})/2,    
\end{eqnarray}
while other flavors are kept the same.
Keep in mind the positive (negative) sign for neutrino (antineutrino) cross sections in Eq.~(\ref{eq:diffXS}). Together with the LO structure functions in Eqs.~(\ref{eq:F23CC}-\ref{eq:F23NC}), we can see that the neutrino-isoscalar scattering cross sections are generally larger than the antineutrino ones, both for the CC and NC cases.
In Sec.~\ref{sec:smallx}, we present the final, absolute isoscalar cross sections in Fig.~\ref{fig:NuNXSec_CT18}
with related discussion; we note that the (anti)neutrino cross sections differ at the lower energies of the plotted range.

We point out that Eq.~(\ref{eq:diffXS}) above additionally assumes the high-energy (massless) limit, corresponding to $m_N\!=\!0$, as is reasonable
for (U)HE neutrino scattering. In the numerical calculations presented below, however, we implement the full expression in Eq.~(\ref{eq:diffXS_mass}), though
the impact of the hadronic mass, $m_N$, is in general negligible.
For NC DIS, higher-order EW corrections can be included through an ``improved" scheme. (See Ref.~\cite{Bertone:2018dse} for details.)
As implied by Eqs.~(\ref{eq:F23CC}-\ref{eq:F23NC}), $F_L$ and $xF_3$ are suppressed in comparison with $F_2$, with respect to the contribution to Eq.~(\ref{eq:diffXS}); as such, the cross section can be approximated with a simplified form only involving $F_2$ as explored in Ref.~\cite{Illarionov:2011wc}. For the sake of precision, however, we stay with the full expression given by
Eq.~(\ref{eq:diffXS_mass}) throughout this work. 

\subsection{Treatment of low-$x$ and low-$Q$ PDFs}
\label{sec:xQ}
\begin{figure}
    \centering
    \includegraphics{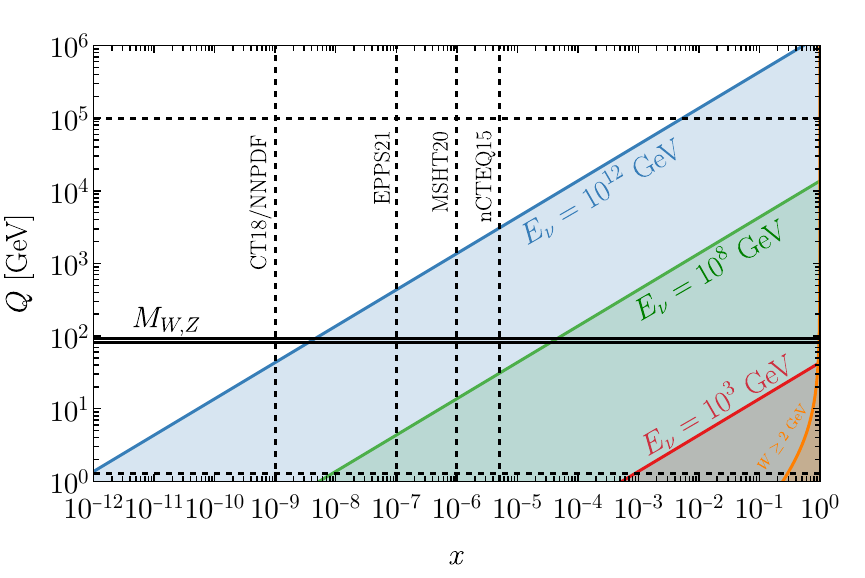}
    \caption{The kinematic $(x,Q)$ ranges contributing to the inclusive neutrino DIS cross section for a given initial neutrino energy, $E_\nu$. The vertical
    dashed lines represent the lower boundary in $x$ for which each (nuclear) PDF group conventionally provides interpolation tables {\it by default}. This should not be
    misconstrued as the lowest value of $x$ directly probed by the fitted data sets in each of these cases, which would generally lie at significantly higher $x$ than
    that indicated.
}
\label{fig:xQ}
\end{figure}
The total cross section can be integrated in terms of the differential cross section in Eq.~(\ref{eq:diffXS_mass}) over $x$ and $Q^2$. That is,
\begin{equation}\label{eq:intXS}
    \sigma=
    \int_{Q_{\min}^{2}}^{2m_NE_{\nu}}\dd Q^{2}N(Q^2)
    \int_{x_{\min}}^{1} \frac{dx}{x}\mathcal{F}(x,Q^2).
\end{equation}
where $x_{\min} = Q^{2}/(2m_NE_{\nu})$. 
The specific functional forms of $N(Q^2)$ and $\mathcal{F}(x,Q^2)$ can be deduced directly from the structure functions $F_{i}$ in terms of Eq.~(\ref{eq:diffXS_mass}).
In an experimental measurement, the neutrino DIS events are selected with a $Q$ cut, such as $Q\geq Q_{\min}=1~\GeV$ in MINERvA~\cite{MINERvA:2016oql}, which can be adopted here in Eq.~(\ref{eq:intXS}).

In Fig.~\ref{fig:xQ}, we show the integrated kinematic $(x,Q)$ region for a few representative neutrino energies, $E_\nu$. We include two dashed horizontal lines corresponding to the CT18 starting scale, $Q_0=1.3~\GeV$, and the upper bound of the $Q$ grids in the LHAPDF format~\cite{Buckley:2014ana}, $Q_{\rm up}=10^{5}~\GeV$. 
In our practical treatment, the phase space below $Q_0$ and above $Q_{\rm up}$ can be obtained through either LHAPDF extrapolation(s)~\cite{Buckley:2014ana} or APFEL's backward/forward DGLAP evolution~\cite{Bertone:2013vaa}. Fig.~\ref{fig:xQ} plots several vertical lines indicating the lower $x$ bounds for the corresponding PDF grids used in this work. The PDFs below these $x$ bounds rely on the LHAPDF or APFEL extrapolation.

\begin{figure}[h]
\centering
\includegraphics[width=0.45\textwidth]{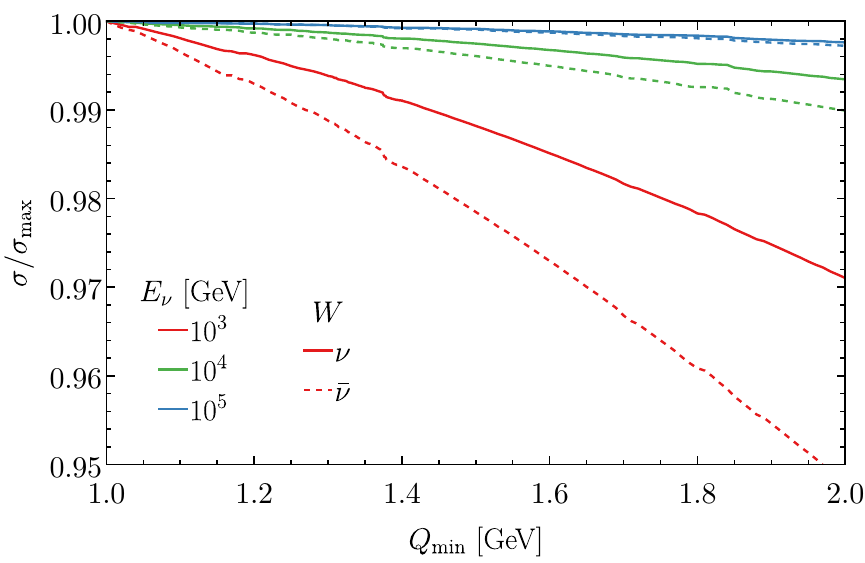}
\includegraphics[width=0.45\textwidth]{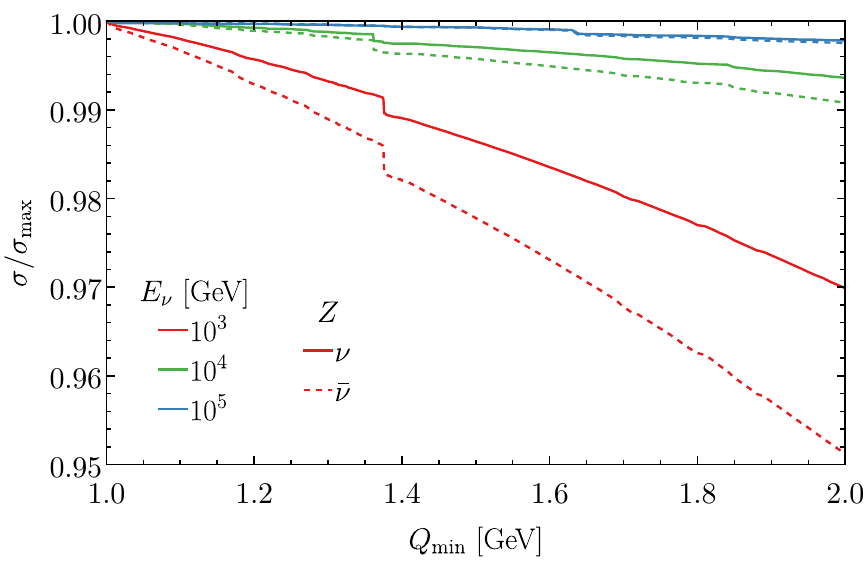}
    \caption{The dependence of the neutrino-isoscalar nucleon charged-current (left) and neutral-current (right) DIS cross sections $(\sigma)$ on the lower integration limits $Q_{\min}$, based on the CT18 NNLO PDFs. The upper integration limit is fixed by collision energy $Q_{\max}=\sqrt{2m_NE_\nu}$. $\sigma_{\max}$ indicates the maximal cross section with $Q_{\min}=1~\GeV$.}
    \label{fig:scanQ}
\end{figure}

In Fig.~\ref{fig:scanQ}, we explore the dependence of neutrino-isoscalar nucleon charged- and neutral-current DIS cross sections on the choice of $Q_{\rm min}$ in Eq.~(\ref{eq:xQ2}).
We plot results for representative neutrino energies spanning the range $10^{3}\!\sim\!10^{5}~\GeV$, where the dependence on Bjorken $x$ is fully covered by the CT18 grids. The cross sections are normalized to the maximal result, $\sigma_{\max}$, where $Q_{\min}=1~\GeV$. The PDFs below $Q_0=1.3~\GeV$ are obtained with the APFEL's backward evolution~\cite{Bertone:2013vaa}. 
The structure functions are calculated in the zero-mass variable-flavor-number (ZM-VFN) scheme at NNLO with CT18 PDFs. In order to define the charged-current DIS scattering consistently beyond the leading order, we take the maximum number of active quark flavors to be $n_f=4$, similarly to Ref.~\cite{Gao:2021fle}. 
The details about the heavy-quark mass as well as the flavor number dependence are left to Sec.~\ref{sec:ACOT}. 
In terms of Fig.~\ref{fig:scanQ}, we see that the backward evolution region $Q\in[1,1.3]~\GeV$ contributes at most approximately 0.8\%(1.2\%) to the (anti)neutrino cross sections, with this contribution peaking for $E_\nu=10^{3}~\GeV$. 
With increasing neutrino energy, this low-$Q$ contribution quickly becomes negligible.
A detailed exploration of the dependence of our calculation on $Q_{\min}$ is presented in App.~\ref{app:xQlimit}. In the end, we find that the kinematic region around $Q\sim M_{W,Z}$ contributes most significantly to the total cross section, and we designate this {\it the important $(x,Q)$ kinematics}, as investigated in App.~\ref{app:xQlimit}.

\begin{figure}[h]
\centering
\includegraphics[width=0.45\textwidth]{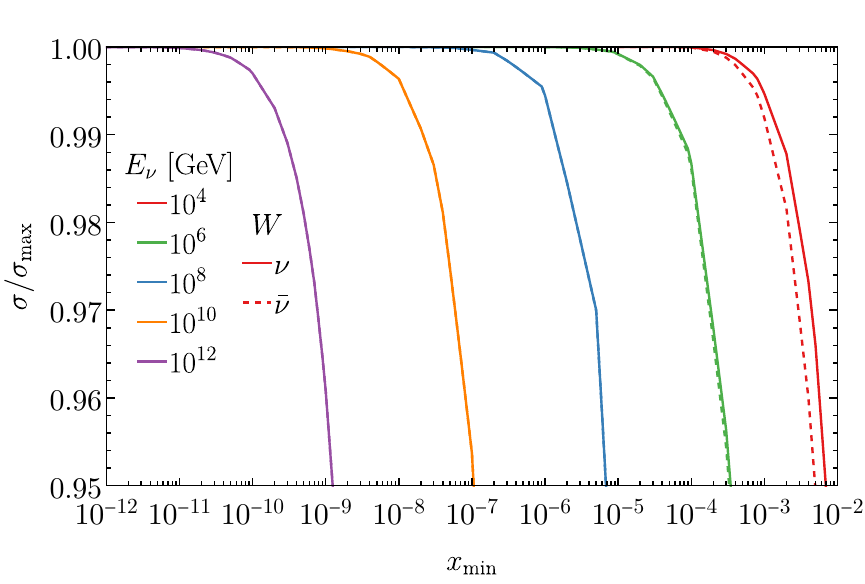}
\includegraphics[width=0.45\textwidth]{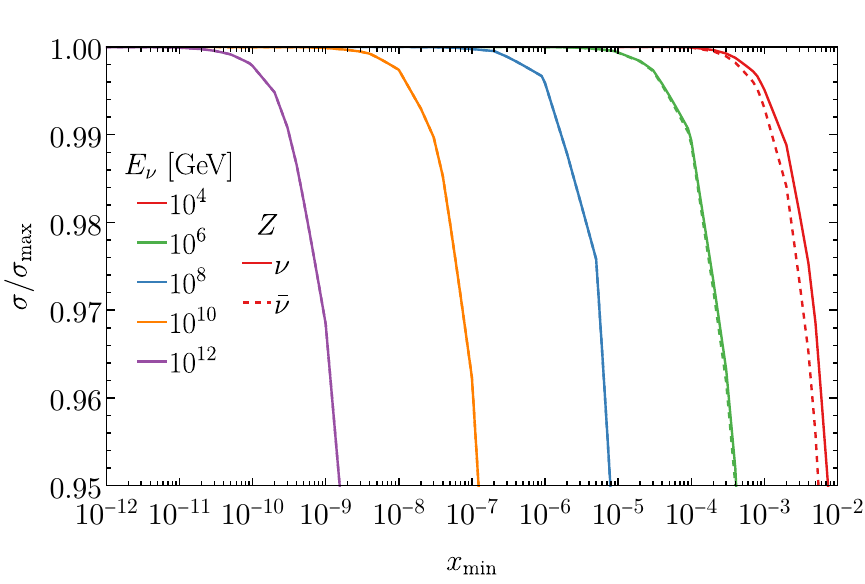}
    \caption{Similar to Fig.~\ref{fig:scanQ}, but for integration limit $x_{\min}$. The upper integration limit is fixed at $x_{\max}=1$.
    $\sigma_{\max}$ represents the maximal $\sigma$ with $x_{\min}=Q_{\min}^2/(2m_NE_\nu)$.
    }
    \label{fig:scanx}
\end{figure}

With increasing neutrino energies, we see that the relevant values of $x$ can fall below the region probed by existing data, and even below the $x$ grid provided by different PDF groups, as indicated by the vertical lines in Fig.~\ref{fig:xQ}.
At ultra-high neutrino energies, \emph{e.g.}, $E_\nu>10^{9}~\GeV$, the kinematics will even cover a momentum fraction below the lowest $x$ value provided by various PDF groups, such as $x_{\min}=10^{-9}$ in CT18~\cite{Hou:2019efy} and NNPDF3.1~\cite{NNPDF:2017mvq}/4.0~\cite{NNPDF:2021njg}. In the CSMS calculation~\cite{Cooper-Sarkar:2011jtt}, two treatments for low $x$ have been performed, based on ({\it i}) extrapolation or ({\it ii}) freeze-in, $ f(x<x_{\min},Q)=f(x_{\min},Q)$. 
We have also explored another option in which we first extrapolate the provided LHAPDF grids to obtain PDFs below $x_{\min}$ at the starting scale, $Q_0$, and then evolve these to cover the whole phase space shown in Fig.~\ref{fig:xQ}; this option is denoted as the ``Evolution'' method below. In this approach, we first show the fraction of the extrapolated region $x\!<\!x_{\min}$ contributing to the total integrated cross section in Fig.~\ref{fig:scanx}. Similarly to the $Q_{\min}$ case, we normalize the cross sections with different $x_{\min}$ choices to the maximal scenario, where $x_{\min}=Q_{\min}^2/(2m_NE_\nu)$, for a few representative neutrino energies. 
We see that this extrapolation only contributes at most 3\% for $E_\nu=10^{12}~\GeV$. For smaller energies, the extrapolation region $x<x_{\min}=10^{-9}$ contributes negligibly to the total integrated cross section for CT18 PDFs.
This result gives us confidence that the PDF grids provided in the CT18 global analysis~\cite{Hou:2019efy} are sufficient to explore the neutrino cross section for neutrino energies reaching $E_\nu\lesssim10^{12}~\GeV$. 
Similarly to the $Q_{\min}$ scan, we also explore the impact of the choice of $x_{\min}$ in the full range $[Q_{\min}^2/(2m_NE_\nu),1]$ in App.~\ref{app:xQlimit}, 
which shows the important $(x,Q)$ kinematics around $x\sim M_{W,Z}^2/(2m_NE_\nu)$ and $Q\sim M_{W,Z}$

\begin{figure}
    \centering
\includegraphics[width=0.49\textwidth]{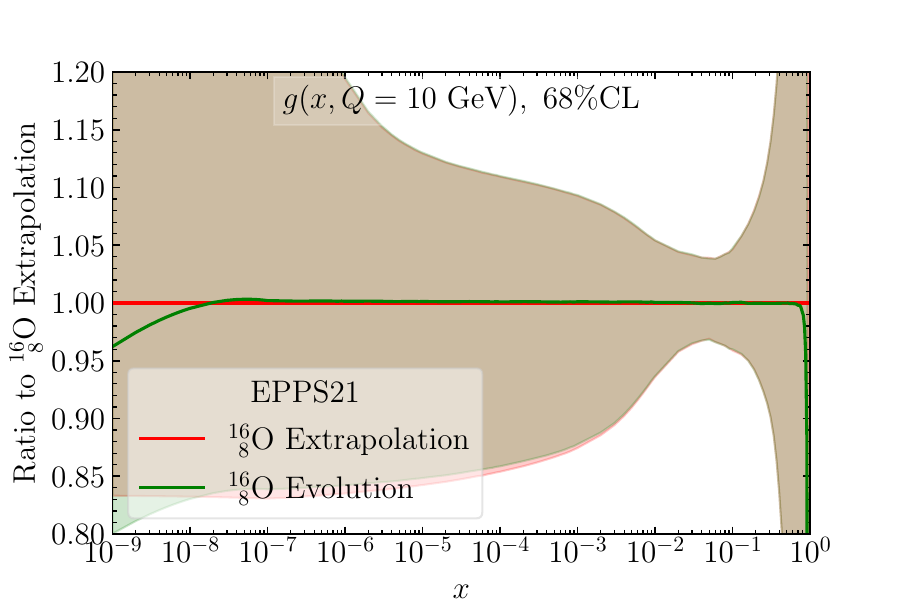}
\includegraphics[width=0.49\textwidth]{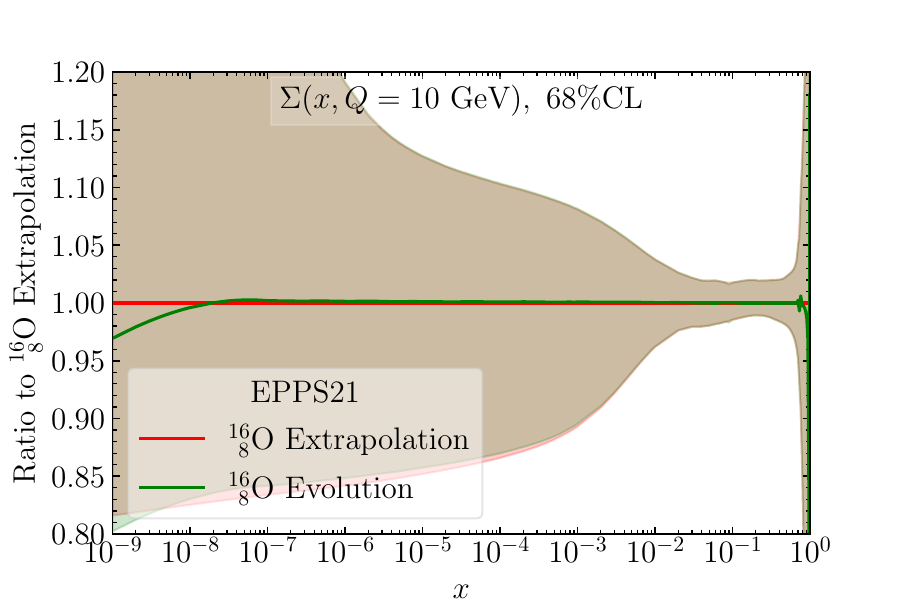}
\includegraphics[width=0.49\textwidth]{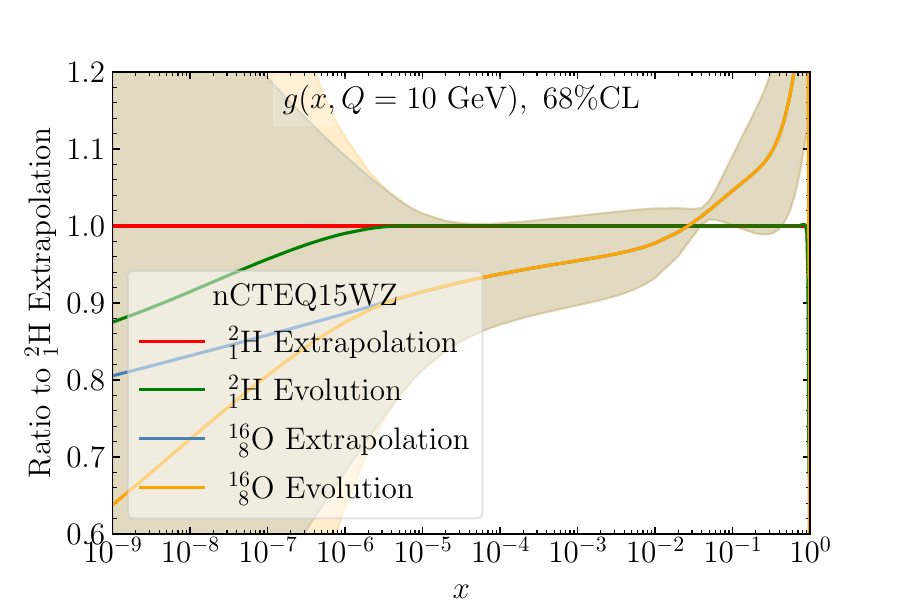}
\includegraphics[width=0.49\textwidth]{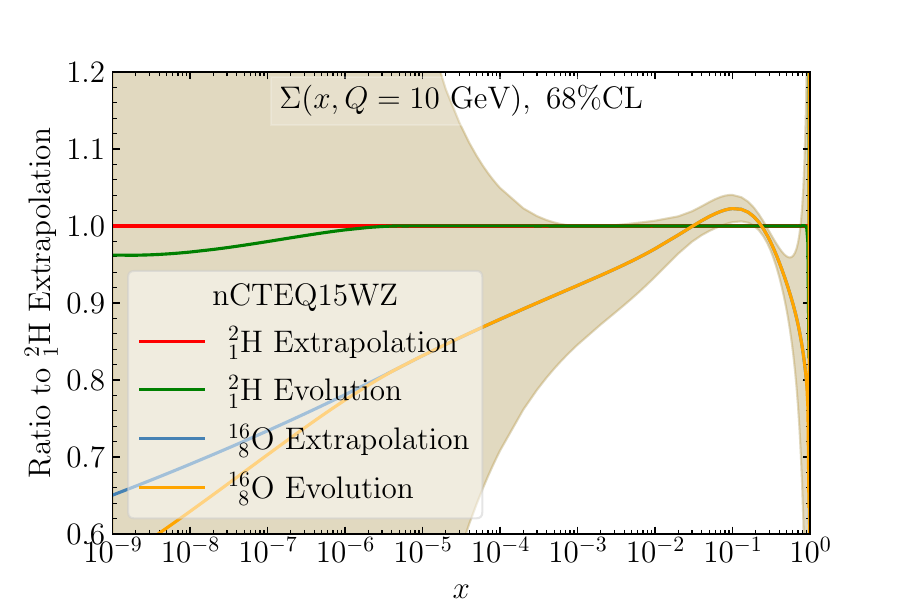}
    \caption{Comparisons of the gluon (left) and flavor-singlet $\Sigma=\sum_{i}^{n_f}(q_i+\bar{q}_i)$ (right) PDFs at $Q=10~\GeV$ for the nuclear $_{~8}^{16}$O with ``extrapolation" and ``evolution" of the EPPS21 (upper) and nCTEQ15WZ (lower) PDF grids.}
    \label{fig:extraPDFs}
\end{figure}

\begin{figure}
    \centering
    \includegraphics[width=0.49\textwidth]{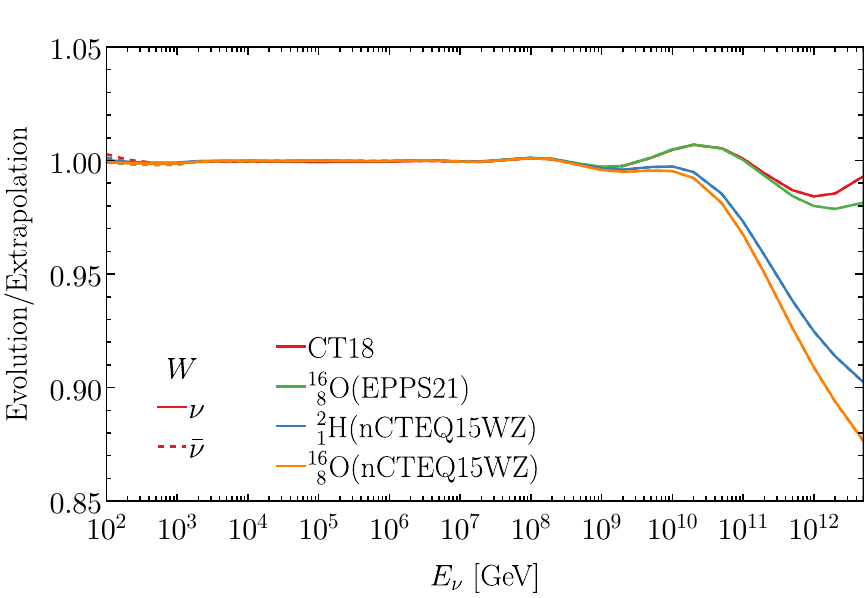}
    \includegraphics[width=0.49\textwidth]{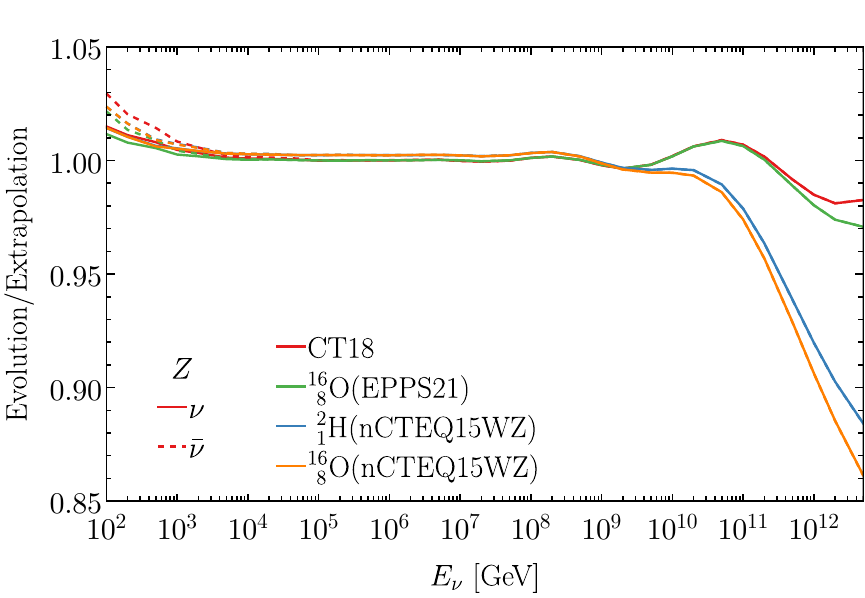}    
    \caption{Comparison of the neutrino-isoscalar charged $(W)$ and neutral $(Z)$ current DIS cross sections, obtained with the ``extrapolation" and ``evolution" methods for the CT18NNLO~\cite{Hou:2019efy}, EPPS21~\cite{Eskola:2021nhw}, and nCTEQ15WZ~\cite{Kusina:2020lyz} (n)PDFs.}
    \label{fig:evolve2extrap}
\end{figure}

We point out that the lower boundary in $x$ of the interpolation grids typically provided by nuclear PDF fits tends to be higher than the corresponding value in CT18 or other free-proton analyses; it should be stressed that this feature reflects both conventions as well as the reality that high-energy data have not yet constrained $A$-dependent nuclear PDFs to as low $x$ as HEP analyses of proton PDFs.\footnote{Here, $A$ denotes the nucleon numbers of the scattered nuclei.}
In Fig.~\ref{fig:xQ}, we indicate the lowest value in $x$ reached by grids of the EPPS21~\cite{Eskola:2021nhw} and nCTEQ~\cite{Kusina:2020lyz} nuclear PDFs, which we examine below to study the possible role of nuclear effects in DIS cross sections on nuclei.
It generally stands to reason that larger extrapolations over $x$ will potentially introduce correspondingly larger uncertainties on the integrated neutrino DIS cross sections.
In Fig.~\ref{fig:extraPDFs}, we compare the gluon and flavor-singlet $\Sigma=\sum_{i}^{n_f}(q_i+\bar{q}_i)$ nuclear PDFs for $_{~8}^{16}$O at $Q=10~\GeV$ based on the extrapolation and evolution approaches applied to EPPS21 (upper panels) and nCTEQ15WZ (lower panels). 
We see that the evolution and extrapolation methods agree exactly for the nuclear PDFs
in those regions of $x$ covered by interpolation grids, as shown. In contrast, at lower 
values of $x$ below those covered by public grids ({\it i.e.}, in the extrapolation 
region) the evolution method can induce a small variation with respect to the LHAPDF extrapolation~\cite{Buckley:2014ana}. As a consequence, the induced neutrino-nucleus cross section normalized to the pure extrapolation is shown in Fig.~\ref{fig:evolve2extrap}.
The corresponding variation for ultra-high energy neutrinos, with $E_\nu\gtrsim10^{8}~\GeV$, can be viewed as the uncertainty due to the low-$x$ extrapolation.
We see that the EPPS21 nuclear PDFs give a similar size for the extrapolation uncertainty in comparison with that based on CT18. For other PDFs, such as MSHT20~\cite{Bailey:2020ooq} or nCTEQ15WZ(SIH)~\cite{Kusina:2020lyz}, the extrapolation uncertainty can be larger.
In addition, Fig.~\ref{fig:evolve2extrap} also tells us the variation at low neutrino energy, such as $E_\nu\lesssim10^{3}~\GeV$, which results from the difference between the APFEL backward evolution approach and LHAPDF extrapolation below the DGLAP starting scale, \emph{i.e.}, $1<Q<Q_0=1.3~\GeV$.
We see that this variation at $E_\nu=100~\GeV$ can be negligible for charged-current DIS, while $1\%\sim3\%$ for the neutral current case.

\begin{figure}
    \centering
    \includegraphics[width=0.49\textwidth]{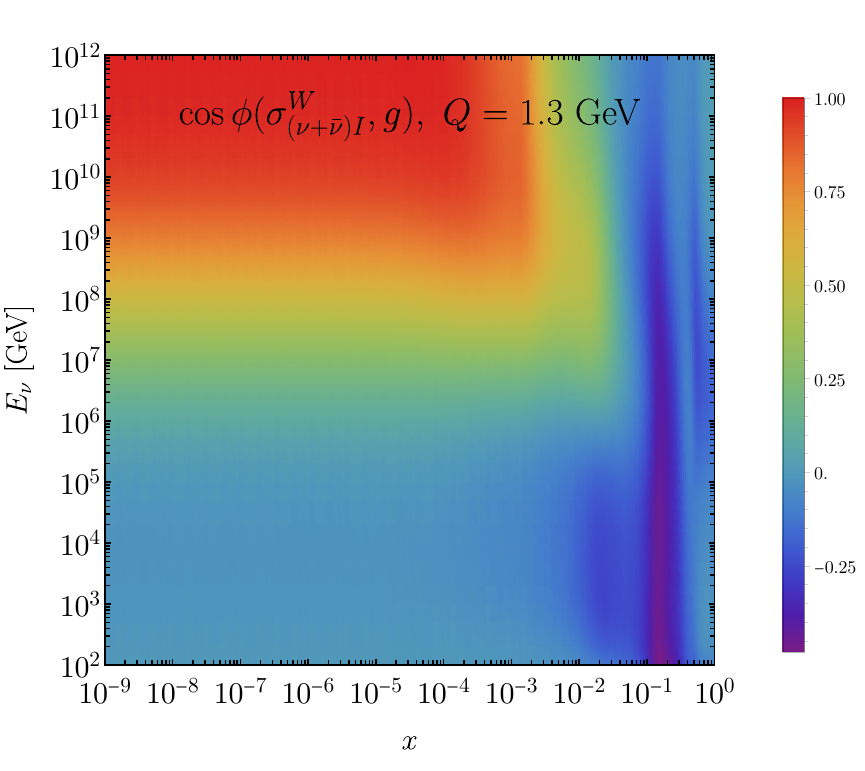}
    \includegraphics[width=0.49\textwidth]{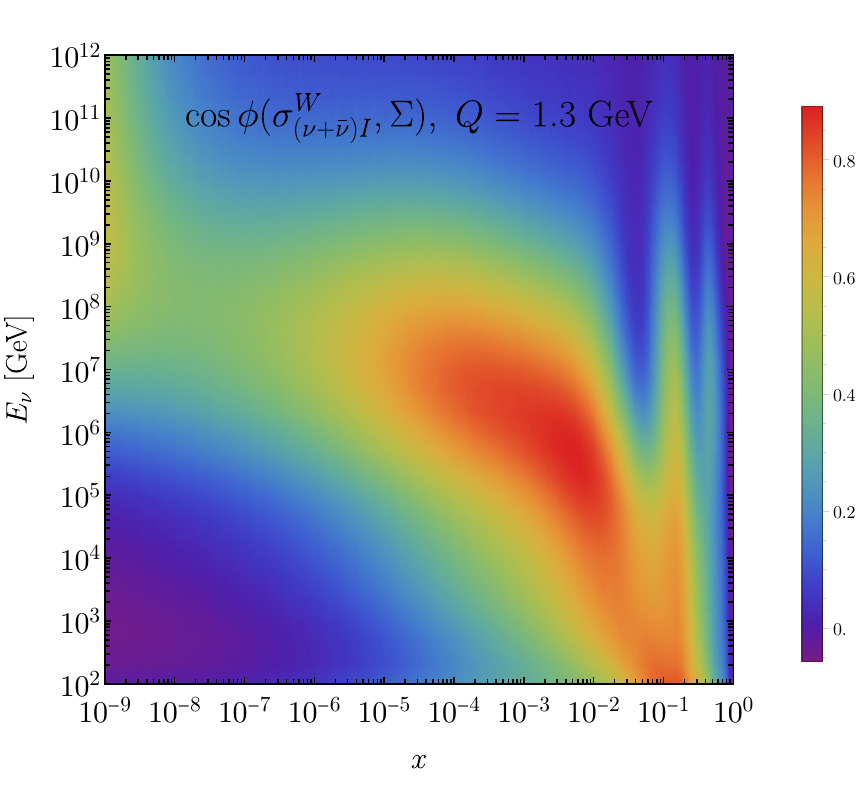}
    \includegraphics[width=0.49\textwidth]{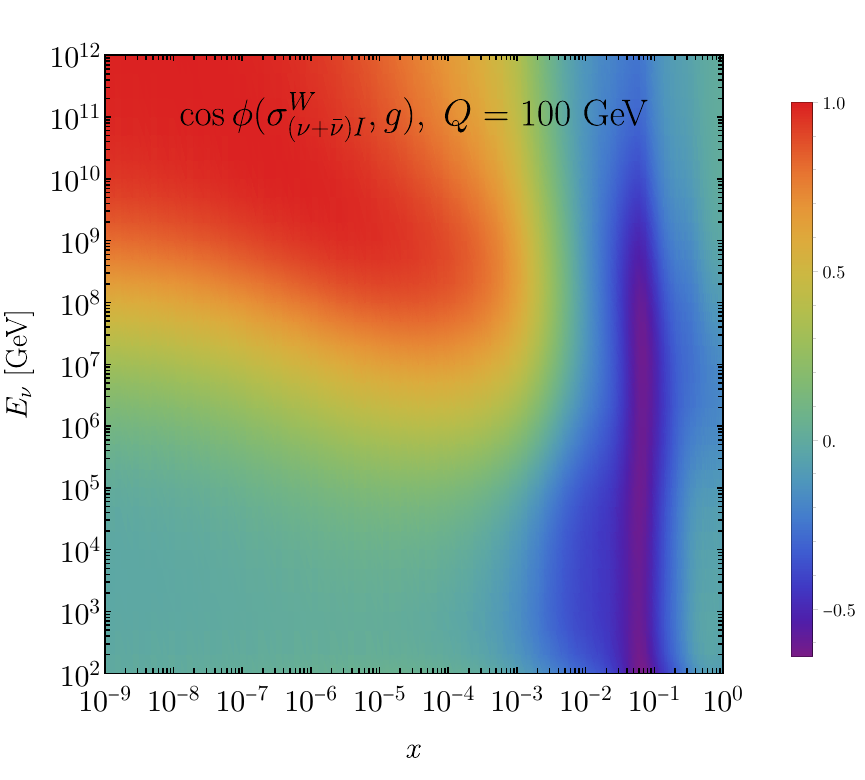}
    \includegraphics[width=0.49\textwidth]{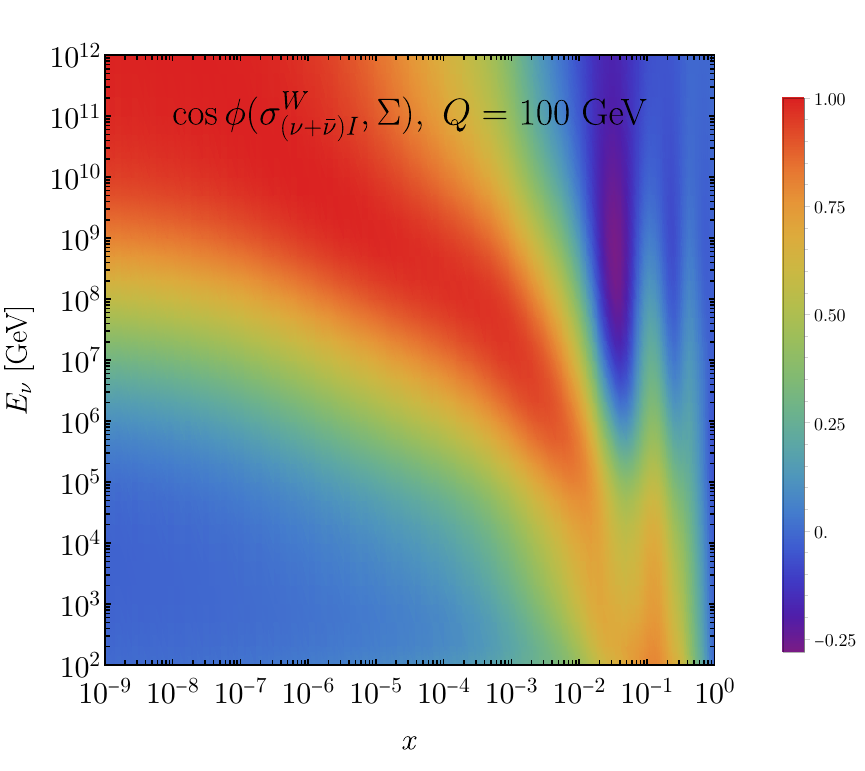}
    \caption{Correlation cosine angles between the neutrino-isoscalar charge-current cross sections (sum of neutrino and antineutrino) and the CT18 gluon (left) and flavor-singlet (right) PDFs at $Q=1.3~\GeV$ (upper) and $100~\GeV$ (lower). The color scale, indicated at the right of each panel, provides the value of $\cos \phi$; note that these scales differ slightly
    between the two panels.}
    \label{fig:corr}
\end{figure}

Until this point, we have been able to include the relevant phase space in the inclusive neutrino scattering cross section as, in practice, 
\begin{equation}\label{eq:xQ2}
    Q\in[Q_{\min},\sqrt{2m_NE_\nu}]\ \ \mathrm{and}\ \ x\in[Q^2/(2m_NE_\nu),1]\, ,
\end{equation}
demonstrated in Fig.~\ref{fig:xQ}.
Based on this setup, we also show the Hessian correlation~\cite{Gao:2013bia}, dubbed as correlation cosine $\cos\phi$ between the neutrino-isoscalar cross sections and the gluon and flavor-singlet PDFs at $Q=1.3$ and 100 GeV in Fig.~\ref{fig:corr} for charged-current DIS.
Here, the neutrino and antineutrino cross sections are summed.
We see that, at relatively low energy, a strong correlation appears in the singlet PDFs near $x\sim10^{-2}\!-\!10^{-1}$. With increasing energy, the sensitive momentum fraction $x$ decreases accordingly, roughly scaling as $x\sim x_{W}=M_{W}^2/(2m_NE_\nu)$, which confirms the important kinematics as examined in App.~\ref{app:xQlimit}.
Similar features have been found for the neutral-current DIS as well.
When the neutrino energy goes above $E_\nu\gtrsim10^{9}~\GeV$, 
the largest sensitivity comes from the gluon PDFs, below the region $x\lesssim10^{-4}$, reflecting the important contribution from the gluon partons. As also shown in Fig.~\ref{fig:corr}, at $Q=100~\GeV$, far above the $Q_0$ scale, we also see a correlation with the flavor-singlet PDFs at $x<10^{-4}$, reflecting the co-evolution of the singlet and gluon PDFs in the DGLAP evolution.

We point out that experiments usually require the invariant mass, $W$, of the recoil system to be in the perturbative region, \emph{e.g.}, $W\!\geq\! W_0\!=2~\GeV$ as for MINERvA~\cite{MINERvA:2016oql}, in order to minimize the contributions from quasielastic scattering and the excitation of hadronic resonances. This $W$ cut excludes a small region in the otherwise permissible phase space as shown in the lower-right corner of Fig.~\ref{fig:xQ}, since
\begin{equation}
    x\leq\frac{Q^2}{W_0^2-m_N^2+Q^2} \leq 1
\end{equation}
for $W^2_0\! \geq\! m_N^2$. However, the removal of this small region only negligibly affects the (anti)neutrino cross section, since PDFs in the high $x\to1$ limit must rapidly vanish for reasons of momentum conservation as typically parametrized by assuming an overall $(1-x)^{\beta}$ (with $\beta>0$) polynomial behavior. Therefore, we take the upper integration limit $x_{\max}=1$ in Eq.~(\ref{eq:xQ2}) in practice.

\subsection{Perturbative orders up to an approximate N3LO}
\label{sec:aN3LO}
In the demonstration above, we used the zero-mass scheme for structure functions at next-to-next-to-leading order (NNLO). 
In this section, we explore possible variations arising from performing the calculation at different orders in perturbation theory.

\begin{figure}
    \centering
    \includegraphics[width=0.49\textwidth]{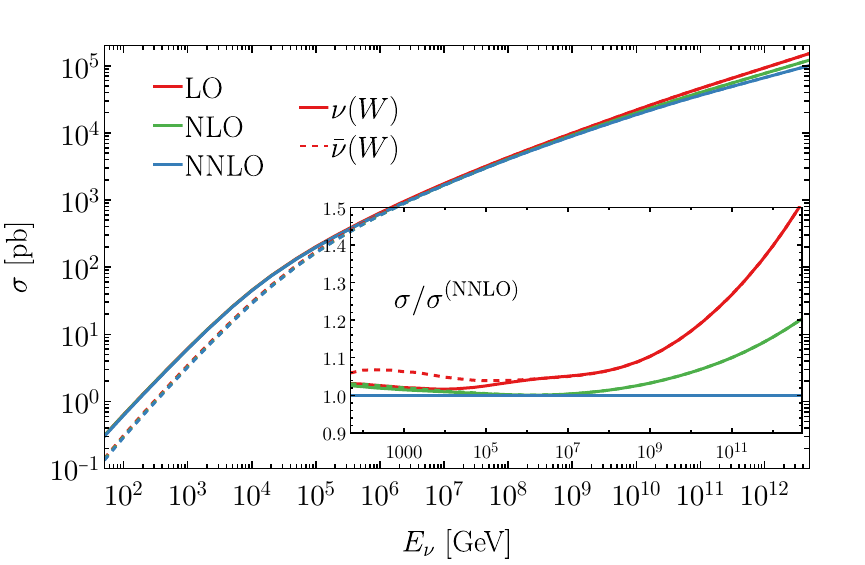}
    \includegraphics[width=0.49\textwidth]{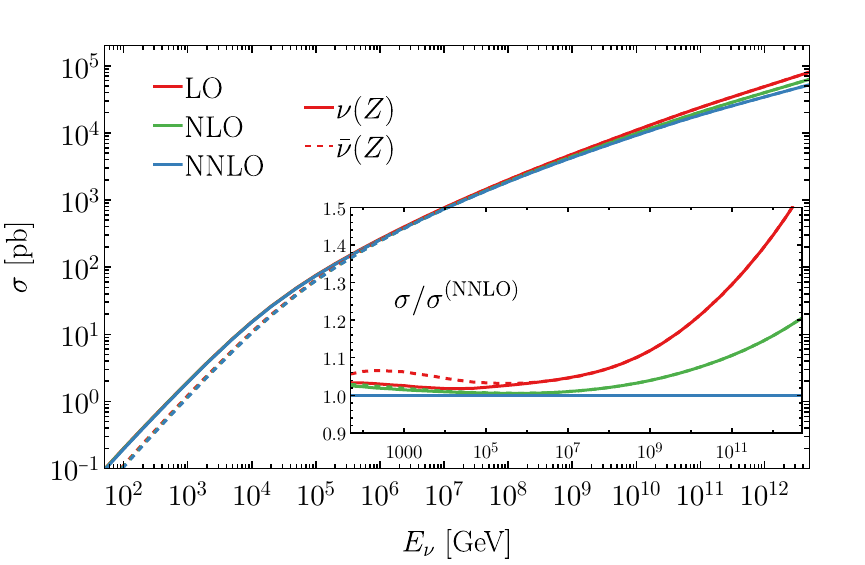}
    \caption{Comparison of the neutrino-isoscalar charged (left) and neutral (right) current DIS cross-sections at different orders of QCD perturbation theory and the ratios with respect to the corresponding NNLO predictions. The LO and NLO cross sections are calculated with the corresponding CT18 PDFs at LO~\cite{Yan:2022pzl} and NLO~\cite{Hou:2019efy}.
    }\label{fig:RLO}
\end{figure}

The CT18 global analysis is performed at both NLO and NNLO, with the corresponding error sets determined with the Hessian method~\cite{Hou:2019efy}. Afterward, the leading order analysis was released based on a few special considerations to improve the quality of fit~\cite{Yan:2022pzl}. In Fig.~\ref{fig:RLO}, we show the CT18 predictions for the neutrino-isoscalar charged and neutral current DIS cross sections at different perturbative QCD orders, with the numerical values of the NNLO predictions presented in Tabs.~\ref{tab:ccscheme}-\ref{tab:ncscheme} of App.~\ref{app:num}. Here, the LO~\cite{Yan:2022pzl} and NLO~\cite{Hou:2019efy} PDFs are adopted with the same corresponding order of Wilson coefficients in the LO and NLO cross sections.
Compared to NNLO, we see that the LO and NLO predictions are larger, and the increments become increasingly significant when $E_\nu>10^7~\GeV$. These differences are mainly driven by two factors: the strong coupling $\alpha_s$ at different orders and higher-order corrections to the structure functions. In the CT18LO analysis~\cite{Yan:2022pzl}, the strong coupling is chosen to be a larger value, $\alpha_s(M_Z)=0.135$ in order to compensate for the missing higher order corrections, especially for the Drell-Yan data. In the CT18 NLO fit, the strong coupling is determined as $\alpha_s(M_Z)=0.118$, similar to the NNLO fit. 
However, when the scale $Q$ runs down to lower values, which contribute to the majority of the total integrated neutrino-nucleon cross section as shown in Fig.~\ref{fig:scanQ2} in App.~\ref{app:xQlimit}, the corresponding strong couplings are larger than that obtained at NNLO. On the other hand, the higher-order corrections contribute negatively to the structure functions. (See Fig.~11 of Ref.~\cite{Xie:2021equ} for a specific example.) As a consequence, we obtain NNLO cross sections smaller than the LO and NLO ones, as shown in Fig.~\ref{fig:RLO}. 

\begin{figure}
\centering
\includegraphics[width=0.49\textwidth]{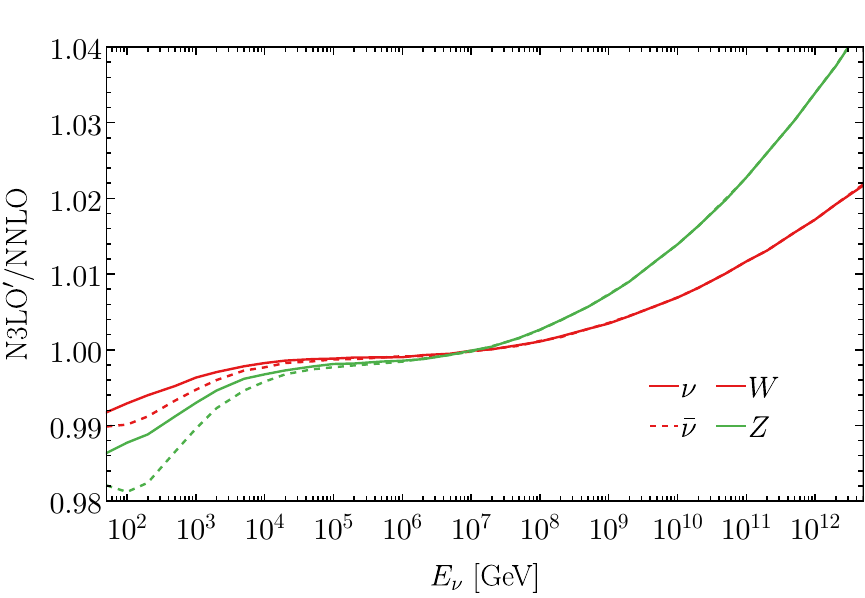}
\caption{The N3LO$'$ ZM-VFN predictions to the (anti)neutrino-isoscalar charged $(W)$ and neutral $(Z)$ current DIS cross sections, with respect to the NNLO ones.}
\label{fig:N3LOPDFs}
\end{figure}

As we see above, the neutrino DIS cross section decreases as the perturbative order increases. A natural question one may ask is whether our result shows evidence of convergence in the perturbative expansion. 
Similarly to our previous work~\cite{Gao:2021fle}, we estimate an \emph{approximate} N3LO contribution (denoted as N3LO$'$) with the ZM N3LO structure functions implemented in the \texttt{v1.2.0-struct-func-devel} version of HOPPET~\cite{Salam:2008qg}.
The resulting cross sections normalized to the NNLO ones as $K_{\textrm{N3LO}'}$ factors (N3LO$'$/NNLO ratios) are displayed in Fig.~\ref{fig:N3LOPDFs} and the numerical value tabulated in Tabs.~\ref{tab:ccscheme}-\ref{tab:ncscheme}. 
We see that N3LO$'$ corrections give a small decrease at low energy, while a slightly larger increase at high energy. The size of the difference is much smaller than that of the NNLO/NLO one as shown in Fig.~\ref{fig:RLO}.
This result gives us confidence that, in terms of the perturbative expansion, our predictions are already showing strong evidence of convergence at NNLO. In the rest of this work, we will adopt the NNLO calculation as our baseline, with N3LO$'$ corrections included with the $K_{\textrm{N3LO}'}$ factors.

\subsection{A general-mass variable-flavor-number scheme}
\label{sec:ACOT}
Up to now, we have adopted the ZM-VFN scheme with CT18 PDFs, which include $n_f=4$ parton flavors. One may wonder about the heavy-quark mass effects as well as the potential contribution from the third-generation quarks, especially when $\sqrt{s}\gg m_{b,t}^2$, which we will examine carefully in this subsection.

The heavy-quark mass corrections to NC DIS at NNLO have been obtained by several groups within corresponding general-mass schemes, such as the ACOT scheme employed in the CTEQ-TEA group~\cite{Guzzi:2011ew}, the FONLL scheme by the NNPDF group~\cite{Forte:2010ta}, and the optimal TR scheme in the MSHT group~\cite{Thorne:2012az}.
For CC DIS, the asymptotic heavy-quark corrections at large momentum transfers ($Q\gg m$) to the structure functions are known up to $\mathcal{O}(\alpha_s^2)$~\cite{Blumlein:2014fqa,Blumlein:2016xcy} and even $\mathcal{O}(\alpha_s^3)$~\cite{Behring:2015roa,Behring:2016hpa}.
Recently, the complete mass corrections to CC DIS have been achieved up to NNLO~\cite{Berger:2016inr,Gao:2017kkx}. The general-mass corrections to the structure functions have been implemented in the ACOT framework~\cite{Gao:2021fle}, which we will mainly rely on in this work.

\begin{figure}
    \centering
    \includegraphics[width=0.48\textwidth]{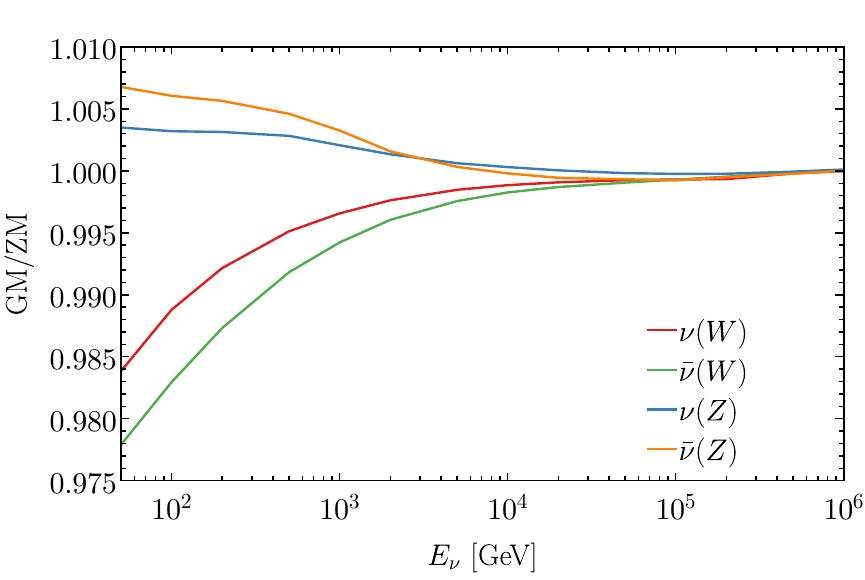}
    \includegraphics[width=0.48\textwidth]{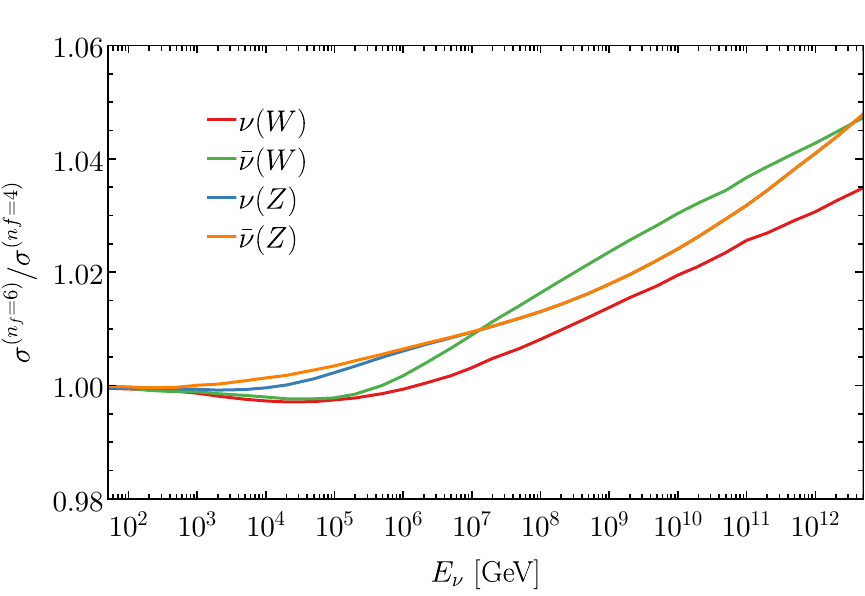}
    \caption{The cross-section ratios of (anti)neutrino-isoscalar charged-current DIS in the General-Mass and Zero-Mass schemes (left) and different flavors (right).}
    \label{fig:ACOT}
\end{figure}

In Fig.~\ref{fig:ACOT} (left), we show the charm mass correction to the (anti)neutrino cross sections, with the ACOT general-mass scheme normalized to the ZM-VFN scheme up to $n_f=4$ flavors, with the numerical values as $K_{\rm GM}$ factors (GM/ZM ratios) presented in Tabs.~\ref{tab:ccscheme}-\ref{tab:ncscheme}.
At lower neutrino energies, \emph{e.g.}, $E_\nu=10^{2}~\GeV$, the full charm-mass dependent structure functions reduce the antineutrino (neutrino) CC DIS cross section by about 2\% (1\%).
In comparison, the massive corrections to the NC DIS cross section can be positive, mainly driven by an enhancement of the $F_2$ structure function at low scales, such as $Q\lesssim2m_c$.
(See Fig.~5 of Ref.~\cite{Guzzi:2011ew} for details.)
The size of the impact is smaller than that in the CC case, with 0.6\% (0.3\%) for antineutrino (neutrino) NC DIS cross sections, respectively. 
We remind the reader that the charm-mass corrections to antineutrino cross sections are always larger than the neutrino ones, mainly because of the relatively smaller absolute antineutrino cross section with respect to the analogous neutrino calculation.\footnote{We note that the apparent equality of the NC neutrino and antineutrino GM/ZM ratios near $E_\nu\sim3~\TeV$ is an artificial smoothing effect due to numerical imprecision at the permille level.}
Meanwhile, the mass effect vanishes very quickly with increasing neutrino energy. With $E_\nu\gtrsim10^{4}~\GeV$, the general-mass results are almost identical to those obtained in the ZM scheme. 
For this reason, we will assume the ZM-VFN scheme for the remainder of this work, with heavy-quark mass effects folded in via $K_{\rm GM}$-factors in Tabs.~\ref{tab:ccscheme}-\ref{tab:ncscheme}.

The CT18 PDFs adopt $n_f=5$ as their default~\cite{Hou:2019efy}. For $n_f\!\geq\!5$, the CC scatterings involve the $gW^\pm\to t\bar{b}(\bar{t}b)$ partonic sub-processes beginning at NLO; these contain a collinear singularity in the $g\to b\bar{b}(t\bar{t})$ splittings in the zero-mass limit.
The $g\to b\bar{b}$ collinear divergence can be absorbed into a redefinition of $b$-quark PDF. Similarly, we have to introduce $t$-quark PDF to absorb the $g\to t\bar{t}$ collinear divergence when the top quark becomes massless.
For this reason, we take $n_f=6$ to include the third-generation quarks in the zero-mass scheme beyond the leading order consistently. The $n_f=6$ cross sections normalized to the $n_f=4$ ones are shown in Fig.~\ref{fig:ACOT} (right), with the numerical values as $K_{n_f}$ factors in Tabs.~\ref{tab:ccscheme}-\ref{tab:ncscheme}.
We remind that the $n_f=4$ and $n_f=6$ PDFs are obtained with the same CT18 parameterization as well as strong coupling at the starting scale $Q_0$. In such a way, the light-flavor PDFs remain identical below the heavy-flavor mass. The difference only appears once the factorization scale crosses the corresponding partonic threshold when the heavy quarks become active. 

In Fig.~\ref{fig:ACOT} (right), we see that the $n_f=6$ scheme gives at most about 5\% larger cross sections
at $E_\nu=10^{12}~\GeV$ than the $n_f=4$ ones for the CC (NC) DIS processes. 
For the NC scattering, the $n_f=6$ enhancement from the contribution of $b,t$ partons is identical at high (anti)neutrino energies. In comparison, at low energies, the neutrino enhancement is slightly smaller than the antineutrino one, due to the correspondingly larger absolute NC cross section as mentioned before. Similarly, the larger neutrino CC cross section leads the corresponding $n_f=6$ enhancement to be smaller than the antineutrino one at high energy.

Starting from this point, we will take $n_f=6$ as the default to include the bottom and top parton's contribution in the rest of this work. 
Similarly to the charm mass effect as examined above, the top/bottom mass can slightly reduce the neutrino-nucleon cross section. However, considering the relatively small contribution from the third-general quark PDF even at extremely high energy, we can safely neglect the mass effect without a noticeable effect.

\section{PDF uncertainties: proton and nuclear PDFs at low $x$}
\label{sec:PDFs}
\subsection{Small-$x$ resummation}
\label{sec:smallx}

As we see in Sec.~\ref{sec:xQ}, the neutrino cross section in the ultrahigh-energy region is very sensitive to the PDFs, especially gluon, in the small $x$ region. On general grounds, for Bjorken $x_\mathrm{B}\!<\!10^{-3}$ and momentum transfer $Q$ around a few GeV, we would expect that the small-$x$ logarithms will be enhanced, and eventually enter a partonic saturation phase as $x\to0$. In the NNPDF~\cite{Ball:2017otu} and xFitter~\cite{xFitterDevelopersTeam:2018hym} frameworks, a small-$x$ logarithm has been resummed up to the next-to-leading level (NLLx) based on BFKL dynamics~\cite{Balitsky:1978ic,Balitsky:1979ns,Fadin:1975cb,Lipatov:1976zz,Kuraev:1976ge,Kuraev:1977fs}.
It is found that the description of the low-$x$ DIS data, especially data measured at HERA~\cite{H1:2013ktq,H1:2015ubc}, has been improved. 

\begin{figure}
\centering
\includegraphics[width=0.49\textwidth]{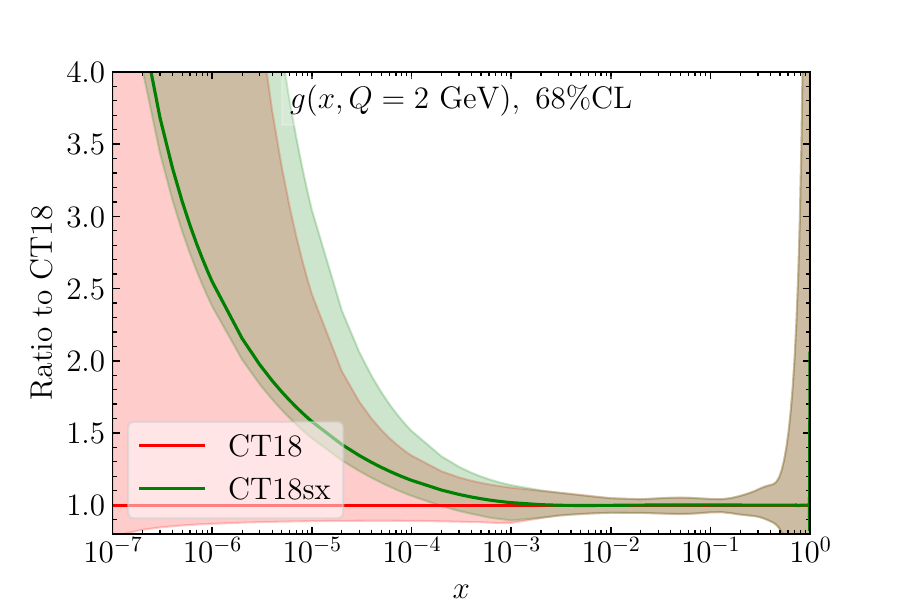}
\includegraphics[width=0.49\textwidth]{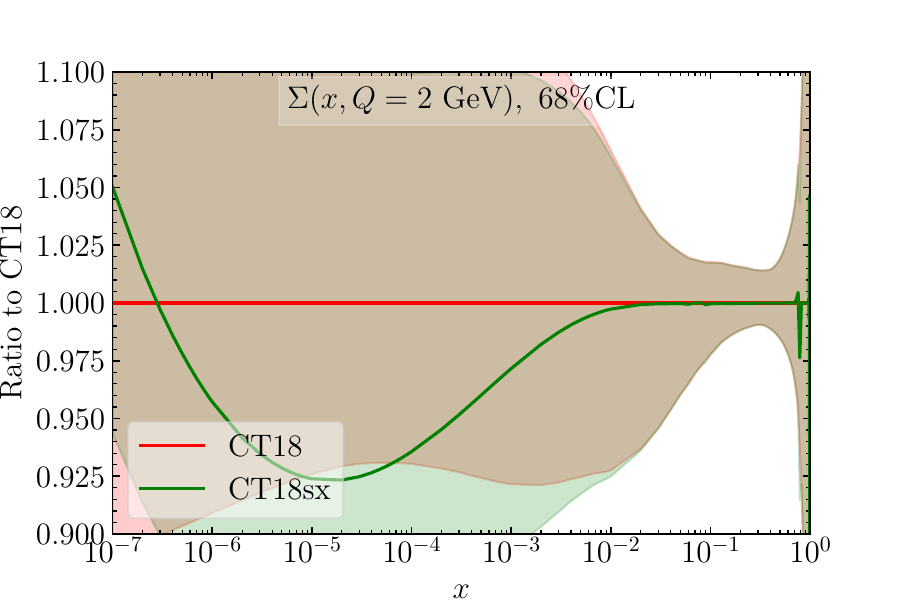}
\includegraphics[width=0.49\textwidth]{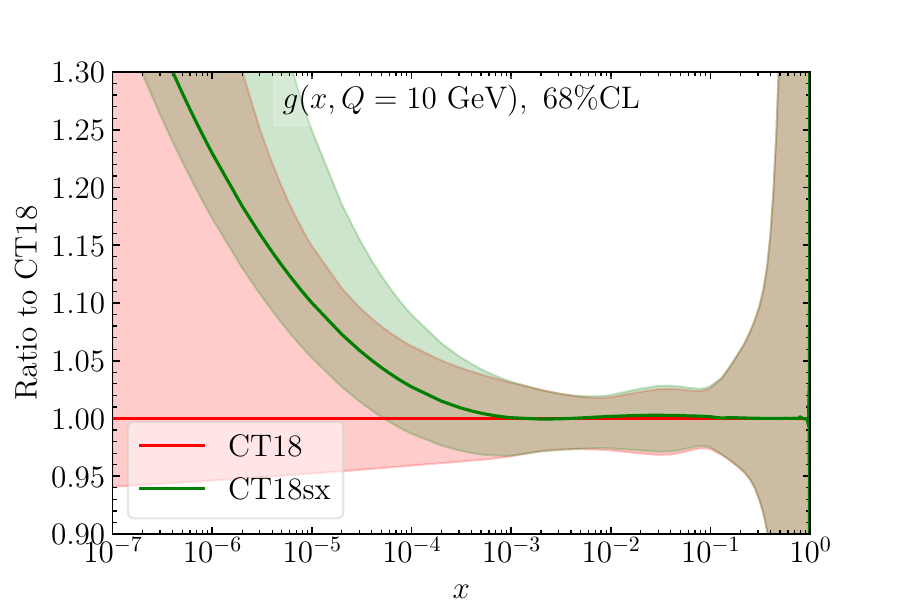}
\includegraphics[width=0.49\textwidth]{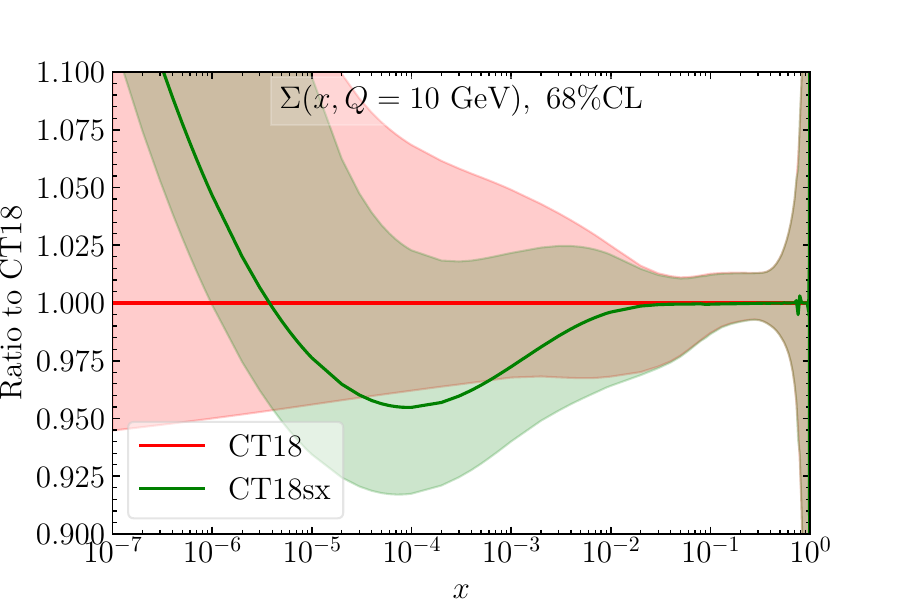}
\includegraphics[width=0.49\textwidth]{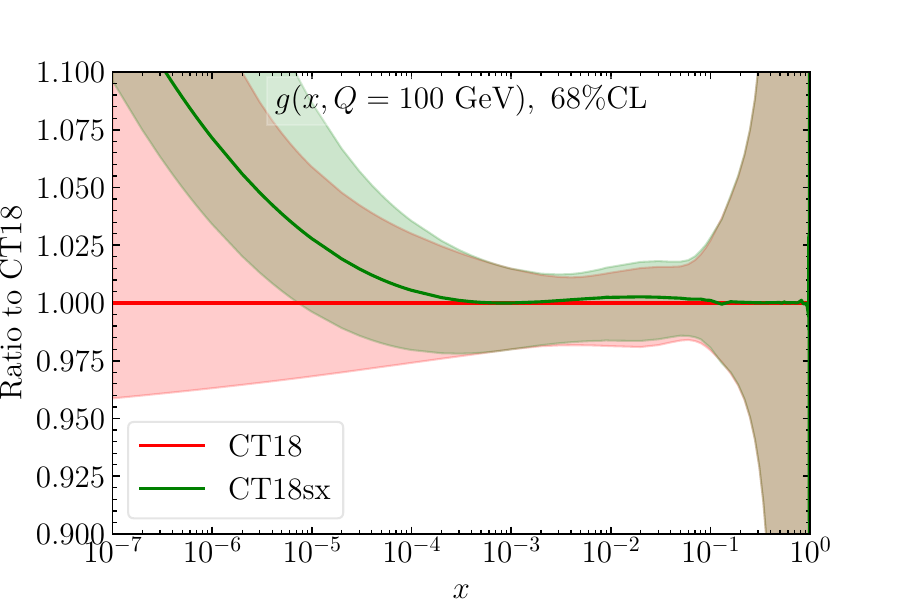}
\includegraphics[width=0.49\textwidth]{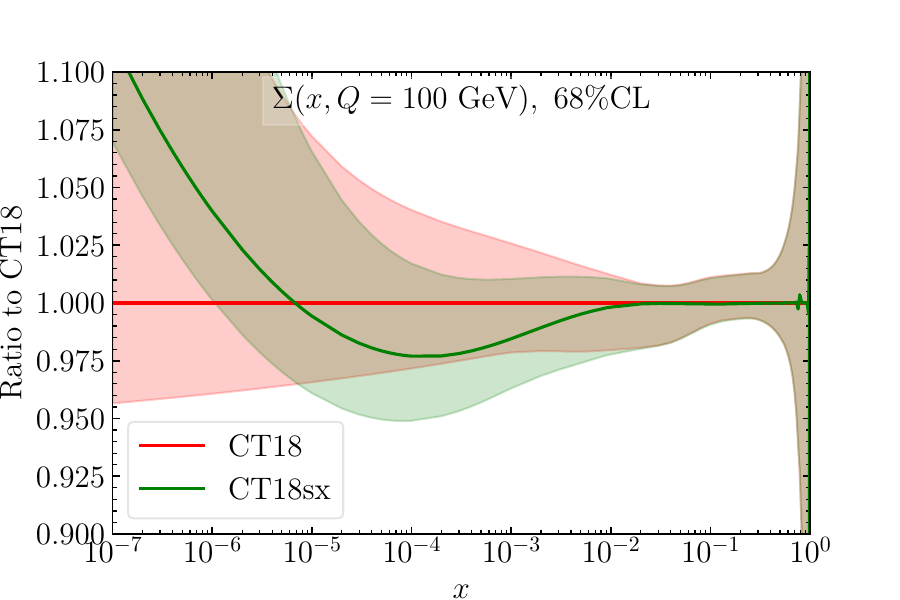}
    \caption{The comparison of gluon and flavor-singlet PDFs for CT18 and CT18sx at the scales $Q=2,10,100$ GeV, respectively.
    }
    \label{fig:PDFs}
\end{figure}

In this work, we have interfaced the HELL~\cite{Bonvini:2016wki,Bonvini:2017ogt} package with APFEL~\cite{Bertone:2013vaa} to resum large $\sim\!\log(1/x)$ logarithmic corrections for parton evolution up to NNLx with matching to the NNLO DGLAP evolution; we dub the resulting distributions as CT18sx PDFs~\cite{Guzzi:2021fre}.
The gluon and singlet PDFs compared to the nominal CT18 ones are displayed in Fig.~\ref{fig:PDFs}.
Small-$x$ resummation enhances the gluon PDF but reduces singlet PDFs in the small-$x$ region at low factorization scales. However, this small-$x$ variation gradually dies out at larger scales, as can be observed by comparing the gluon PDF at $Q=2~\GeV$ up to that at $Q=100~\GeV$. A similar effect can be obtained with an $x$-dependent DIS scale, motivated by the partonic saturation model~\cite{Golec-Biernat:1998zce}.

\begin{figure}[h]
    \centering
    \includegraphics[width=0.48\textwidth]{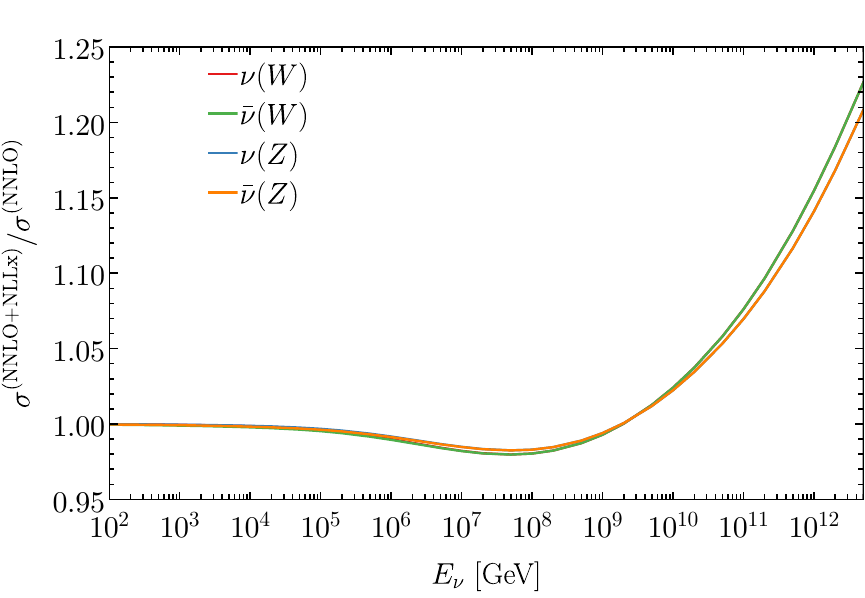}   
    \caption{The (anti)neutrino-isoscalar cross sections for charged $(W)$ and neutral $(Z)$ current DIS, with small-$x$ resummed up to next-to-leading logarithms (NLLx) with respect to NNLO predictions. 
    }
    \label{fig:smallx}
\end{figure}

Applying the correspondingly matched DIS Wilson coefficient functions provided by the HELL framework~\cite{Bonvini:2016wki,Bonvini:2017ogt}, we explored the small-$x$ resummed (anti)neutrino cross sections normalized to the NNLO fixed-order ones, as shown in Fig.~\ref{fig:smallx}, with the numerical values tabulated as $K_{\rm NLLx}$ factors in Tabs.~\ref{tab:ccscheme}-\ref{tab:ncscheme}. At first sight, we see that the small-$x$ resummed cross section is almost identical for the neutrino and antineutrino DIS cross sections.
A closer examination reveals that at low neutrino energies $(E_\nu\lesssim10^{5}~\GeV)$, the small-$x$ resummation has no impact at all, as the corresponding kinematics only cover the intermediate to the large-$x$ region, cf. Fig.~\ref{fig:xQ}.
Starting around $E_\nu\sim10^{6}~\GeV$, we obtain a slight reduction of the resummed cross section, as a result of the smaller quark PDFs, indicated in Fig.~\ref{fig:smallx}.
A turnover appears around $E_\nu\sim10^{8}~\GeV$, which roughly corresponds to $Q=M_W$ and $x=10^{-5}\sim10^{-4}$ according to the important $(x,Q)$ kinematics, as explained in App.~\ref{app:xQlimit}.
When the neutrino energy continues to increase above $E_\nu\gtrsim10^{10}~\GeV$, we get a cross-section \emph{enhancement}, mainly driven by the enlarged gluon PDF in the extremely small-$x$ region. 
When $E_\nu\sim10^{12}~\GeV$, the small-$x$ resummed enhancement can be as large as 10\%. Due to this reason, we will adopt the small-$x$ resummed cross section in the main presentation of this work.

\begin{figure}
\centering
\includegraphics[width=0.49\textwidth]{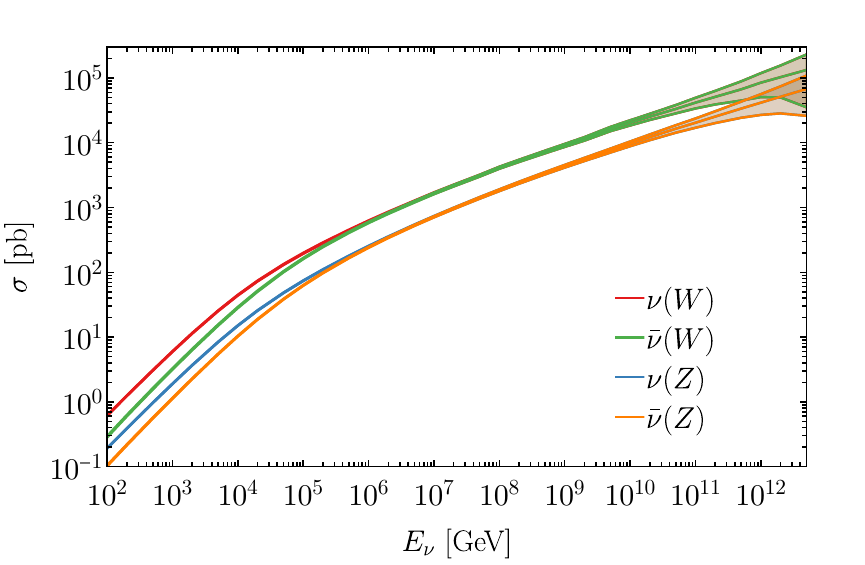}
\includegraphics[width=0.49\textwidth]{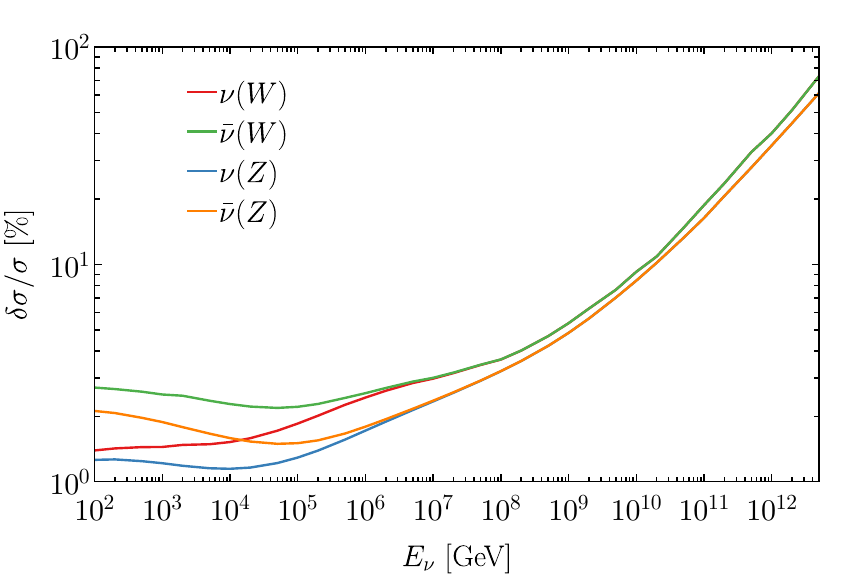}   
    \caption{(Left) The CT18 predictions for neutrino-isoscalar scattering cross sections $\sigma$; and (right) the corresponding PDF uncertainties, $\delta\sigma/\sigma$. 
    }
    \label{fig:NuNXSec_CT18}
\end{figure}

In Fig.~\ref{fig:NuNXSec_CT18}, we display the final CT18 predictions on the neutrino-isoscalar scattering cross sections and the corresponding PDF uncertainties.
As discussed above, we take the ZM-VFN scheme up $n_f=6$ and the small-$x$ resummation as a baseline, with $K$ factors from the approximate N3LO$'$ fixed-order perturbation, the heavy-quark mass corrections in the ACOT scheme, as tabulated in Tab.~\ref{tab:ccscheme}-\ref{tab:ncscheme}. The specific numerical results are presented in Tab.~\ref{tab:ccdis}-\ref{tab:ncdis} of App.~\ref{app:num}.
In general, the charged current-cross sections are a few times larger than the neutral current ones. The high-energy neutrino and antineutrino cross sections converge as a result of the asymptotically identical quark and antiquark densities in the small-$x$ region.
At low (anti)neutrino energy, the neutrino cross sections are larger than the antineutrino ones, as a result of valence contribution to $xF_3$ at large $x$ as indicated in Eqs.~(\ref{eq:F23CC}-\ref{eq:F23NC}). The antineutrino DIS gives larger relative PDF uncertainty than the neutrino one, mainly as a result of the larger antiquark (\emph{i.e.}, sea quark) uncertainty with respect to the quark (mainly valence) one in $xF_3$.

\begin{figure}
\centering
\includegraphics[width=0.49\textwidth]{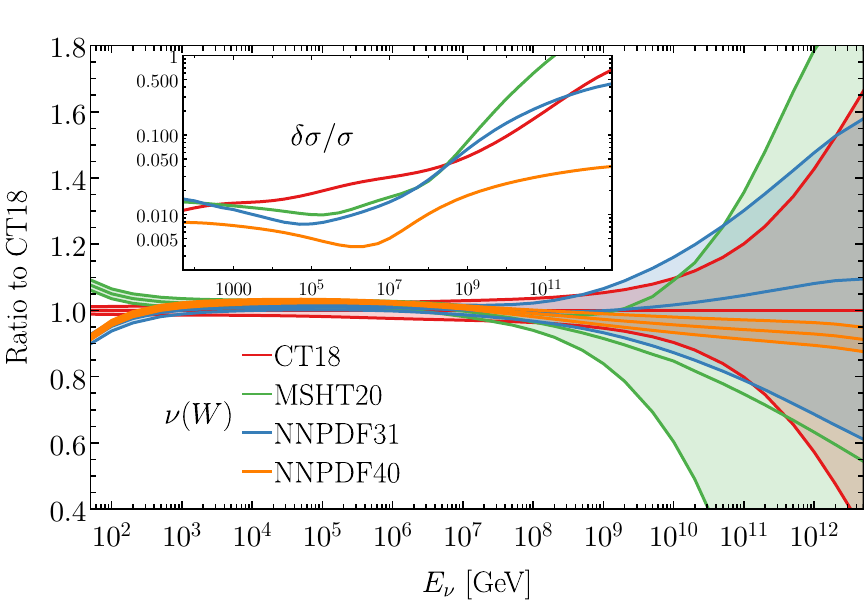}
\includegraphics[width=0.49\textwidth]{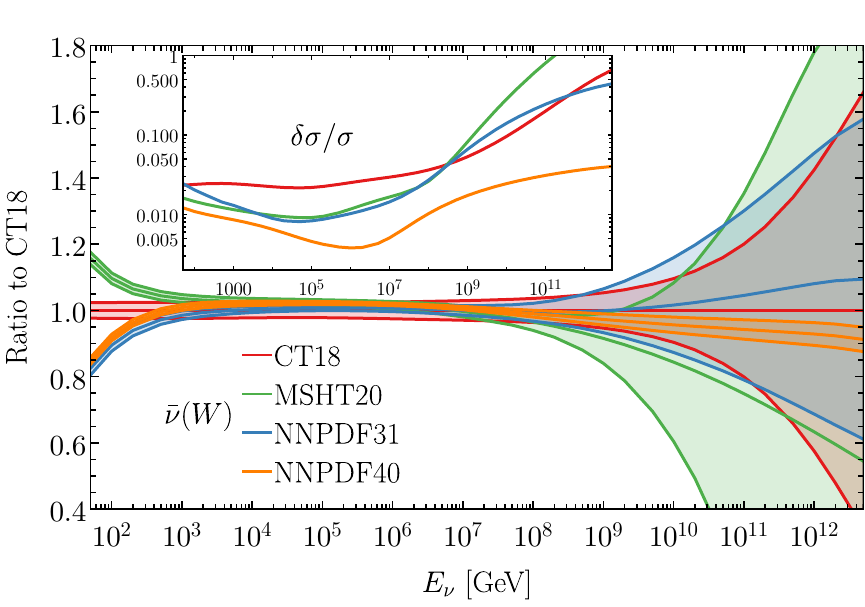}
\includegraphics[width=0.49\textwidth]{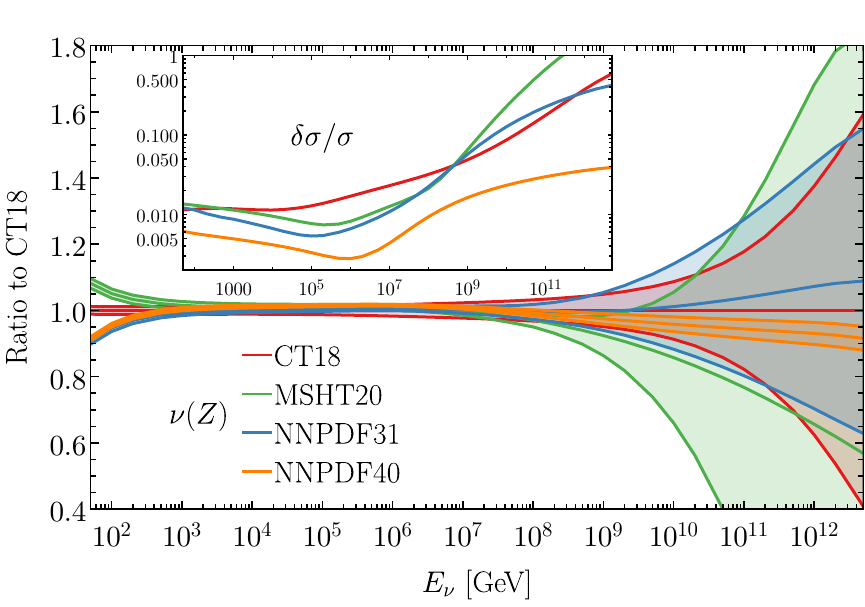}
\includegraphics[width=0.49\textwidth]{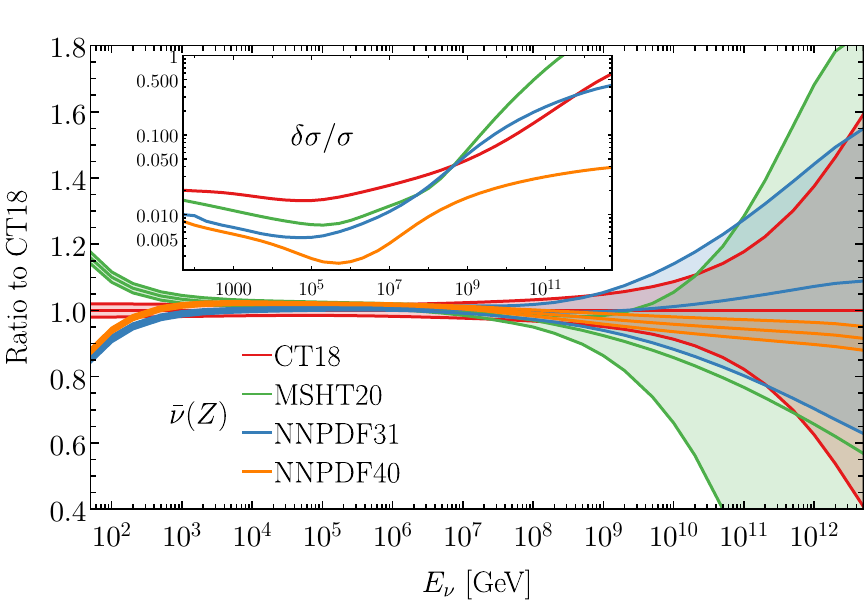}  
\caption{The comparison of the predictions of neutrino-isoscalar scattering cross sections as well as the uncertainties of modern PDFs from CT18, MSHT20, and NNPDF3.1/4.0, respectively.
}
\label{fig:PDFunc}
\end{figure}

In Fig.~\ref{fig:PDFunc}, we compare the predictions of neutrino-isoscalar scattering cross sections of modern PDFs, CT18~\cite{Hou:2019efy}, MSHT20~\cite{Bailey:2020ooq}, NNPDF3.1~\cite{NNPDF:2017mvq} and NNPDF4.0~\cite{NNPDF:2021njg}, as well as the corresponding PDF uncertainties.
We see that all four modern PDFs give overall consistent predictions when $10^{3}\lesssim E_\nu\lesssim10^{8}~\GeV$. 
At a low neutrino energy when $E_\nu\lesssim10^{3}~\GeV$, MSHT20 gives larger cross sections while the NNPDF3.1 and NNPDF4.0 give smaller ones, with respect to the CT18 predictions. At an ultrahigh energy when $E_\nu\gtrsim10^{8}~\GeV$, the NNPDF3.1 and 4.0 cross sections are more or less comparable to the CT18 ones. In comparison, MSHT20 gives sizably smaller predictions, reflecting the extrapolated low-$x$ PDF behavior when $x<10^{-6}$. In the insets of Fig.~\ref{fig:PDFunc}, we normalize the PDF uncertainties to the corresponding central sets to show the relative error sizes. The CT18 PDF uncertainties are shown in Fig.~\ref{fig:NuNXSec_CT18}, already. In comparison with the CT18 ones, we see that MSHT20 and NNPDF3.0 give slightly smaller error bands in the neutrino energy range $E_\nu\lesssim10^{8}~\GeV$, resulted from the corresponding smaller PDF uncertainties. Many PDF comparisons can be found in Refs.~\cite{PDF4LHCWorkingGroup:2022cjn,Amoroso:2022eow}.
In the ultrahigh energy region $E_\nu\gtrsim10^{8}~\GeV$, MSHT20 gives larger PDF error bands, also driven by the low-$x$ extrapolation, as discussed in Sec.~\ref{sec:xQ}. As a final remark, the NNPDF4.0 gives noticeably smaller uncertainties than others, while the reliability of the PDF error quantification still remains as a puzzle~\cite{Courtoy:2022ocu}.

\subsection{Nuclear PDFs and uncertainties}
\label{sec:nuclear}
\begin{figure}
\centering
\includegraphics[width=0.49\textwidth]{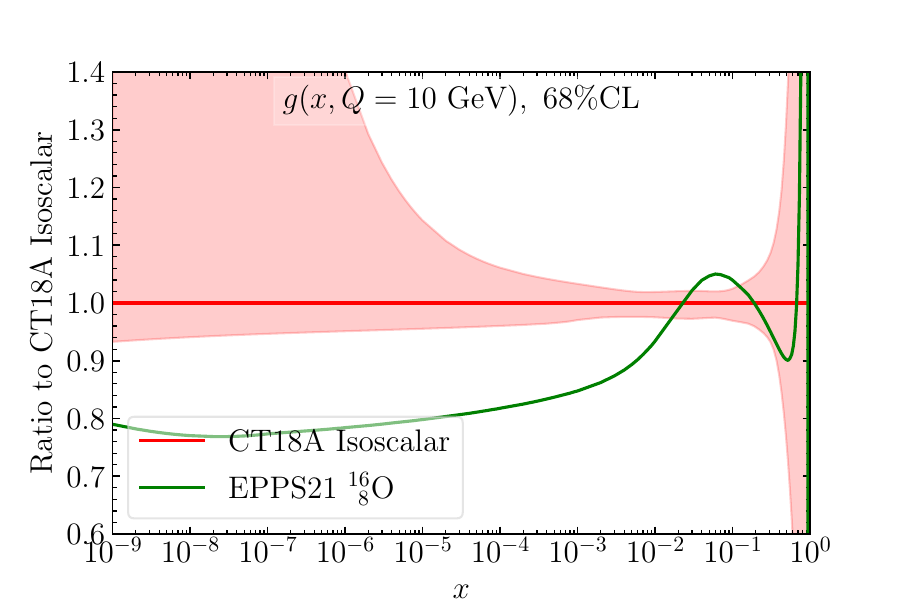}
\includegraphics[width=0.49\textwidth]{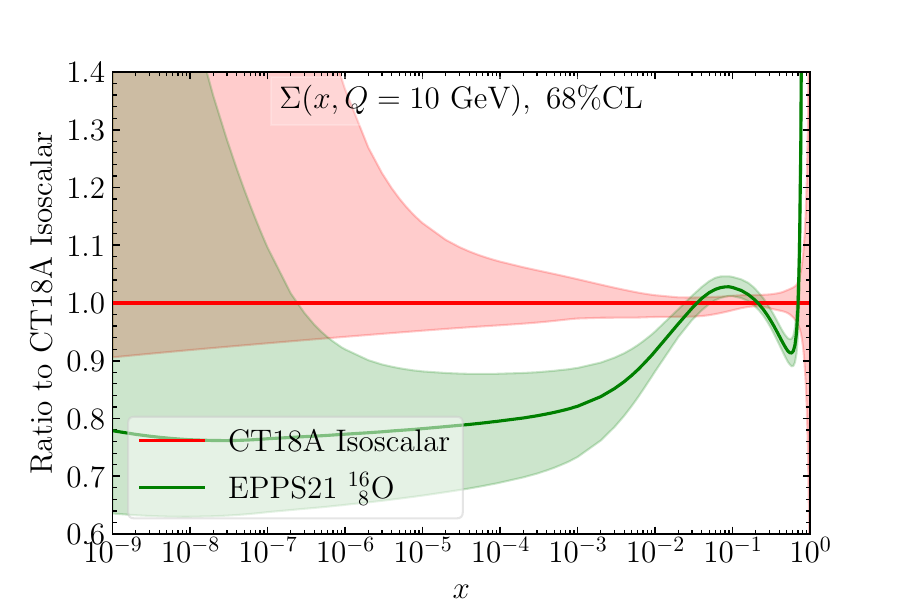}
\includegraphics[width=0.49\textwidth]{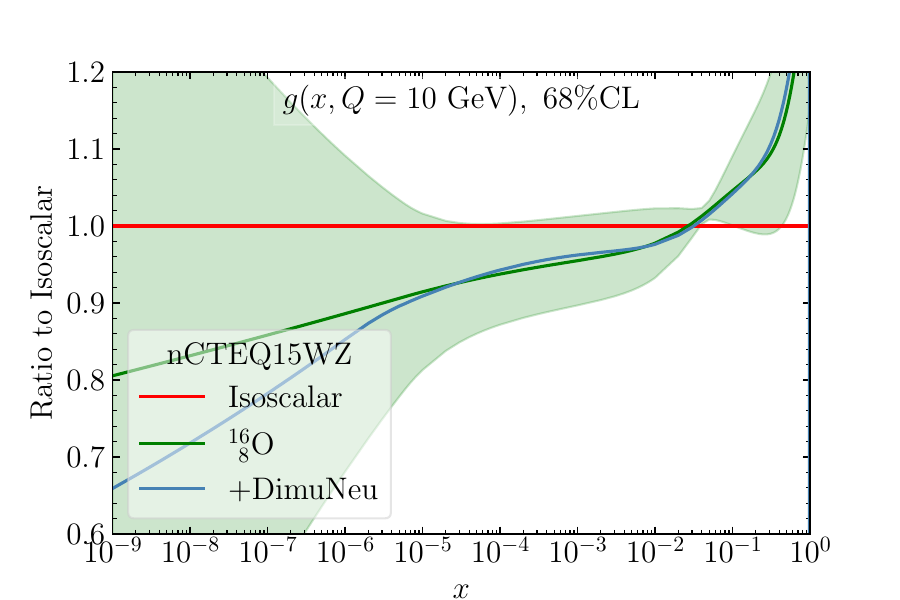}
\includegraphics[width=0.49\textwidth]{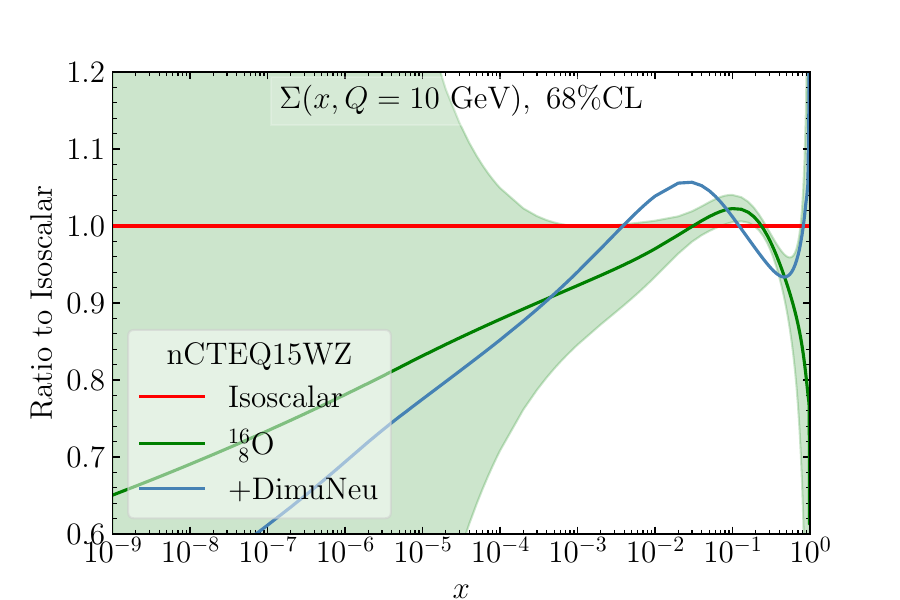}    
\caption{Comparison of the gluon and flavor-singlet PDFs of $_{~8}^{16}$O at $Q=10~\GeV$,
with the EPPS21 and nCTEQ15WZ nPDF sets. The low-$x$ PDFs are obtained with the ``extrapolation" method.
}
\label{fig:nuclPDFs}
\end{figure}

Owing to the small magnitude of $G_F$, neutrino-scattering experiments have historically relied on nuclear targets to maximize the relevant cross sections. This has been true for terrestrial neutrino oscillation and DIS measurements, which typically involve heavy nuclei
such as $^{40}\mathrm{Ar}$ and $^{56}\mathrm{Fe}$. In the meantime, present or planned neutrino observatories, including IceCube
and KM3NeT, entail Cherenkov detection of the charged lepton from CC neutrino reactions with ice or water, such that the predominant nuclear interaction is with H$_2$O or the isoscalar nuclei O.
In all such experiments, incident neutrinos resolve the partonic substructure of {\it nuclei} rather than of free nucleons;
thus, such nuclear DIS events are subject to modifications, relative to scattering from free nucleons, due to the influence of the nuclear medium. See Refs.~\cite{Kulagin:2004ie,Kulagin:2014vsa,Alekhin:2017fpf} and references therein for various nuclear effects. 

In this work, we follow the BGR calculation~\cite{Bertone:2018dse} to take the nuclear parton distribution functions (nPDFs) to account for the effects of nuclear binding on the free-nucleon PDFs --- physics which is known to possess rich phenomenology from very low to high $x$. Much like the proton PDFs at very low $x$, nuclear PDFs remain essentially unconstrained for $x\! <\! 10^{-4}$, driving significant uncertainties which can, in turn, propagate to high-energy nuclear cross sections. This lack of constraints comes not only from the challenge, familiar from studies of proton PDFs, of probing QCD bound states at very high energies, but from the additional complication of gathering such information from a sufficient variety of nuclear species as to allow a detailed unfolding of the nPDFs' $A$ dependence; such knowledge is required to leverage the world's nuclear data to improve nPDF predictions for specific nuclei like O relevant to IceCube.
See the reviews~\cite{Paakkinen:2017jpo,Ethier:2020way} for the latest progress of nuclear PDFs.

In our present study of neutrino DIS at ultra-high energy, the prevailing nuclear corrections are those at very low $x$, where the nuclear medium produces a relative suppression of the nPDFs in a phenomenon known as {\it nuclear shadowing}, as shown in Fig.~\ref{fig:nuclPDFs}.
In contrast, the mild enhancement of nuclear PDFs --- so-called {\it anti-shadowing} --- occurs at substantially larger $x\! \sim\! 0.1$ but remains incompletely
determined, especially for neutrino scattering. As a result, a sizable contribution to the full uncertainty in ultra-high energy
neutrino-nuclear DIS cross section originates with the incomplete knowledge of the exact size of nPDFs and $x$ dependence of the low-$x$ nuclear shadowing corrections.

As these considerations are important in deriving realistic uncertainty estimates for the neutrino-nuclear cross section at very high energies, we consider two main scenarios for the nPDFs in this work. Specifically, in Fig.~\ref{fig:nuclPDFs} we compare the gluon and flavor-singlet nPDFs for O based on two recent extractions: EPPS21~\cite{Eskola:2021nhw} and nCTEQ15WZ~\cite{Kusina:2020lyz}.
We remind the reader that the EPPS21 nPDFs start with the CT18A NLO fits of the free proton PDFs~\cite{Hou:2019efy} and include the global nuclear data, fitting a nuclear correction factor, $R$, on top of this free-proton baseline. In contrast,
the nCTEQ PDFs implement an alternative philosophy of fitting nPDFs which are directly parametrized with an explicit $A$ dependence but yielding an assumed free-proton baseline
as an $A\!=\!1$ boundary condition of the parametrization.
In either case, additional theoretical uncertainty comes from the longstanding issue of the applicability and limits of QCD factorization in nuclear DIS and
the question of whether there may be differences between electromagnetic vs. weak interactions with the nuclear medium. Were there differences, these might be
realized as distinct nuclear corrections to the DIS structure functions measured in the scattering of charged leptons vs.~neutrinos from nuclear targets, including
the degree to which (anti)shadowing corrections equally apply to such interactions at low $x$.

Since the behavior of nuclear corrections in neutrino scattering at
low $x$ may influence the high-energy neutrino-nuclear cross sections investigated in this study, we also consider an alternative nPDF scenario (dubbed as ``DimuNeu'') based only
on fits to neutrino data~\cite{Muzakka:2022wey} as a means of cross-checking our primary calculations based on the recent nPDFs of EPPS21~\cite{Eskola:2021nhw} and nCTEQ15WZ~\cite{Kusina:2020lyz}.
We emphasize that the (heavy) nuclear corrections computed on the basis of nuclear PDFs as discussed above are in addition to the
nonperturbative QCD considerations that are already present for free nucleons and arise from target mass effects and sub-leading terms in the twist expansion (so-called higher twist);
notably, these effects have been explored for the proton and have analogous realizations for nuclei~\cite{Ruiz:2023ozv}, but their
impacts are generally limited to large $x$ and modest values of $Q^2$.
For this reason, we do not give them special attention in this study,
although they form a marginal contribution to the full neutrino-nuclear uncertainty.

\begin{figure}
    \centering
\includegraphics[width=0.49\textwidth]{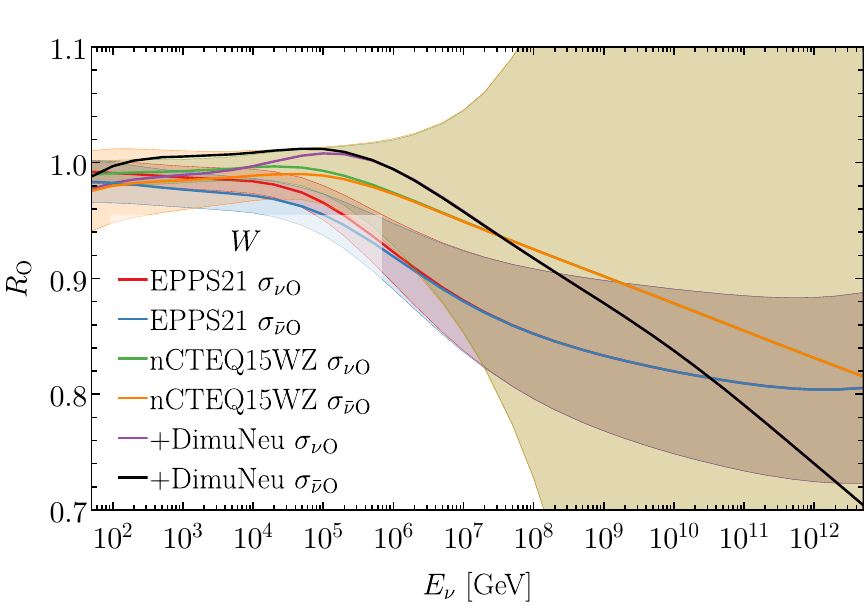}
\includegraphics[width=0.49\textwidth]{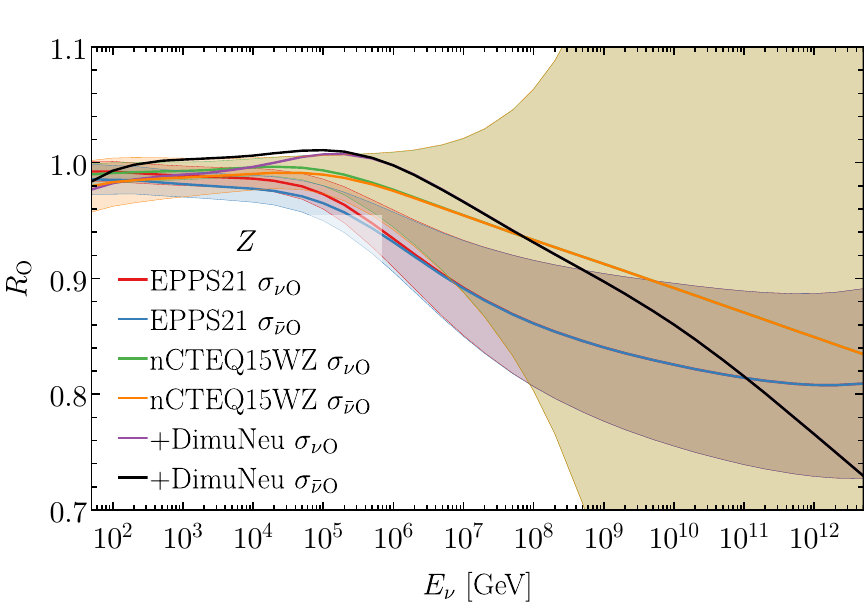}
    \caption{The nuclear correction ratios, $R_{\rm O}$ defined in Eq.~(\ref{eq:RO}), of the neutrino-nuclear charged (left) and neutral (right) current DIS cross sections.}
    \label{fig:nucl}
\end{figure}
An array of LHC measurements involving nuclei --- including
inclusive $W$ and $Z$ production in $p \mathrm{Pb}$ collisions --- have been recorded by the ALICE, ATLAS, CMS, and LHCb experiments~\cite{TheATLAScollaboration:2015lnm,ATLAS:2015mwq,CMS:2015ehw,CMS:2015zlj,CMS:2019leu,ALICE:2016rzo,Senosi:2015omk,LHCb:2014jgh,ALICE:2020jff}.
Despite this progress, the current global data sets constrain nPDFs only at somewhat higher values of $x\!\gtrsim\! 10^{-4}$ relative to free-proton analyses; that is, the lower bound in $x$ of nPDF fits is considerably larger than the corresponding low-$x$ frontier in modern proton PDF determinations, which are constrained by high-energy data down to $x\!\gtrsim\! 10^{-5}$.
Given this present situation, we rely on extrapolations of existing nPDFs.
We note that these polynomial extrapolations to small $x$ are unavoidable on the grounds of numerical stability, despite the naive nature
of the extension of interpolation grids to very small $x$ based on parametrizations that have not been constrained by data to such low $x$; again,
we showed the extrapolated free-nucleon PDFs in Fig.~\ref{fig:extraPDFs} and the corresponding neutrino-nucleon cross sections in Fig.~\ref{fig:evolve2extrap}.

In the context of this behavior observed for the free-nucleon case, we present in Fig.~\ref{fig:nucl} the nuclear corrections 
as the (anti)neutrino cross section ratios of nuclear to isoscalar scatterings
\begin{equation}
    R_{\textrm{O}}=\frac{\sigma_{\nu(\bar{\nu})\textrm{O}}}{\sigma_{\nu(\bar{\nu}) I}}\ ,
\label{eq:RO}
\end{equation}
with the corresponding nuclear PDF uncertainties, where O and $I$ indicate the $_{~8}^{16}$O and isoscalar targets. 
We also collect associated numerical values in Tabs.~\ref{tab:ccdis} and \ref{tab:ncdis} of App.~\ref{app:num} for
the cross sections and nuclear corrections ratios $R_{\rm O}$.
As done for earlier plots, the left and right panels of Fig.~\ref{fig:nucl} correspond to scattering mediated by
$W$ and $Z$ exchange, respectively, and we show the nuclear correction ratio for both $\nu$ and $\bar{\nu}$ cross
sections based on the EPPS21, nCTEQ15WZ, and the recent nCTEQ DimuNeu analysis. For the former two analyses, the plotted
error band represents the nPDF uncertainty obtained in those studies; for the DimuNeu calculations, in contrast, we
simply display the best fit for the sake of comparison against EPPS21 and nCTEQ15WZ as baselines.
%

As can be seen in both panels of Fig.~\ref{fig:nucl}, the nuclear correction ratio is generally consistent with unity --- up to nPDF uncertainties --- at the lower energies that
we plot (for $E_\nu\! \lesssim\! 10^4\, \mathrm{GeV}$), with little evidence of an enhancement above $R_{\textrm{O}}\!=\!1$.
Intriguingly, the $E_\nu$ dependence in the calculation based only on fits to neutrino data, DimuNeu, suggests a slight enhancement consistent
with nuclear anti-shadowing for $E_\nu\! \sim 10^{5}\!-\!10^{6}~\GeV$, although this behavior is within the 1$\sigma$ error band
determined in nCTEQ15WZ. However, EPPS21 reflects no such enhancement, suggesting a need to further investigate the presence of anti-shadowing
in nPDFs for neutrino scattering. In all cases, with growing neutrino energy, the neutrino-nucleus cross section is increasingly suppressed relative to the free-nucleon cross section as nuclear shadowing becomes ever more significant at low $x$. In addition, the nuclear corrections at ultrahigh energy become universal in terms of flavor $\nu/\bar{\nu}$ as well as the charged/neutral current, much as we had observed for scattering
from free nucleons.
We see that in a large $E_\nu$ range, the nuclear uncertainty is the dominant one, until at an extremely high energy, such as $10^{12}~\GeV$. 

\begin{figure}
    \centering
    \includegraphics[width=0.52\textwidth]{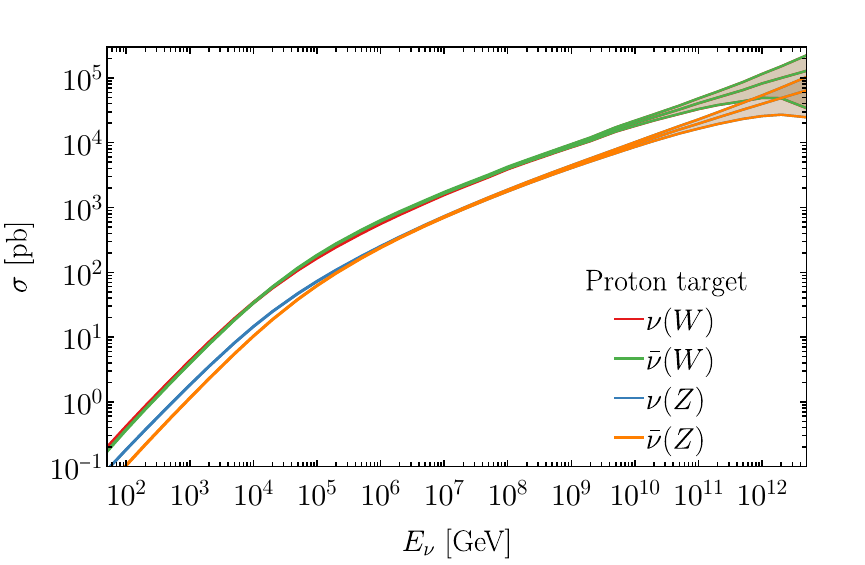}  
    \includegraphics[width=0.47\textwidth]{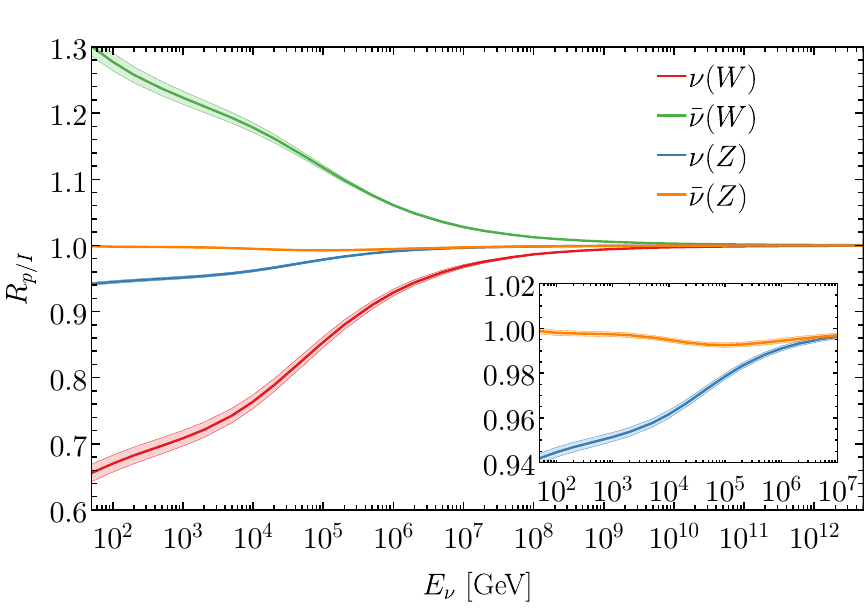}
    \includegraphics[width=0.49\textwidth]{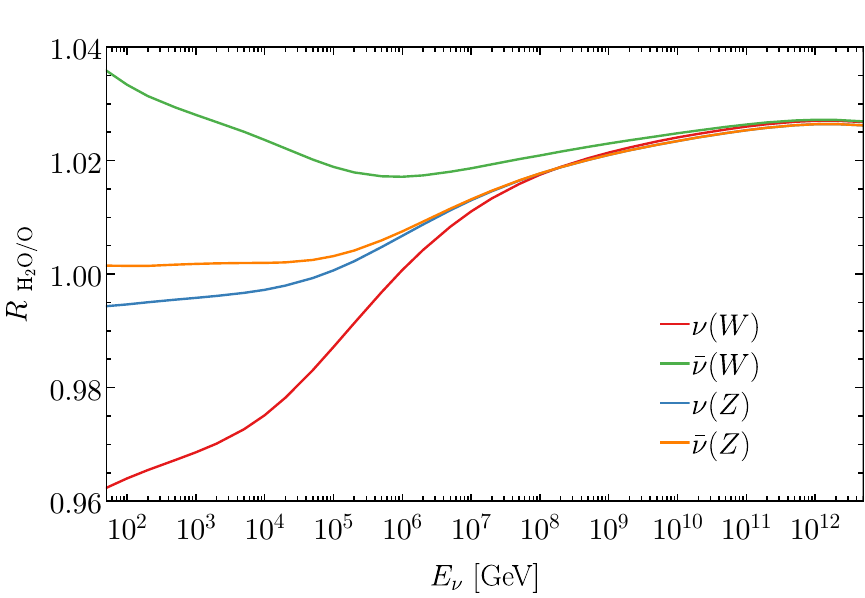}
    \caption{The neutrino-proton DIS cross-section (upper left), and the cross-section ratios of proton to isoscalar (upper right), and the ratio of H$_2$O to O nuclei (lower). The bands in the upper right plot indicate proton Hessian uncertainties of CT18 PDFs.}
    \label{fig:H2OtoO}
\end{figure}

In the IceCube experiment, the high-energy neutrinos are scattered by ice/water. The mass-averaged structure functions of ice can be written as
\begin{equation}
F_i^{\rm H_2O}=\frac{1}{2+A}(2F_i^p+AF_i^{\rm O}),
\end{equation}
where $A=16$ and $Z=N=8$ for the O nucleus. 
Correspondingly, the averaged-nucleon cross section can be expressed as
\begin{equation}
\sigma_{\nu\rm H_2O}=\frac{1}{2+A}(2\sigma_{\nu p}+A\sigma_{\nu\rm O}).
\end{equation}

We show the neutrino-proton DIS cross sections as well as the (anti-)neutrino cross-section ratios of proton to isoscalar $R_{p/I}$ in the upper panels of Fig.~\ref{fig:H2OtoO}.
In the low energy region, the charged-current DIS gives smaller (larger) cross sections for (anti-)neutrino-proton scattering than isoscalar scattering.
This can be understood in terms of the leading partonic subprocesses
\begin{equation}
 \nu+d\to\ell^-+u, ~\bar{\nu}+u\to\ell^++d.
\end{equation}
In terms of its valence content, the free proton contains fewer down but more up quarks than the isoscalar nucleon, leading to a corresponding difference in the CC DIS cross sections.  
The neutral-current scattering gives $R_{p/I}^{\bar{\nu}(Z)}\sim1$, which is resulted from an accidental numerical cancellation between the spin correlation (\emph{i.e.}, in the minus sign in Eq.~(\ref{eq:diffXS})) and the difference between $d,u$ PDFs. 
A similar accidental cancellation happens to the CC scattering of the proton target, which gives almost identical cross sections for neutrino and antineutrino.
In contrast, $R_{p/I}^{\nu(Z)}\lesssim1$, as a result of the accumulation of both effects. 
We also notice that the proton PDF uncertainty is largely canceled in this ratio, especially in the ultrahigh-energy limit, which is 
slightly more pronounced in the NC current case than the CC one.

In the right panel of Fig.~\ref{fig:H2OtoO}, we show the averaged neutrino-water cross section normalized to the oxygen nucleus as
\begin{equation}\label{eq:RH2O}
R_{\textrm{H}_2\textrm{O/O}}=\frac{2\sigma_{\nu p}+A\sigma_{\nu\textrm{O}}}{(2+A)}/\sigma_{\nu\textrm{O}}.
\end{equation}
The corresponding numerical values are listed in Tabs.~\ref{tab:ccdis} and \ref{tab:ncdis} in App.~\ref{app:num}.
Taking the universal nuclear correction in the high-energy limit to be $R_{\rm O}\sim0.81$ [cf.~Eq.~(\ref{eq:RO}) and Fig.~\ref{fig:nucl} based on EPPS21], 
we determine that the averaged molecular water-to-oxygen ratio approaches a constant $R_{\rm H_2O/O}\sim1.026$ at high energy, as shown in Fig.~\ref{fig:H2OtoO}.
The final H$_2$O-averaged cross section can then be obtained with
\begin{equation}\label{eq:H2O}
\sigma_{\nu\textrm{H}_2\textrm{O}}=
\sigma_{\nu I}R_{\rm O}R_{\textrm{H}_2\textrm{O/O}}\ ,
\end{equation}
with the corresponding uncertainty propagated as
\begin{equation}\label{eq:H2Ounc}
\frac{\delta\sigma_{\nu\rm H_2O}}{\sigma_{\nu\rm H_2O}}=
\sqrt{\left(\frac{\delta\sigma_{\nu I}}{\sigma_{\nu I}}\right)^2
+\left(\frac{A}{2+A}\frac{\delta R_{\rm O}}{R_{\rm O}}\right)^2
}\ .
\end{equation}

\section{Predictions for neutrino scattering experiments}
\label{sec:compare}

So far, we have explained our theoretical calculation of the (anti)neutrino-nucleus scattering
in great detail, including both the cross section on free nucleons as well as nuclear corrections. 
Our predicted cross sections, tabulated in Tab.~\ref{tab:ccdis}-\ref{tab:ncdis} in App.~\ref{app:num}, can be directly compared to experimental measurements, both current and future. Here we mainly focus on the IceCube experiment~\cite{IceCube:2020rnc} and the proposed future IceCube-Gen2~\cite{IceCube-Gen2:2020qha}.
We will comment on the collider neutrino energy gap filled by the FASER$\nu$ experiment~\cite{FASER:2020gpr} at the LHC as well.

\subsection{Comparisons with CSMS and other calculations}
\label{sec:CSMS}
For the IceCube measurements~\cite{IceCube:2020rnc,IceCube:2017roe}, the calculation of Cooper-Sarkar, Mertsch, and Sarkar (CSMS)~\cite{Cooper-Sarkar:2011jtt} has been adopted for both neutrino-flux calibrations as well the comparison with the experimental measurement. 
In addition, a number of similar calculations exist in the literature, such as the Gandhi, Quigg, Reno, and Sarcevic (GQRS)~\cite{Gandhi:1998ri}, Connolly, Thorne, and Waters (CTW)~\cite{Connolly:2011vc}, as well as Bertone, Gauld, and Rojo (BGR)~\cite{Bertone:2018dse,Garcia:2020jwr}, and two recent updates from Jeong and Reno (JR)~\cite{Jeong:2023hwe} as well as NNSF$\nu$~\cite{Candido:2023utz}. Similar to our framework, all these calculations adopt the structure-function approach, but with different schemes at different orders, as well as different PDFs. Both the GQRS and CTW took the LO structure functions, while PDFs are based on CTEQ4M~\cite{Lai:1996mg} and MSTW08~\cite{Martin:2009iq}, respectively. The CSMS calculation is performed with the Thorne-Roberts scheme~\cite{Thorne:1997ga,Thorne:2006qt} at NLO with the HERAPDF1.5 PDFs~\cite{Cooper-Sarkar:2010yul}. The BGR calculation~\cite{Bertone:2018dse,Garcia:2020jwr} is the most closed one to our framework, with the heavy-flavor structure function in the FONLL scheme up to $n_f=6$ flavors~\cite{Forte:2010ta} together with the small-$x$ resummation~\cite{Ball:2017otu} included. 
The recent JR calculation~\cite{Jeong:2023hwe} includes contribution from the shallow inelastic scattering. In comparison, the NNSF$\nu$ approach~\cite{Candido:2023utz} adopts a neural network to fit the neutrino DIS structure functions (SFs) while the low-$Q^2$ $F_3$ is taken from the Bodek-Yang parameterization~\cite{Yang:1998zb,Bodek:2002vp,Bodek:2003wd,Bodek:2004pc,Bodek:2010km,Bodek:2021bde}.
In this work, we have gone beyond in two aspects. We have included the complete heavy-quark effect up to NNLO in the ACOT scheme~\cite{Gao:2021fle}. It turns out the heavy-flavor effect is negligible at an ultrahigh energy, while a negative 2\% when $E_\nu\sim100~\GeV$, as examined in Sec.~\ref{sec:th}. We also extend our calculation up to approximate N3LO, with zero-mass Wilson coefficient functions. 

\begin{figure}
\centering
\includegraphics[width=0.49\textwidth]{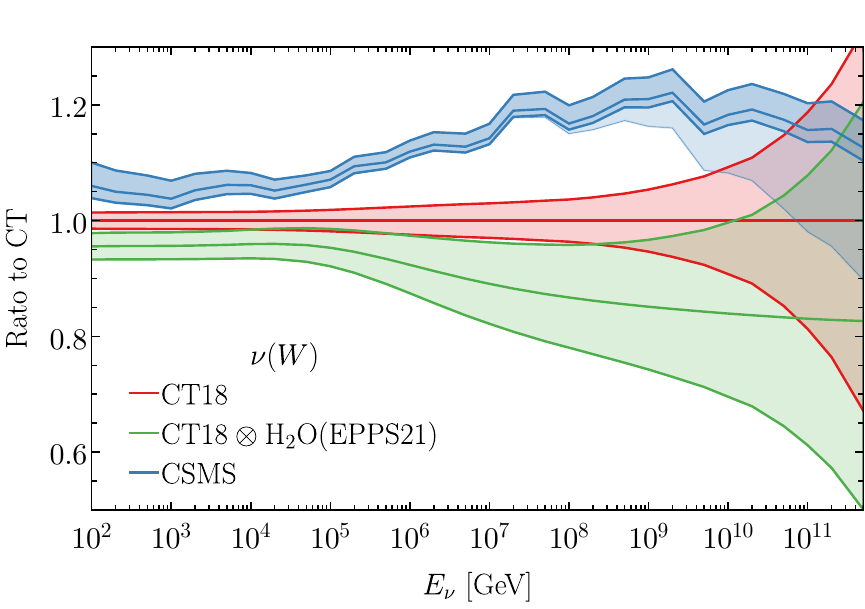}
\includegraphics[width=0.49\textwidth]{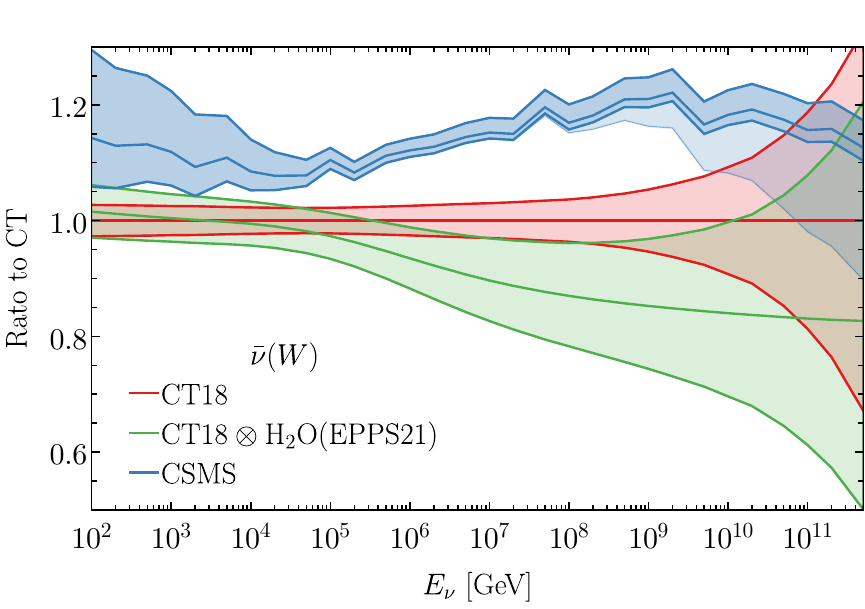}
\includegraphics[width=0.49\textwidth]{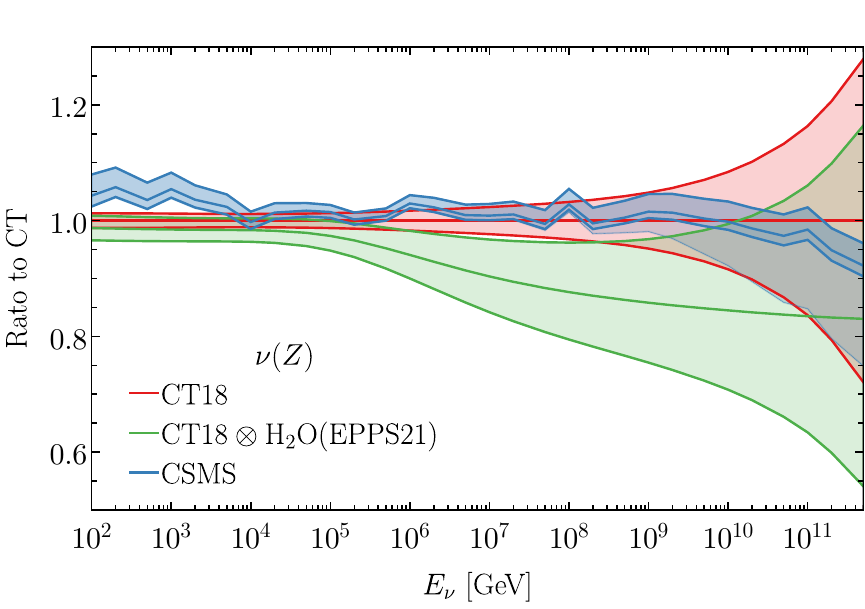}
\includegraphics[width=0.49\textwidth]{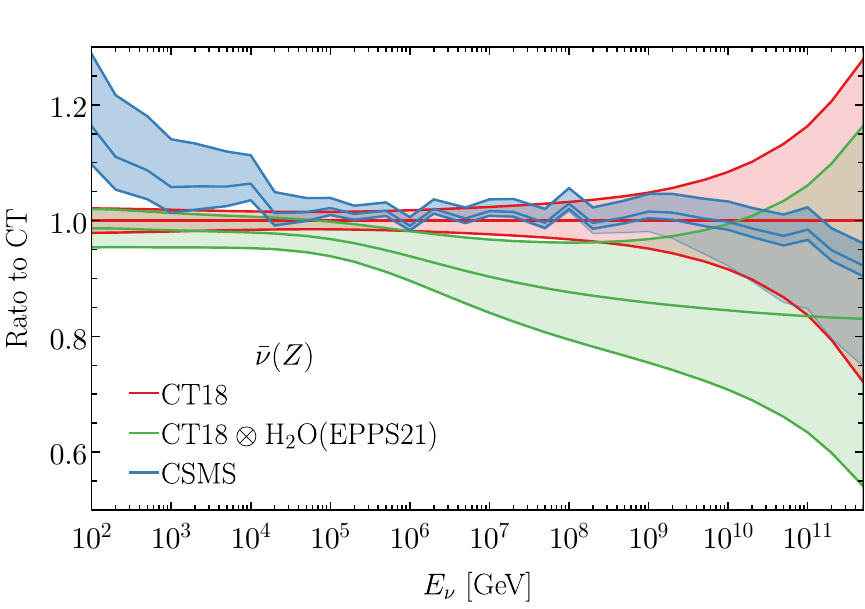}
\caption{Comparison of the neutrino-isoscalar cross sections with the existing calculations between the CT18 and CSMS~\cite{Cooper-Sarkar:2011jtt}, for the CC (upper) and NC (lower) scattering processes. 
}
\label{fig:compare}
\end{figure}
\begin{figure}
    \centering
    \includegraphics[width=0.49\textwidth]{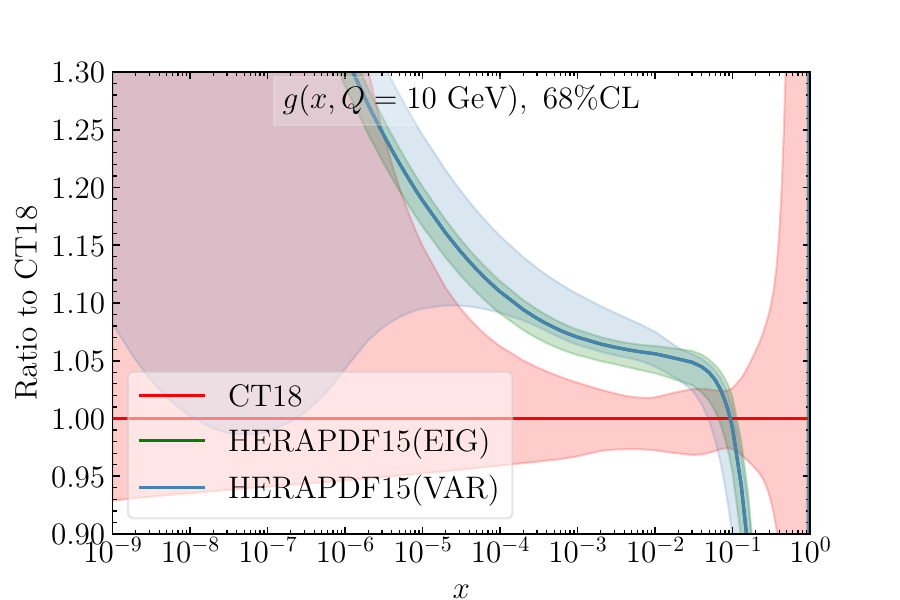}
    \includegraphics[width=0.49\textwidth]{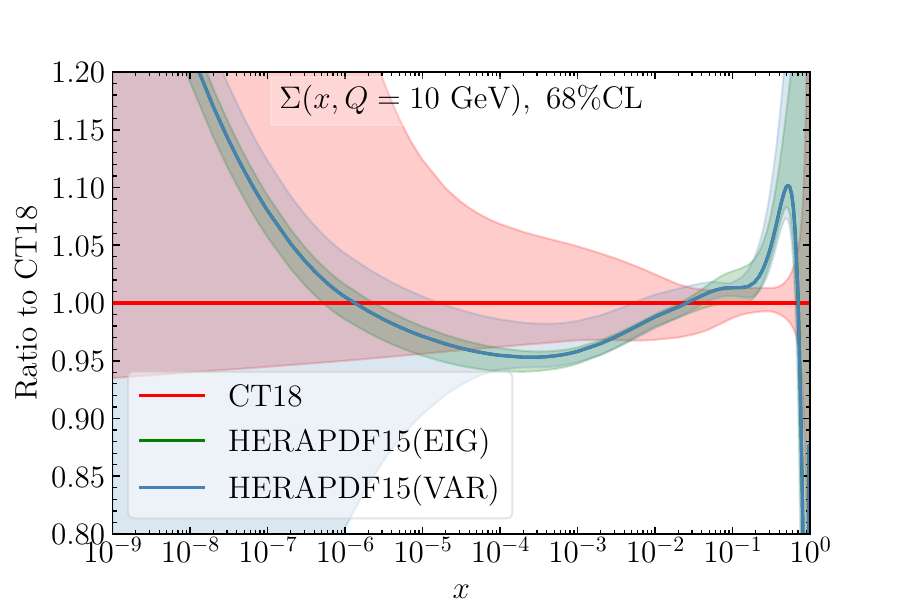}
    \caption{The comparison of gluon and flavor-singlet PDFs at $Q=10~\GeV$ between the CT18 NNLO and HERAPDF1.5 NLO PDFs.}
    \label{fig:HERAPDF15}
\end{figure}

In this work, we are mainly targeting a state-of-the-art prediction for the high-energy neutrino cross section measured at IceCube. 
In Fig.~\ref{fig:compare}, we compare our calculation with the CSMS result --- the theoretical prediction adopted in IceCube~\cite{IceCube:2020rnc,IceCube:2017roe}, with our numerical results tabulated in App.~\ref{app:num}.
Compared to the free-nucleon (isoscalar) CT18 predictions, the CSMS calculation gives overall larger cross sections for charged-current DIS, while the neutral-current cross section is in good agreement. 
The larger CC cross section can be understood in terms of the corresponding larger small-$x$ PDFs, as shown in Fig.~\ref{fig:HERAPDF15}. 
Relative to the free-isoscalar cross section, the H$_2$O nucleon-averaged calculation receives a negative nuclear correction as examined in Sec.~\ref{sec:nuclear}.

The HERAPDF1.5 PDFs, adopted in the CSMS calculation, give smaller PDF error bands, reflecting the different criteria adopted in two sets. In addition, CSMS realized that member 9 of HERAPDF1.5 gives the largest deviation of neutrino cross sections from the central set, shown as the lower boundary line in Fig.~\ref{fig:compare}. This particular member was argued to be excluded when quantifying PDF uncertainty in Ref.~\cite{Cooper-Sarkar:2011jtt} as it gives negative gluon at low $x$ and low $Q$. 
However, it has been shown that negative PDFs should be acceptable as long as the physical observables, such as cross section, remain positive~\cite{Collins:2021vke}.
%
%
In comparison, HERAPDF adopts the $\Delta\chi^2=1$ criterion in generating its 68\% confidence level (CL) Hessian eigenvector sets, while CT18 takes $\Delta\chi^2\!=\!100$ for the 90\% CL uncertainty in order to capture variations in the non-perturbative parameterization forms, selection of data sets as well as the variation of theoretical setups~\cite{Pumplin:2001ct,Pumplin:2002vw}.
The 68\% CL error band is obtained by dividing the 90\% CL one by a factor of 1.645.
We remind that HERAPDF also released \texttt{VAR} sets, in order to capture variations of some theoretical parameters, such as strong coupling, heavy-quark mass, evolution starting scale, etc~\cite{Cooper-Sarkar:2010yul}. In Fig.~\ref{fig:HERAPDF15}, we also show the error band including the HERAPDF15VAR set, which indicates its comparable uncertainty with the CT18 prediction.

\subsection{IceCube high-energy neutrino measurements}
\label{sec:IceCube}

The IceCube Collaboration has observed high-energy astrophysical neutrinos since 2013~\cite{IceCube:2013gge}.
Two measurements of neutrino-nucleon cross sections have been reported. 
In Ref.~\cite{IceCube:2017roe}, the Earth absorption cross section is analyzed for energies between 6.3 TeV and 980 TeV, by using the upward-going neutrino-induced muons with 10784 events.
In Ref.~\cite{IceCube:2020rnc}, neutrino cross sections
including all three neutrino flavors were reported for neutrino energies from 60 TeV to 10 PeV, 
based on the 60 high-energy starting events (HESE) with 7.5 years of data.
We also note that IceCube has recently reported the measurement of a possible Glashow resonance event~\cite{IceCube:2021rpz}, providing further motivation to quantify
the neutrino-nucleon cross section at corresponding kinematics so as to understand backgrounds to neutrino-electron scattering. We discuss this aspect in further detail in App.~\ref{sec:Glashow}.
%

\begin{figure}[h]
    \centering
\includegraphics[width=0.49\textwidth]{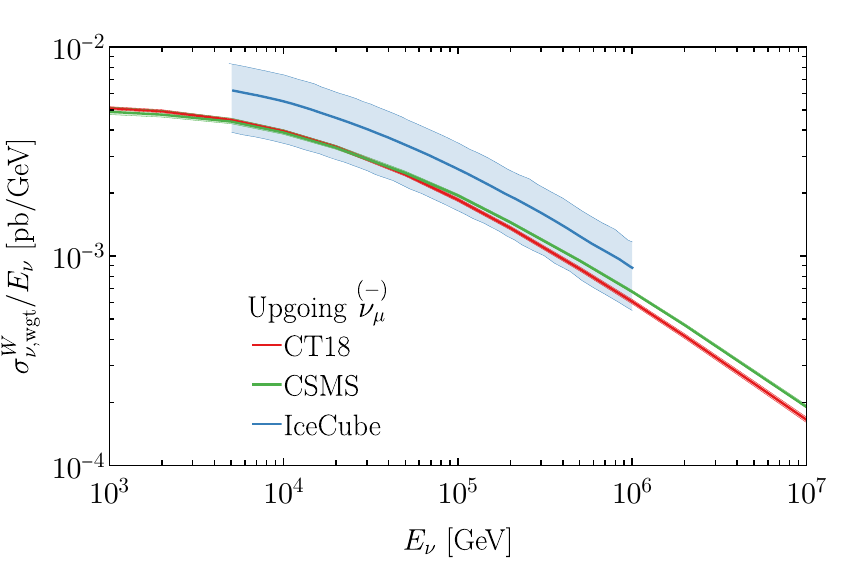}    
\includegraphics[width=0.49\textwidth]{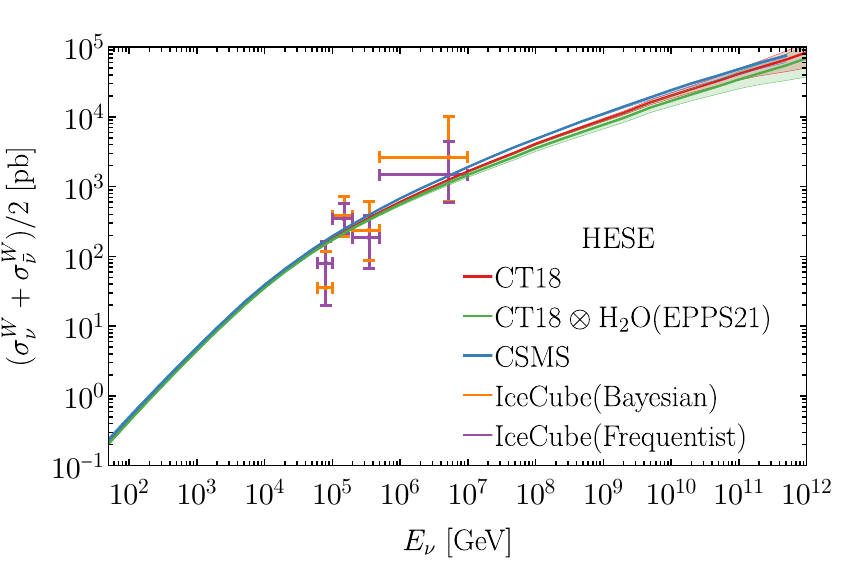}
    \caption{Comparison of the theoretical predictions of charged-current neutrino
    cross sections with the IceCube measurements~\cite{IceCube:2017roe,IceCube:2020rnc}. In the left panel for measurements with up-going muon neutrinos~\cite{IceCube:2017roe}, the neutrino and antineutrino cross sections are weighted with flux as Eq.~(\ref{eq:wgt}). In the right panel for the HESE sample, the neutrino and antineutrino cross sections are averaged, with both Bayesian and Frequentist analyses shown here~\cite{IceCube:2020rnc}. The CT18 error bands indicate the PDF uncertainty, while the H$_2$O averaged band combines the EPPS21 nuclear uncertainty in terms of Eq.~(\ref{eq:H2Ounc}).
    }
    \label{fig:ICresults}
\end{figure}

In Fig.~\ref{fig:ICresults}, we show the CT18 and CSMS predictions for the neutrino
charged-current DIS cross sections, compared with IceCube data. 
Similar to the IceCube simulation~\cite{Miarecki:2016kku}, in the earth absorption (up-going muon) case, neutrino and antineutrino cross section was flux weighted as
\begin{equation}\label{eq:wgt}
\sigma_{\nu,\rm wgt}=
\frac{\Phi_{\nu}\sigma_{\nu}+\Phi_{\bar{\nu}}\sigma_{\bar{\nu}}}{\Phi_{\nu}+\Phi_{\bar{\nu}}},
\end{equation}
with the corresponding flux ratio $\Phi_{\nu}/\Phi_{\bar{\nu}}$ taken from Refs.~\cite{Gaisser:2014eaa,OPERA:2010cos}.
Similarly to the IceCube analysis, nuclear corrections are not included in this comparison due to the complication of the Earth's nuclear abundance.
In the all-flavor HESE cross section, the average of neutrino and antineutrino CC DIS events was reported. We include the data with both Bayesian and Frequentist analyses~\cite{IceCube:2020rnc}. 
A previous analysis based on 33 events~\cite{Bustamante:2017xuy} has been superseded and is not shown here. 
Our theoretical predictions with and without nuclear corrections based on the H$_2$O nucleon average are provided in this case. 

We include the CSMS predictions as well, which were used in the IceCube analysis.
The difference between our predictions and CSMS, including relative uncertainties, can be inferred from Fig.~\ref{fig:compare}. Both the CSMS and CT18 (both with and without nuclear effects estimated according to EPPS21) predictions provide good descriptions of the IceCube data, considering the large experimental uncertainty, while the CT18 incorporates more comprehensive effects, as detailed in Sec.~\ref{sec:th}-\ref{sec:PDFs}.
In comparison with CSMS NLO calculation, the CT18 baseline is at NNLO, which receives negative a few percent corrections at low (anti)neutrino energy, with size growing up to $20\%$ when $E_\nu=10^{12}~\GeV$.
The approximate N3LO contribution has been estimated with the zero-mass structure functions, which can give 2\% corrections for CC DIS and 4\% to NC DIS.
The heavy-quark mass effect has been examined with the recently developed ACOT scheme at the NNLO~\cite{Gao:2017kkx,Gao:2021fle}, which turns out to be $-2\%$ at low (anti)neutrino energy around $E_\nu\sim100~\GeV$ and diminishes very quickly with energy increasing. 
The quark flavors have been included up to $n_f=6$, and top quarks can contribute $2-4\%$ to the total cross section when $E_\nu=10^{12}~\GeV$.
In addition, we have also included the small-$x$ resummation based on the BFKL matched to the DGLAP evolution, which enhances the neutrino cross section up to 20\% at high (anti)neutrino energy. 
Finally, based on nuclear PDF sets, EPPS21 and nCTEQ15WZ, we have estimated the nuclear corrections to be negative $-20\%$ as a result of the nuclear shadowing effect.
The comparison between the CT18 and CSMS results can be found in Sec.~\ref{sec:CSMS}.


As shown in Fig.~\ref{fig:ICresults}, the experimental uncertainties are overwhelming in comparison with the theoretical ones, both driven by the statistic errors and the complicated systematics, such as the neutrino flux and the earth model. 
Therefore, a better understanding of systematics as well as higher statistics will be critical in future measurements.
IceCube Collaboration has proposed a significant upgrade of the IceCube Antarctic neutrino observatory, to be called IceCube-Gen2~\cite{IceCube-Gen2:2020qha}.
Two distinct features of Gen2, (i) a larger (by a factor of 10) volume of the ice Cherenkov detector and (ii) an additional shallow radio array to detect higher-energy neutrinos with $E_{\nu} > 10^{11}$ GeV, will provide better measurements for the neutrino cross sections.
Our theoretical predictions apply to IceCube-Gen2 as well.

\textbf{Neutral-current (NC) cross sections.}
One goal of the IceCube Collaboration is to measure the flux of cosmic neutrinos as a function of neutrino energy~\cite{IceCube:2021keu}. 
The IceCube Observatory uses Cherenkov radiation to detect charged particles as a proxy for UHE neutrinos. High-energy neutrinos that undergo charged current (CC) interactions in the detector create high-energy charged leptons, which radiate Cherenkov light. Obviously, the theoretical CC cross section is important for interpreting the IceCube Observatory results.

The theoretical {\em neutral current} (NC) cross section is also important because the IceCube Collaboration uses {\em the absorption of neutrinos by the Earth} to estimate the flux of neutrinos incident on the Earth. Neutrinos incident on the Earth from the Northern Hemisphere must pass through a fraction of the Earth, depending on their direction, before reaching the IceCube Observatory at the South Pole. As they travel through the Earth, some neutrinos will be absorbed or scattered by interacting with matter, including both CC and NC interactions. 
In such a way, a better determination of the NC cross section will play a role in the earth absorption rate, as discussed further in App.~\ref{app:abspt}.

\subsection{Accelerator neutrinos and the energy gap}
\begin{figure}[h]
    \centering
    \includegraphics[width=0.49\textwidth]{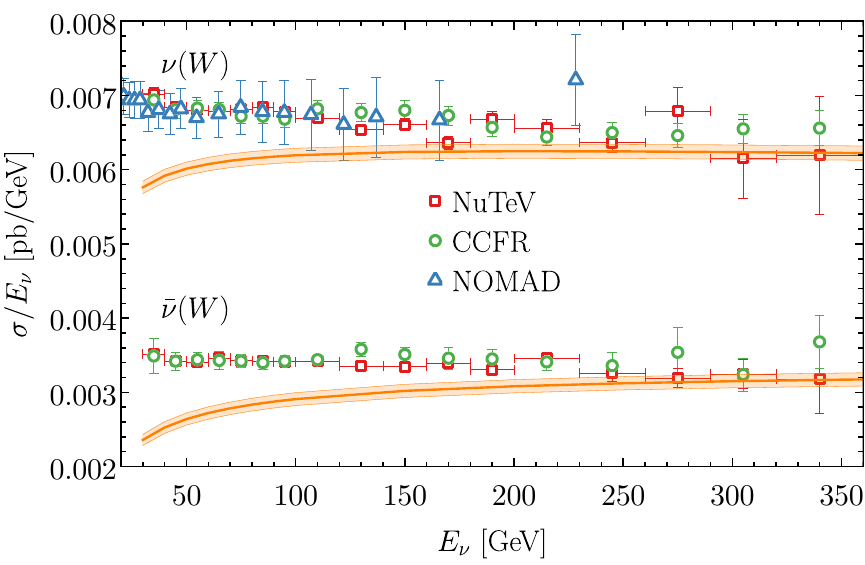}
    \includegraphics[width=0.49\textwidth]{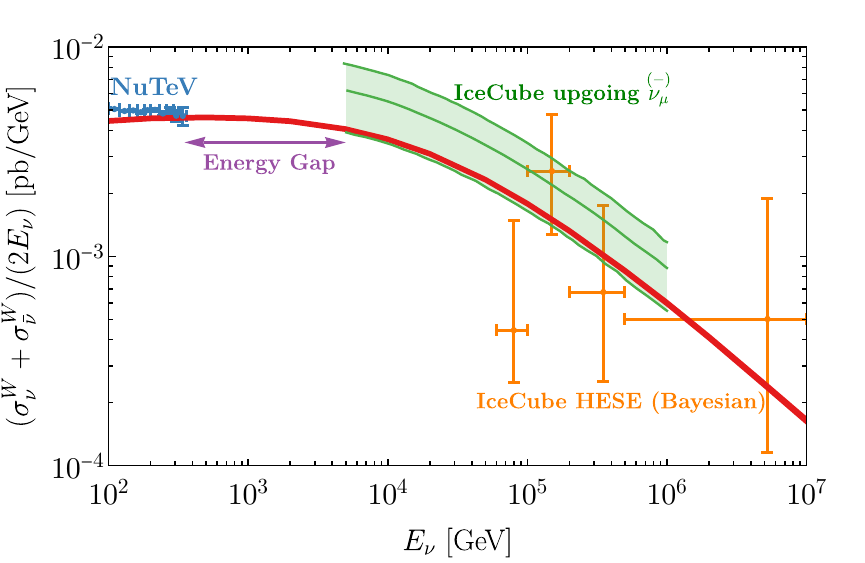}    
    \caption{Left: The CT18 predictions for the neutrino-isoscalar charged-current cross sections divided by the (anti)neutrino energy, $\sigma/E_{\nu}$, in comparison with data measured at accelerator-based experiments~\cite{NuTeV:2005wsg,Seligman:1997fe,NOMAD:2007krq}.
    Right: The CT18 prediction of the averaged neutrino-isoscalar charged-current cross sections divided by neutrino energy in the energy gap ($360~\GeV\lesssim E_\nu\lesssim6~\TeV$), which can be measured by the FASER and other experiments at the FPF at the LHC~\cite{Anchordoqui:2021ghd,Feng:2022inv}. We included cross sections below 360 GeV measured by NuTeV~\cite{NuTeV:2005wsg} and above 6.3 (60) TeV by IceCube upgoing $\overset{(-)}{\nu_\mu}$~\cite{IceCube:2017roe} (HESE Bayesian~\cite{,IceCube:2020rnc}) analyzes.}
    \label{fig:collider}
\end{figure}

In Fig.~\ref{fig:collider}~(left), we compare our theoretical predictions for the neutrino-isoscalar charged current cross section in comparison with the data measured at accelerator experiments, such as NuTeV~\cite{NuTeV:2005wsg}, CCFR~\cite{Seligman:1997fe} and NOMAD~\cite{NOMAD:2007krq}.
We remind the reader that the NOMAD has released only the neutrino rather than antineutrino CC inclusive cross sections so far~\cite{NOMAD:2007krq}.
Due to the complication of different target materials used in different experiments, we don't include nuclear corrections here, which deserve future dedicated studies. 
In general, we see that these three experiments give consistent results for both the neutrino and antineutrino beams. In comparison, our theoretical calculations give a good description of the high-energy data. The low-energy predictions are smaller than the experimental measurements. This behavior was already noticed in our previous work~\cite{Gao:2021fle}, which indicates the importance of missing contributions from other nuclear scattering processes, such as quasi-elastic scattering and the hadronic resonance production, which dominate at the neutrino-nuclear cross section at low energies~\cite{Formaggio:2012cpf}. 

In Fig.~\ref{fig:collider}~(right), we show the averaged (anti)neutrino-isoscalar charged-current cross section divided by the (anti)neutrino energy $\sigma(E_{\nu})/E_{\nu}$. We also include the cross section measured by the NuTeV Collaboration, as one of the highest energies reached for accelerator neutrinos~\cite{NuTeV:2005wsg}. The ``energy gap" between the NuTeV ($\lesssim360~\GeV$) and the IceCube data ($\gtrsim6~\TeV$) can be bridged by ongoing and future measurements at the LHC, such as the FASER (ForwArd Search ExpeRiment)~\cite{FASER:2019dxq} as well as other experiments at the Forward Physics Facility (FPF)~\cite{Anchordoqui:2021ghd,Feng:2022inv}.
Meanwhile, the first neutrino interaction candidates were reported in 2021~\cite{FASER:2021mtu}, and the first direct observation was made very recently~\cite{FASER:2023zcr}.
The maximum neutrino energy produced at the LHC can potentially reach the order of 10 TeV.
As an optimistic consequence, the energy gap can be completely closed by these experiments.
Our theoretical cross section is also shown as the red line in Fig.~\ref{fig:collider}~(right), which provides state-of-the-art predictions for the neutrino scattering measured at the FPF~\cite{Anchordoqui:2021ghd,Feng:2022inv}.

\section{Conclusions}
\label{sec:conc}
In this work, we have presented up-to-date theoretical calculations for neutrino-nucleon deeply-inelastic scattering (DIS) mediated by both charged- and neutral-current interactions over a wide range of neutrino energies, spanning $E_\nu\! \sim\! 50~\GeV$ up to $10^{12}~\GeV$. We have also
generalized these results to neutrino DIS from nuclei by estimating high-energy nuclear corrections to interactions with free
(isoscalar) nucleons.
As is typical, we separate the cross section into a (weak) leptonic tensor as well as a corresponding hadronic tensor.
This latter quantity may be expanded on a complete basis of allowed Lorentz structures, such that folded into the hadronic tensor are the structure functions of the target nucleon or nucleus; these in turn can be computed according to
well-established QCD factorization theorems
that facilitate the separation of short-distance matrix elements, which are now calculable to high perturbative order (at present,
N3LO$'$), from the
nonperturbative parton distribution functions (PDFs) of the nucleon or nucleus.
On the basis of this formalism, we predict neutrino DIS cross sections, leveraging recent developments in
QCD perturbation theory and the most current PDF determinations provided by global QCD analyses, which systematically
unfold the PDFs from diverse high-energy data taken at collider and fixed-target experiments.
This work, therefore, provides the current best theoretical description for both accelerator-based neutrino experiments at higher
energies, as well as astrophysical neutrino observatories, which can potentially observe events in the ultra-high energy regime. 

Using the recent CT18 NNLO proton PDFs~\cite{Hou:2019efy} together with the Wilson coefficients calculated up to N3LO$'$, which includes the exact NNLO heavy-parton mass effect~\cite{Gao:2017kkx,Gao:2021fle} and N3LO contribution with zero-mass partons~\cite{Salam:2008qg},
we carefully examine each of the stages of the neutrino-nucleon DIS cross section calculation.
As the total neutrino cross sections involve broad integrations over the $(x, Q)$ phase space, knowledge of the structure
functions, including those at low scales, is a prerequisite. In the low-$Q$ region, below the starting scale of the CT18 PDFs ($Q_0=1.3~\GeV$), we have explored the impact on the total DIS cross section of taking either LHAPDF extrapolation~\cite{Buckley:2014ana} or APFEL backward evolution~\cite{Bertone:2013vaa} to lower $Q$. We find that the dependence on this selection produces approximately percent-level variations in the cross section at neutrino energies of $E_\nu=10^{3}~\GeV$; at larger values of $E_\nu$, this difference quickly vanishes.
On the other hand, as the neutrino energy enters the ultra-high energy region, $E_\nu\!>\!10^{8}~\GeV$,
the cross section increasingly gains contributions from very small parton momentum fractions, $x\! \ll\! 1$, extending the dependence
of the cross section to values of $x$ which have not been directly probed experimentally, and potentially even beyond the lower $x$ bounds of available PDF interpolation grids. We have quantified the contribution(s) from this low-$x$ extrapolation region, which ultimately amounts to $\sim\!3\%$ of the total cross section at $E_\nu=10^{12}~\GeV$.
The corresponding uncertainty is quantified with the LHAPDF extrapolation and APFEL extrapolation together with evolution, which is found to be at most 1\% for CT18 NNLO. This uncertainty is larger for other PDF sets, such as the nCTEQ15WZ nuclear PDFs~\cite{Kusina:2020lyz}, due to a larger extrapolation region from a larger $x$ value. 
Furthermore, we have identified the important $(x,Q)$ kinematics, which is around $(x,Q)\sim(M_{W,Z}^2/(2m_NE_\nu),M_{W,Z})$, based on separate scans of integration limits $x_{\min}$ and $Q_{\min}$ as well as the joint two-dimensional scan.

We have also explored higher-order effects by comparing the LO, NLO, and NNLO calculations implemented alongside the corresponding CT18 PDFs consistently determined at the same orders. These corrections up to NNLO are generally negative, about a few percent at low neutrino energies, but increasing to 20\% percent at $E_\nu=10^{12}~\GeV$.
The missing higher-order effect is estimated with zero-mass N3LO Wilson coefficients together with NNLO PDFs and found to be $2\%$ for CC DIS, and $4\%$ for NC DIS. 
We have also investigated heavy-quark effects using the recent NNLO ACOT calculation for CC DIS~\cite{Gao:2021fle}, which we find to be at most 2\% for $E_\nu=10^2~\GeV$ before rapidly becoming negligible at higher energies. Moreover, the contribution from the third quark generation is examined within a variable-flavor-number scheme. 
We find that the bottom quark and top quarks contribute negligibly at low (anti)neutrino energy when $E_\nu<10^{6}~\GeV$. With increasing (anti)neutrino energy, each can contribute at most $\sim\!2\!-\!4\%$ for $E_\nu=10^{12}~\GeV$.
As the small-$x$ phase space contribution becomes important with increasing neutrino energy, we have included small-$x$ resummation effects based on the BFKL evolution using the HELL~\cite{Bonvini:2016wki,Bonvini:2017ogt} framework interfaced to APFEL~\cite{Bertone:2013vaa}, finding the associated effect can be as large as 10\% for $E_\nu=10^{12}~\GeV$.
Moreover, the effects of nuclear corrections were estimated using two of the latest nuclear PDF extractions, EPPS21~\cite{Eskola:2021nhw} and nCTEQ15WZ~\cite{Kusina:2020lyz}. For example, if we base our default nuclear correction estimates on EPPS21, we obtain a negative shift in the neutrino-oxygen cross section relative to isoscalar (free-nucleon) scattering up to $-20\%$ when $E_\nu\gtrsim10^{10}~\GeV$, a result of the nuclear shadowing effect at small $x$.
We also find the uncertainties associated with these nuclear-medium effects, especially the low-$x$ shadowing, to be particularly sizable and an important limitation to the precision of ultra-high energy neutrino-nuclear scattering predictions. 

We have compared our predictions to those obtained in the CSMS model~\cite{Cooper-Sarkar:2011jtt},
which is based on an NLO QCD calculation using HERAPDF1.5 PDFs. While our calculation agrees with CSMS for NC DIS, our results are $\sim\!10\!-\!20\%$ smaller for CC DIS. This difference is a consequence of two primary factors: the negative NNLO corrections included in our calculation as well as the (comparatively) smaller gluon PDF obtained by CT18 in the small-$x$ region relative to HERAPDF1.5. 
We also provide the PDF uncertainty based on the Hessian eigenvector sets~\cite{Pumplin:2001ct}. It is found to be a few percent at intermediate high energy when $E_\nu\lesssim10^{7}~\GeV$, which grows up to $60\%\sim70\%$ at $E_\nu=10^{12}~\GeV$. Our PDF uncertainties are generally larger than the CSMS result, mainly because of different uncertainty criteria. In the CT18 PDFs, the 90\% CL Hessian eigenvector sets are determined by a 100-unit increase, $\Delta \chi^2\! =\! 100$, in the global $\chi^2$ (out of a total global $\chi^2\! \approx\! 4300$); this extended tolerance captures the full PDF error stemming from tensions among fitted data as well as uncertainties from the non-perturbative parameterization and other QCD parameters~\cite{Pumplin:2001ct,Pumplin:2002vw}.
In comparison, HERAPDF assumes a $\Delta\chi^2=1$ criteria to determine its error bands at the 68\% CL, while other QCD parameter uncertainties are quantified by corresponding \texttt{VAR} sets~\cite{Cooper-Sarkar:2010yul}, which are comparable with the CT18 ones. 

In addition to other theoretical calculations, we have also compared our results to existing measurements, including the high-energy events at IceCube as well as the accelerator-based neutrino experiments.
Our theoretical predictions give a good description of the IceCube results, both for the earth absorption cross sections~\cite{IceCube:2017roe} and the so-called high-energy starting events~\cite{IceCube:2020rnc}.
Compared to the theoretical calculations, current experimental uncertainties are much larger, driven by complicated systematics as well as limited statistics. 
As for the accelerator neutrino cross sections~\cite{NOMAD:2007krq,NuTeV:2005wsg,Seligman:1997fe}, our theoretical calculations agree well with the high-energy tails with $E_\nu\!\gtrsim\!250~\GeV$ but under-predict the data at more modest energies, indicating the importance of missing contributions from low-energy nuclear processes like quasi-elastic scattering and resonance production, which are beyond the scope of this work. 
In between the accelerator neutrinos and the IceCube Observatory, an energy gap exists for $360~\GeV\lesssim E_\nu\lesssim6~\TeV$, which may potentially be filled by the ongoing and future FASER~\cite{FASER:2018bac} as well as other Forward Physics Facilites~\cite{Anchordoqui:2021ghd,Feng:2022inv} experiments at the LHC.

A number of improvements to the present work can be pursued in the future. For instance, very recently, approximate N3LO PDFs have been released based on the global analysis of the MSHT group~\cite{McGowan:2022nag}.
A complete N3LO PDF set is needed to obtain the full N3LO contribution.
As we have seen, theoretical uncertainties increase significantly with neutrino energy, mainly induced by the large free-nucleon and nuclear PDF uncertainties in the small-$x$ region. Constraining small-$x$ PDFs, especially the gluon PDF, better can greatly enhance the accuracy of high-energy theoretical predictions. This might be furthered by incorporating LHCb data on the forward production of $D$ and $B$ mesons~\cite{LHCb:2015swx,LHCb:2017vec} into global QCD analyses of PDFs.
In addition, obtaining improved theoretical control over gluon saturation effects at small $x$~\cite{Kovchegov:2019atj}, further investigating the systematics of PDF extrapolation to low $x$, and pursuing
studies of nuclear PDF issues related to shadowing at $x\! \ll\! 1$ 
would all be invaluable to next-generation precision for neutrino-nuclear scattering.

\begin{acknowledgments}
We would like to thank Valerio Bertone and Rhorry Gauld for the instruction about the BGR calculation, Donna Naples for providing the NuTeV data, and our CTEQ-TEA colleagues for many helpful discussions. We also thank Faiq Muzakka, Fred Olness, and the larger nCTEQ collaboration for providing grid files
corresponding to the recent DimuNeu nPDF fit as well as for useful discussions related to
neutrino-nuclear interactions, and Tyce DeYoung for reading a part of this manuscript.
Meanwhile, we are grateful to Johannes Bluemlein, Amanda Cooper-Sarkar, Peter Denton, Juan Rojo, Victor B. Valera, and Bei Zhou for various constructive comments.
The work of KX is supported by the U.S. Department of Energy under grant No. DE-SC0007914, the U.S. National Science Foundation under Grants No. PHY-2112829 and PHY-2310497, and also in part by the PITT PACC. 
The work of JG is sponsored by the National Natural Science Foundation of China under Grant No.12275173 and No.11835005.
The work of TJH at Argonne National Laboratory was supported by the U.S.~Department of Energy under contract DE-AC02-06CH11357.
The work of CPY is supported by the U.S. National Science Foundation under Grant No. PHY-2013791. CPY is also grateful for the support from the Wu-Ki Tung endowed chair in particle physics.
The work of KX was performed partly at the Aspen Center for Physics, which is supported by
the U.S. National Science Foundation under Grant No. PHY-1607611 and No. PHY-2210452.
This work used resources of high-performance computing clusters from SMU M2/M3, MSU HPCC, as well as Pitt CRC.
\end{acknowledgments}

\appendix

\section{Neutrino cross sections}
\label{app:num}
In this section, we collect the numerical values of the CT18 predictions
for the neutrino-nucleus charged- and neutral-current deep inelastic scattering cross sections
in Tabs.~\ref{tab:ccscheme}-\ref{tab:ncdis}, with the detailed explanations in Sec~\ref{sec:th}-\ref{sec:PDFs}.

As discussed in Sec.~\ref{sec:aN3LO}, we first present the CT18 next-to-next-to-leading order (NNLO) predictions to the neutrino-isoscalar scattering cross sections in the Zero-Mass Variable-Flavor-Number (ZM-VFN) scheme up to $n_f=4$, in Tabs.~\ref{tab:ccscheme}-\ref{tab:ncscheme}. 
In the corresponding blocks, the second columns indicate to the $K$-factors from the approximate N3LO corrections (N3LO$'$), calculated with the massless N3LO Wilson coefficients with the CT18 NNLO PDFs. In the third column, we show the heavy-quark mass corrections in the ACOT general-mass scheme as the $K$-factor ratios $K_{\rm GM}=\sigma_{\textrm{GM}}/\sigma_{\textrm{ZM}}$, with discussion in Sec.~\ref{sec:ACOT}.
Afterwards, the flavor-number dependence is displayed as the ratios $K_{n_f}=\sigma^{(n_f=6)}/\sigma^{(n_f=4)}$ in the fourth column. Finally, the small-$x$ resummation effect is shown as the fifth column, with details in Sec.~\ref{sec:smallx}. The antineutrino cross sections follow after as another block.

\begin{table}
	\centering
	\begin{tabular}{c||c|c|c|c|c||c|c|c|c|c}
		\hline
		\multirow{2}{*}{$E_\nu~[\GeV]$} &\multicolumn{5}{c||}{$\sigma_{\nu I}^W$} & \multicolumn{5}{c}{$\sigma_{\bar{\nu}I}^W$}\\
		\cline{2-11}
		& NNLO  & $K_{\textrm{N3LO}'}$ & $K_{\rm GM}$  & $K_{n_f}$ & $K_{\textrm{NLLx}}$ 
		& NNLO  & $K_{\textrm{N3LO}'}$ & $K_{\rm GM}$  & $K_{n_f}$ & $K_{\textrm{NLLx}}$ \\
		\hline
		5e1  &  0.301 & 0.992 & 0.984 & 1.000 & 1.000 & 0.133 & 0.990 & 0.978 & 1.000 & 1.000 \\
		1e2  &  0.620 & 0.993 & 0.989 & 0.999 & 1.000 & 0.292 & 0.990 & 0.983 & 1.000 & 1.000 \\
		2e2  &  1.25  & 0.994 & 0.992 & 0.999 & 1.000 & 0.617 & 0.991 & 0.987 & 0.999 & 0.999 \\
		5e2  &  3.09  & 0.995 & 0.995 & 0.999 & 1.000 & 1.60  & 0.993 & 0.992 & 0.999 & 0.999 \\
		1e3  &  6.02  & 0.996 & 0.997 & 0.999 & 0.999 & 3.24  & 0.995 & 0.994 & 0.999 & 0.999 \\
		2e3  &  1.15e1& 0.997 & 0.998 & 0.998 & 0.999 & 6.45  & 0.996 & 0.996 & 0.999 & 0.999 \\
		5e3  &  2.56e1& 0.998 & 0.998 & 0.998 & 0.999 & 1.54e1& 0.997 & 0.998 & 0.998 & 0.998 \\
		1e4  &  4.46e1& 0.998 & 0.999 & 0.997 & 0.999 & 2.87e1& 0.998 & 0.998 & 0.998 & 0.998 \\
		2e4  &  7.37e1& 0.999 & 0.999 & 0.997 & 0.998 & 5.14e1& 0.998 & 0.999 & 0.998 & 0.997 \\
		5e4  &  1.33e2& 0.999 & 0.999 & 0.997 & 0.997 & 1.03e2& 0.998 & 0.999 & 0.998 & 0.996 \\
		1e5  &  1.97e2& 0.999 & 0.999 & 0.997 & 0.996 & 1.64e2& 0.999 & 0.999 & 0.998 & 0.995 \\
		2e5  &  2.85e2& 0.999 & 0.999 & 0.998 & 0.995 & 2.51e2& 0.999 & 1.000 & 0.998 & 0.994 \\
		5e5  &  4.49e2& 0.999 & 1.000 & 0.999 & 0.992 & 4.16e2& 0.999 & 1.000 & 1.000 & 0.992 \\
		1e6  &  6.22e2& 0.999 & 1.000 & 0.999 & 0.990 & 5.92e2& 0.999 & 1.000 & 1.002 & 0.989 \\
		2e6  &  8.47e2& 0.999 & 1.000 & 1.000 & 0.988 & 8.21e2& 0.999 & 1.000 & 1.004 & 0.987 \\
		5e6  &  1.25e3& 0.999 & 1.000 & 1.002 & 0.984 & 1.23e3& 0.999 & 1.000 & 1.007 & 0.984 \\
		1e7  &  1.68e3& 1.000 & 1.000 & 1.003 & 0.982 & 1.66e3& 1.000 & 1.000 & 1.009 & 0.982 \\
		2e7  &  2.20e3& 1.000 & 1.000 & 1.005 & 0.981 & 2.18e3& 1.000 & 1.000 & 1.011 & 0.980 \\
		5e7  &  3.12e3& 1.001 & 1.000 & 1.007 & 0.980 & 3.09e3& 1.001 & 1.000 & 1.014 & 0.980 \\
		1e8  &  4.13e3& 1.001 & 1.000 & 1.008 & 0.980 & 4.09e3& 1.001 & 1.000 & 1.016 & 0.980 \\
		2e8  &  5.25e3& 1.002 & 1.000 & 1.010 & 0.982 & 5.20e3& 1.002 & 1.000 & 1.019 & 0.982 \\
		5e8  &  7.13e3& 1.003 & 1.000 & 1.012 & 0.987 & 7.06e3& 1.003 & 1.000 & 1.021 & 0.987 \\
		1e9  &  8.93e3& 1.003 & 1.000 & 1.014 & 0.993 & 8.84e3& 1.004 & 1.000 & 1.024 & 0.993 \\
		2e9  &  1.11e4& 1.004 & 1.000 & 1.016 & 1.000 & 1.10e4& 1.004 & 1.000 & 1.026 & 1.000 \\
		5e9  &  1.56e4& 1.006 & 1.000 & 1.018 & 1.013 & 1.54e4& 1.006 & 1.000 & 1.028 & 1.013 \\
		1e10 &  1.91e4& 1.007 & 1.000 & 1.019 & 1.024 & 1.89e4& 1.007 & 1.000 & 1.030 & 1.024 \\
		2e10 &  2.34e4& 1.008 & 1.000 & 1.021 & 1.037 & 2.31e4& 1.008 & 1.000 & 1.032 & 1.037 \\
		5e10 &  3.01e4& 1.010 & 1.000 & 1.023 & 1.058 & 2.98e4& 1.010 & 1.000 & 1.034 & 1.058 \\
		1e11 &  3.68e4& 1.012 & 1.000 & 1.026 & 1.076 & 3.64e4& 1.012 & 1.000 & 1.037 & 1.076 \\
		2e11 &  4.42e4& 1.013 & 1.000 & 1.027 & 1.097 & 4.37e4& 1.013 & 1.000 & 1.039 & 1.097 \\
		5e11 &  5.60e4& 1.015 & 1.000 & 1.029 & 1.128 & 5.53e4& 1.016 & 1.000 & 1.041 & 1.128 \\
		1e12 &  6.91e4& 1.017 & 1.000 & 1.031 & 1.155 & 6.83e4& 1.017 & 1.000 & 1.043 & 1.155 \\
		2e12 &  8.16e4& 1.019 & 1.000 & 1.033 & 1.184 & 8.07e4& 1.019 & 1.000 & 1.045 & 1.184 \\
		5e12 &  1.01e5& 1.022 & 1.000 & 1.035 & 1.226 & 1.00e5& 1.022 & 1.000 & 1.047 & 1.226\\
		\hline
	\end{tabular}
	\caption{The charged-current neutrino-isoscalar scattering cross sections $\sigma_{\nu(\bar{\nu})I}^W$ and the corresponding $K$ factors. The first column in each block denotes the next-to-next-to leading order (NNLO) cross section in the Zero-Mass Variable-Flavor-Number (ZM-VFN) scheme up to $n_f=4$. The rest ones are $K$ factors from approximate N3LO $K_{\textrm{N3LO}'}$, the ACOT general-mass scheme $K_{\rm GM}$ (up to $n_f=4$), the six-flavor scheme $K_{n_f}=\sigma^{(n_f=6)}/\sigma^{(n_f=4)}$, as well as the small-$x$ logarithms resummed up to the next-to-leading level $K_{\rm NLLx}$.}
	\label{tab:ccscheme}
\end{table}

\begin{table}
	\centering
	\begin{tabular}{c||c|c|c|c|c||c|c|c|c|c}
		\hline
		\multirow{2}{*}{$E_\nu~[\GeV]$} &\multicolumn{5}{c||}{$\sigma_{\nu I}^Z$} & \multicolumn{5}{c}{$\sigma_{\bar{\nu}I}^Z$}\\
		\cline{2-11}
		& NNLO  & $K_{\textrm{N3LO}'}$ & $K_{\rm GM}$  & $K_{n_f}$ & $K_{\textrm{NLLx}}$ 
		& NNLO  & $K_{\textrm{N3LO}'}$ & $K_{\rm GM}$  & $K_{n_f}$ & $K_{\textrm{NLLx}}$ \\
		\hline
		5e1  & 0.0946  & 0.986 & 1.004 & 1.000 & 1.000 & 0.0487 & 0.982 & 1.007 & 1.000 & 1.000 \\
		1e2  &  0.195  & 0.988 & 1.003 & 1.000 & 1.000 & 0.105  & 0.981 & 1.006 & 1.000 & 1.000 \\
		2e2  &  0.393  & 0.989 & 1.003 & 1.000 & 1.000 & 0.221  & 0.982 & 1.006 & 1.000 & 1.000 \\
		5e2  &  0.977  & 0.991 & 1.003 & 0.999 & 1.000 & 0.569  & 0.987 & 1.005 & 1.000 & 0.999 \\
		1e3  &  1.92   & 0.993 & 1.002 & 0.999 & 1.000 & 1.15   & 0.990 & 1.003 & 1.000 & 0.999 \\
		2e3  &  3.70   & 0.995 & 1.001 & 0.999 & 0.999 & 2.29   & 0.992 & 1.002 & 1.000 & 0.999 \\
		5e3  &  8.46   & 0.996 & 1.001 & 0.999 & 0.999 & 5.51   & 0.995 & 1.000 & 1.001 & 0.999 \\
		1e4  &  1.51e1 & 0.997 & 1.000 & 0.999 & 0.999 & 1.04e1 & 0.996 & 1.000 & 1.001 & 0.998 \\
		2e4  &  2.58e1 & 0.997 & 1.000 & 1.000 & 0.998 & 1.88e1 & 0.997 & 0.999 & 1.002 & 0.998 \\
		5e4  &  4.83e1 & 0.998 & 1.000 & 1.001 & 0.998 & 3.85e1 & 0.997 & 0.999 & 1.003 & 0.997 \\
		1e5  &  7.40e1 & 0.998 & 1.000 & 1.002 & 0.997 & 6.27e1 & 0.998 & 0.999 & 1.003 & 0.996 \\
		2e5  &  1.10e2 & 0.998 & 1.000 & 1.003 & 0.996 & 9.80e1 & 0.998 & 0.999 & 1.004 & 0.995 \\
		5e5  &  1.79e2 & 0.998 & 1.000 & 1.005 & 0.994 & 1.67e2 & 0.998 & 1.000 & 1.005 & 0.993 \\
		1e6  &  2.53e2 & 0.999 & 1.000 & 1.006 & 0.992 & 2.42e2 & 0.998 & 1.000 & 1.006 & 0.991 \\
		2e6  &  3.52e2 & 0.999 & 1.000 & 1.007 & 0.990 & 3.43e2 & 0.999 & 1.000 & 1.007 & 0.989 \\
		5e6  &  5.35e2 & 0.999 & 1.000 & 1.008 & 0.987 & 5.29e2 & 0.999 & 1.000 & 1.009 & 0.987 \\
		1e7  &  7.24e2 & 1.000 & 1.000 & 1.009 & 0.985 & 7.19e2 & 1.000 & 1.000 & 1.009 & 0.985 \\
		2e7  &  9.70e2 & 1.000 & 1.000 & 1.010 & 0.983 & 9.66e2 & 1.000 & 1.000 & 1.010 & 0.983 \\
		5e7  &  1.40e3 & 1.002 & 1.000 & 1.012 & 0.983 & 1.40e3 & 1.002 & 1.000 & 1.012 & 0.982 \\
		1e8  &  1.84e3 & 1.003 & 1.000 & 1.013 & 0.983 & 1.84e3 & 1.003 & 1.000 & 1.013 & 0.983 \\
		2e8  &  2.39e3 & 1.004 & 1.000 & 1.014 & 0.985 & 2.39e3 & 1.004 & 1.000 & 1.014 & 0.985 \\
		5e8  &  3.32e3 & 1.006 & 1.000 & 1.016 & 0.989 & 3.32e3 & 1.006 & 1.000 & 1.016 & 0.989 \\
		1e9  &  4.22e3 & 1.007 & 1.000 & 1.018 & 0.994 & 4.22e3 & 1.007 & 1.000 & 1.018 & 0.994 \\
		2e9  &  5.32e3 & 1.009 & 1.000 & 1.020 & 1.000 & 5.32e3 & 1.009 & 1.000 & 1.020 & 1.000 \\
		5e9  &  7.18e3 & 1.012 & 1.000 & 1.022 & 1.012 & 7.18e3 & 1.012 & 1.000 & 1.022 & 1.012 \\
		1e10 &  8.98e3 & 1.014 & 1.000 & 1.024 & 1.022 & 8.98e3 & 1.014 & 1.000 & 1.024 & 1.022 \\
		2e10 &  1.12e4 & 1.016 & 1.000 & 1.026 & 1.035 & 1.12e4 & 1.016 & 1.000 & 1.026 & 1.035 \\
		5e10 &  1.47e4 & 1.020 & 1.000 & 1.029 & 1.053 & 1.47e4 & 1.020 & 1.000 & 1.029 & 1.053 \\
		1e11 &  1.78e4 & 1.023 & 1.000 & 1.032 & 1.070 & 1.78e4 & 1.023 & 1.000 & 1.032 & 1.070 \\
		2e11 &  2.17e4 & 1.026 & 1.000 & 1.034 & 1.088 & 2.17e4 & 1.026 & 1.000 & 1.034 & 1.088 \\
		5e11 &  2.79e4 & 1.030 & 1.000 & 1.038 & 1.117 & 2.79e4 & 1.030 & 1.000 & 1.038 & 1.117 \\
		1e12 &  3.35e4 & 1.034 & 1.000 & 1.041 & 1.141 & 3.35e4 & 1.034 & 1.000 & 1.041 & 1.141 \\
		2e12 &  4.01e4 & 1.037 & 1.000 & 1.044 & 1.168 & 3.01e4 & 1.038 & 1.000 & 1.044 & 1.168 \\
		5e12 &  5.06e4 & 1.043 & 1.000 & 1.048 & 1.208 & 5.06e4 & 1.043 & 1.000 & 1.048 & 1.208 \\
		\hline
	\end{tabular}
	\caption{Similar to Tab.~\ref{tab:ccscheme}, but for the neutral current case.}
	\label{tab:ncscheme}
\end{table}

In Tabs.~\ref{tab:ccdis}-\ref{tab:ncdis}, we summarize our final CT18 predictions to the neutrino-isoscalar cross sections $\sigma_{\nu I}$, with $n_f=6$ parton flavors and the small-$x$ resummation together with the $K$-factors from the approximate N3LO$'$ and the ACOT general-mass scheme as tabulated in Tabs.~\ref{tab:ccscheme}-\ref{tab:ncscheme}. In the corresponding block, the second column indicates the CT18 proton PDF uncertainty quantified with the traditional Hessian error method at the 68\% confidence level (CL)~\cite{Pumplin:2001ct}.
In the third ($R_{\nu\textrm{O}}$) and fourth columns, we present the nuclear corrections to the oxygen ($_{~8}^{16}$O) target and the corresponding uncertainty, based on the EPPS21 nuclear PDF sets~\cite{Eskola:2021nhw}, explored in Sec.~\ref{sec:nuclear}.
The fifth column ($R_{\rm H_2O/O}$) denotes the water (H$_2$O) nucleon-averaged ratio defined in Eq.~(\ref{eq:RH2O}). The final neutrino-water cross section can be obtained with Eq.~(\ref{eq:H2O}) and uncertainty with Eq.~(\ref{eq:H2Ounc}).

\begin{table}
    \centering
    \begin{tabular}{c||cc|ccc||cc|ccc}
    \hline
$E_\nu~[\GeV]$ 
& $\sigma_{\nu I}^{W}$~[pb] & $\delta\sigma_{\nu I}^{W}~[\%]$ 
& $R_{\nu\textrm{O}}^{W}$ &  $\delta R_{\nu\textrm{O}}^{W}~[\%]$
& $R_{\rm H_2O/O}^{\nu(W)}$  
& $\sigma_{\bar{\nu}I}^{W}$~[pb] & $\delta\sigma_{\bar{\nu}I}^{W}~[\%]$ 
& $R_{\bar{\nu}\textrm{O}}^{W}$ &  $\delta R_{\bar{\nu}\textrm{O}}^{W}~[\%]$
& $R_{\rm H_2O/O}^{\bar{\nu}(W)}$  \\
\hline
5e1  & 0.294  & 1.3 &0.992 &1.0 & 0.962 & 0.129  & 2.7 & 0.984 & 1.8 & 1.036 \\
1e2  & 0.608  & 1.4 &0.991 &1.0 & 0.964 & 0.283  & 2.7 & 0.983 & 1.8 & 1.033 \\
2e2  & 1.23   & 1.4 &0.990 &1.0 & 0.965 & 0.602  & 2.7 & 0.981 & 1.7 & 1.031 \\
5e2  & 3.05   & 1.4 &0.989 &1.0 & 0.967 & 1.58   & 2.6 & 0.978 & 1.6 & 1.029 \\
1e3  & 5.97   & 1.4 &0.987 &1.0 & 0.969 & 3.20   & 2.5 & 0.977 & 1.6 & 1.028 \\
2e3  & 1.14e1 & 1.5 &0.986 &1.0 & 0.970 & 6.38   & 2.5 & 0.975 & 1.5 & 1.027 \\
5e3  & 2.54e1 & 1.5 &0.985 &1.0 & 0.973 & 1.53e1 & 2.4 & 0.973 & 1.5 & 1.025 \\
1e4  & 4.43e1 & 1.5 &0.984 &1.1 & 0.975 & 2.85e1 & 2.3 & 0.972 & 1.5 & 1.024 \\
2e4  & 7.32e1 & 1.6 &0.981 &1.1 & 0.978 & 5.10e1 & 2.2 & 0.969 & 1.6 & 1.022\\
5e4  & 1.32e2 & 1.7 &0.974 &1.3 & 0.983 & 1.02e2 & 2.2 & 0.962 & 1.7 & 1.020 \\
1e5  & 1.96e2 & 1.9 &0.965 &1.5 & 0.987 & 1.63e2 & 2.2 & 0.955 & 1.9 & 1.019 \\
2e5  & 2.83e2 & 2.0 &0.954 &1.9 & 0.991 & 2.49e2 & 2.3 & 0.946 & 2.1 & 1.018 \\
5e5  & 4.44e2 & 2.3 &0.937 &2.4 & 0.997 & 4.12e2 & 2.4 & 0.931 & 2.7 & 1.017 \\
1e6  & 6.14e2 & 2.4 &0.923 &2.9 & 1.000 & 5.87e2 & 2.6 & 0.919 & 3.1 & 1.017 \\
2e6  & 8.36e2 & 2.6 &0.909 &3.5 & 1.004 & 8.13e2 & 2.7 & 0.907 & 3.6 & 1.017 \\
5e6  & 1.24e3 & 2.8 &0.892 &4.3 & 1.008 & 1.22e3 & 2.9 & 0.891 & 4.4 & 1.018 \\
1e7  & 1.66e3 & 3.0 &0.881 &4.9 & 1.011 & 1.64e3 & 3.0 & 0.880 & 5.0 & 1.019 \\
2e7  & 2.17e3 & 3.2 &0.871 &5.5 & 1.013 & 2.16e3 & 3.2 & 0.870 & 5.5 & 1.019 \\
5e7  & 3.08e3 & 3.4 &0.860 &6.1 & 1.016 & 3.07e3 & 3.5 & 0.859 & 6.1 & 1.020 \\
1e8  & 4.08e3 & 3.6 &0.852 &6.6 & 1.018 & 4.08e3 & 3.7 & 0.852 & 6.6 & 1.021 \\
2e8  & 5.21e3 & 4.0 &0.846 &7.0 & 1.019 & 5.21e3 & 4.0 & 0.846 & 7.0 & 1.022 \\
5e8  & 7.14e3 & 4.7 &0.838 &7.5 & 1.020 & 7.14e3 & 4.7 & 0.838 & 7.5 & 1.022 \\
1e9  & 9.02e3 & 5.4 &0.833 &7.8 & 1.021 & 9.02e3 & 5.4 & 0.833 & 7.8 & 1.023 \\
2e9  & 1.14e4 & 6.3 &0.829 &8.1 & 1.022 & 1.14e4 & 6.3 & 0.829 & 8.1 & 1.024 \\
5e9  & 1.62e4 & 7.6 &0.823 &8.5 & 1.023 & 1.62e4 & 7.6 & 0.823 & 8.5 & 1.024 \\
1e10 & 2.01e4 & 9.2 &0.820 &8.7 & 1.024 & 2.01e4 & 9.2 & 0.820 & 8.7 & 1.025 \\
2e10 & 2.49e4 & 11  &0.816 &8.9 & 1.025 & 2.49e4 & 11  & 0.816 & 8.9 & 1.025 \\
5e10 & 3.29e4 & 15  &0.812 &9.2 & 1.025 & 3.29e4 & 15  & 0.812 & 9.2 & 1.026 \\
1e11 & 4.11e4 & 19  &0.809 &9.4 & 1.026 & 4.11e4 & 19  & 0.809 & 9.4 & 1.026 \\
2e11 & 5.05e4 & 24  &0.807 &9.5 & 1.026 & 5.05e4 & 24  & 0.807 & 9.5 & 1.027 \\
5e11 & 6.60e4 & 32  &0.805 &9.7 & 1.027 & 6.60e4 & 23  & 0.805 & 9.7 & 1.027 \\
1e12 & 8.36e4 & 40  &0.804 &9.9 & 1.027 & 8.36e4 & 40  & 0.804 & 9.9 & 1.027 \\
2e12 & 1.02e5 & 51  &0.804 &10  & 1.027 & 1.02e5 & 51  & 0.804 & 10  & 1.027 \\
5e12 & 1.31e5 & 73  &0.805 &10  & 1.027 & 1.31e5 & 73  & 0.805 & 10  & 1.027 \\
    \hline
    \end{tabular}
    \caption{Charged current neutrino-isoscalar scattering cross sections $\sigma^W_{\nu(\bar{\nu})I}$, calculated with the CT18 PDFs with flavor up to $n_f=6$, approximate N3LO corrections (N3LO$'$), as well as small-$x$ resummed up to next-to-leading logarithmic level (NLLx). The 
    $\delta\sigma_{\nu(\bar{\nu})I}$ column indicates the isoscalar uncertainty from proton PDFs at 68\% CL, while the nuclear correction as cross-section ratios of O to isoscalar targets $R_{\nu(\bar{\nu})\textrm{O}}$ are obtained with EPPS21 and uncertainty folded in $\delta R_{\nu(\bar{\nu})\textrm{O}}$. The water cross section ratios $R_{\rm H_{2}O/O}$ are defined in Eq.~(\ref{eq:RH2O}), with the final cross section propagated with Eq.~(\ref{eq:H2O}) and uncertainty with Eq.~(\ref{eq:H2Ounc}).}
    \label{tab:ccdis}
\end{table}

\begin{table}
    \centering\small
    \begin{tabular}{c||cc|ccc||cc|ccc}
    \hline
$E_\nu~[\GeV]$ 
& $\sigma_{\nu I}^{Z}$~[pb] & $\delta\sigma_{\nu I}^{Z}~[\%]$ 
& $R_{\nu\textrm{O}}^{Z}$ &  $\delta R_{\nu\textrm{O}}^{Z}~[\%]$
& $R_{\rm H_2O/O}^{\nu(Z)}$  
& $\sigma_{\bar{\nu} I}^{Z}$~[pb] & $\delta\sigma_{\bar{\nu} I}^{Z}~[\%]$
& $R_{\bar{\nu}\textrm{O}}^{Z}$ &  $\delta R_{\bar{\nu}\textrm{O}}^{Z}~[\%]$
& $R_{\rm H_2O/O}^{\bar{\nu}(Z)}$ \\
\hline
5e1  & 0.0935 & 1.2 & 0.992 & 0.9 & 0.994 & 0.0480 & 2.1 & 0.986 & 1.3 & 1.001 \\
1e2  & 0.193  & 1.3 & 0.991 & 0.9 & 0.995 & 0.104  & 2.1 & 0.985 & 1.3 & 1.001 \\
2e2  & 0.389  & 1.3 & 0.990 & 0.9 & 0.995 & 0.217  & 2.1 & 0.985 & 1.2 & 1.001 \\
5e2  & 0.969  & 1.2 & 0.989 & 0.9 & 0.995 & 0.563  & 2.0 & 0.983 & 1.2 & 1.002 \\
1e3  & 1.90   & 1.2 & 0.987 & 0.9 & 0.996 & 1.14   & 1.9 & 0.982 & 1.1 & 1.002 \\
2e3  & 3.67   & 1.2 & 0.986 & 0.9 & 0.996 & 2.27   & 1.8 & 0.980 & 1.1 & 1.002 \\
5e3  & 8.42   & 1.2 & 0.985 & 0.9 & 0.997 & 5.48   & 1.7 & 0.979 & 1.2 & 1.002 \\
1e4  & 1.51e1 & 1.1 & 0.984 & 0.9 & 0.997 & 1.03e1 & 1.6 & 0.978 & 1.2 & 1.002 \\
2e4  & 2.57e1 & 1.2 & 0.981 & 1.0 & 0.998 & 1.88e1 & 1.5 & 0.976 & 1.2 & 1.002 \\
5e4  & 4.82e1 & 1.2 & 0.974 & 1.1 & 0.999 & 3.84e1 & 1.5 & 0.971 & 1.4 & 1.002 \\
1e5  & 7.38e1 & 1.3 & 0.966 & 1.3 & 1.001 & 6.25e1 & 1.5 & 0.965 & 1.6 & 1.003 \\
2e5  & 1.10e2 & 1.4 & 0.954 & 1.6 & 1.002 & 9.77e1 & 1.6 & 0.957 & 1.8 & 1.004 \\
5e5  & 1.78e2 & 1.6 & 0.947 & 2.1 & 1.005 & 1.67e2 & 1.7 & 0.943 & 2.3 & 1.006 \\
1e6  & 2.52e2 & 1.7 & 0.934 & 2.7 & 1.007 & 2.41e2 & 1.8 & 0.931 & 2.8 & 1.008 \\
2e6  & 3.51e2 & 1.9 & 0.921 & 3.2 & 1.009 & 3.42e2 & 1.9 & 0.918 & 3.3 & 1.009 \\
5e6  & 5.33e2 & 2.1 & 0.904 & 4.0 & 1.011 & 5.26e2 & 2.2 & 0.903 & 4.1 & 1.012 \\
1e7  & 7.20e2 & 2.3 & 0.892 & 4.6 & 1.013 & 7.14e2 & 2.4 & 0.891 & 4.6 & 1.013 \\
2e7  & 9.64e2 & 2.6 & 0.881 & 5.2 & 1.015 & 9.60e2 & 2.6 & 0.881 & 5.2 & 1.015 \\
5e7  & 1.40e3 & 2.9 & 0.869 & 5.9 & 1.016 & 1.40e3 & 2.9 & 0.869 & 5.9 & 1.017 \\
1e8  & 1.84e3 & 3.2 & 0.861 & 6.4 & 1.018 & 1.83e3 & 3.2 & 0.861 & 6.3 & 1.018 \\
2e8  & 2.40e3 & 3.6 & 0.854 & 6.8 & 1.019 & 2.41e3 & 3.6 & 0.854 & 6.8 & 1.019 \\
5e8  & 3.36e3 & 4.2 & 0.846 & 7.3 & 1.020 & 3.42e3 & 4.2 & 0.846 & 7.3 & 1.020 \\
1e9  & 4.30e3 & 4.8 & 0.840 & 7.6 & 1.021 & 4.30e3 & 4.8 & 0.840 & 7.6 & 1.021 \\
2e9  & 5.48e3 & 5.6 & 0.835 & 7.9 & 1.022 & 5.48e3 & 5.6 & 0.835 & 7.9 & 1.022 \\
5e9  & 7.51e3 & 7.0 & 0.830 & 8.3 & 1.023 & 7.51e3 & 7.0 & 0.830 & 8.3 & 1.023 \\
1e10 & 9.54e3 & 8.4 & 0.826 & 8.5 & 1.023 & 9.54e3 & 8.4 & 0.826 & 8.5 & 1.023 \\
2e10 & 1.21e4 & 10  & 0.822 & 8.8 & 1.024 & 1.21e4 & 10  & 0.822 & 8.8 & 1.024 \\
5e10 & 1.63e4 & 13  & 0.817 & 9.0 & 1.025 & 1.63e4 & 13  & 0.817 & 9.0 & 1.025 \\
1e11 & 2.01e4 & 16  & 0.814 & 9.2 & 1.025 & 2.01e4 & 16  & 0.814 & 9.2 & 1.025 \\
2e11 & 2.51e4 & 21  & 0.812 & 9.4 & 1.026 & 2.51e4 & 21  & 0.812 & 9.4 & 1.026 \\
5e11 & 3.33e4 & 28  & 0.809 & 9.6 & 1.026 & 3.33e4 & 28  & 0.809 & 9.6 & 1.026 \\
1e12 & 4.11e4 & 35  & 0.808 & 9.7 & 1.026 & 4.11e4 & 35  & 0.808 & 9.8 & 1.026 \\
2e12 & 5.07e4 & 45  & 0.808 & 9.9 & 1.026 & 5.07e4 & 45  & 0.808 & 9.9 & 1.026 \\
5e12 & 6.67e4 & 61  & 0.809 & 10  & 1.026 & 6.67e4 & 61  & 0.809 & 10  & 1.026 \\
    \hline
    \end{tabular}
    \caption{Similar to Tab.~\ref{tab:ccdis}, but for the neutral current case.
    }
    \label{tab:ncdis}
\end{table}

\clearpage

\section{The important $(x,Q)$ kinematics}
\label{app:xQlimit}

\begin{figure}
\centering
\includegraphics[width=0.32\textwidth]{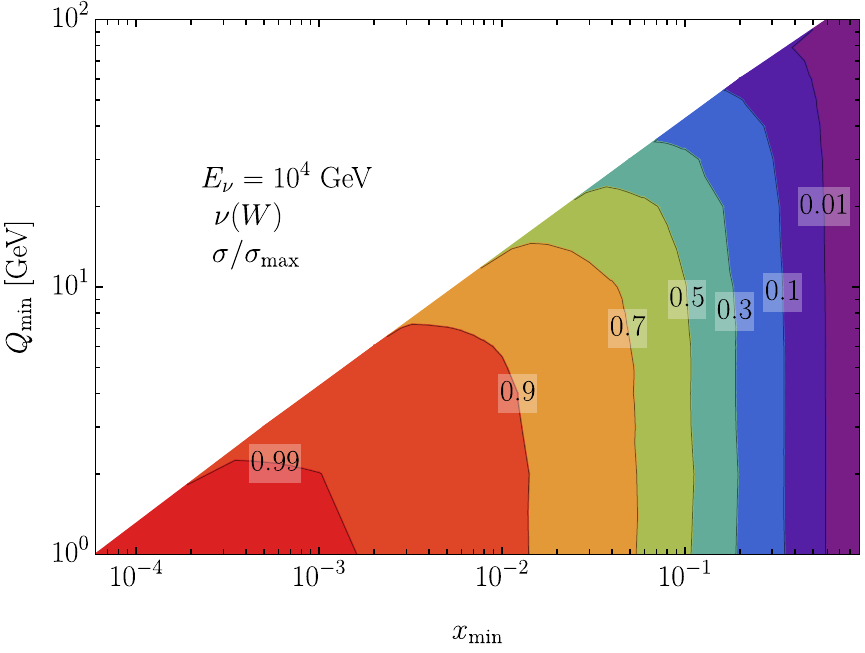}
\includegraphics[width=0.32\textwidth]{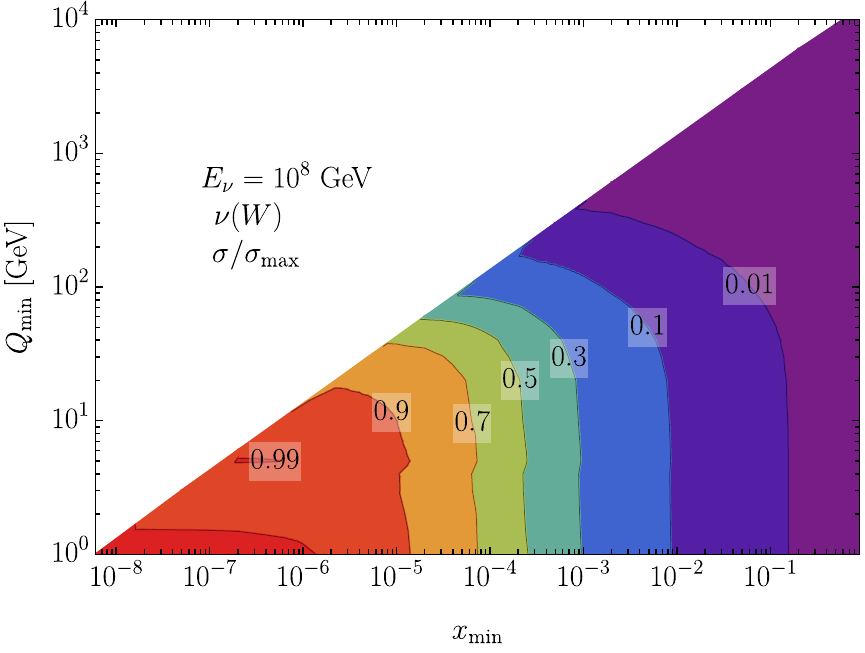}
\includegraphics[width=0.32\textwidth]{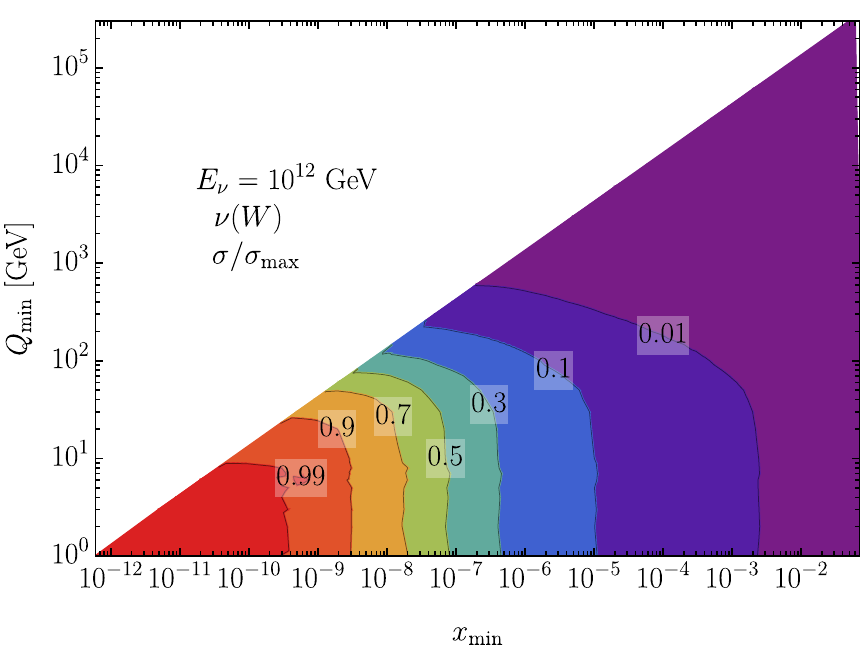}
\includegraphics[width=0.32\textwidth]{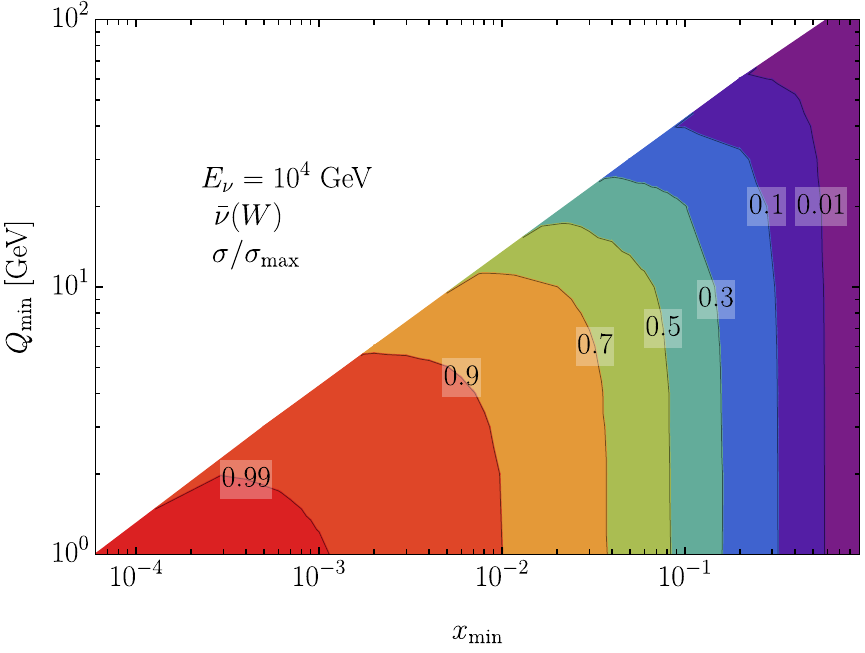}
\includegraphics[width=0.32\textwidth]{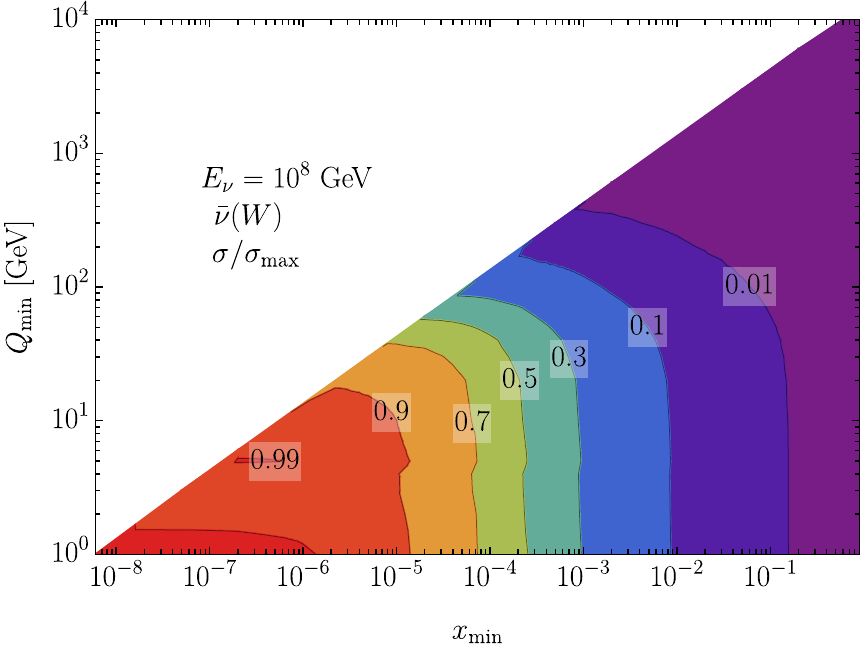}
\includegraphics[width=0.32\textwidth]{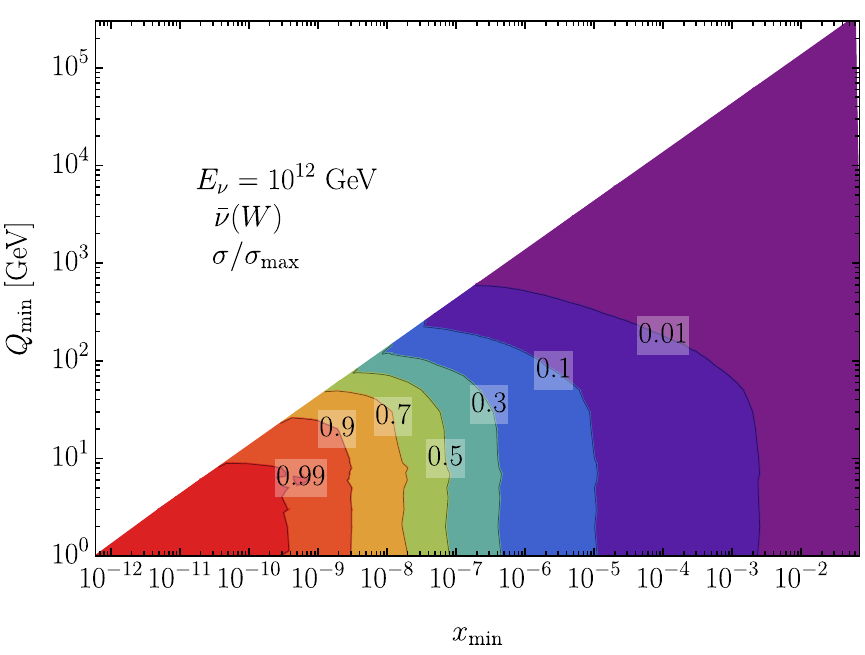}
\includegraphics[width=0.32\textwidth]{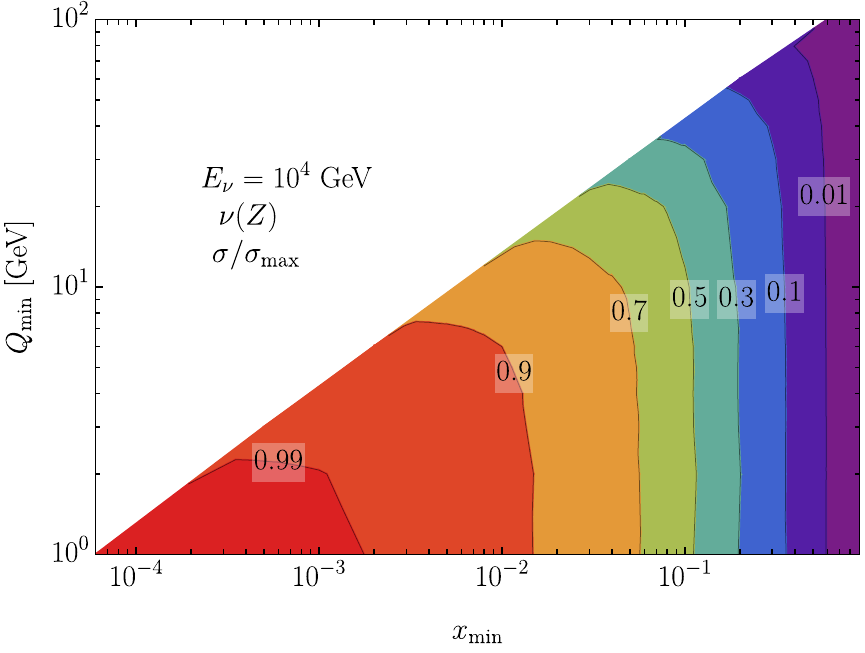}
\includegraphics[width=0.32\textwidth]{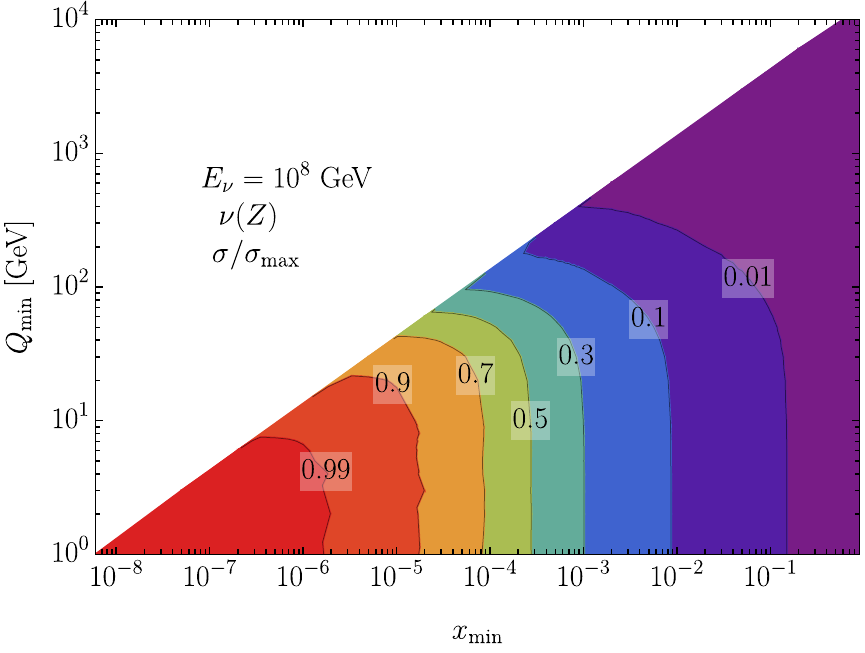}
\includegraphics[width=0.32\textwidth]{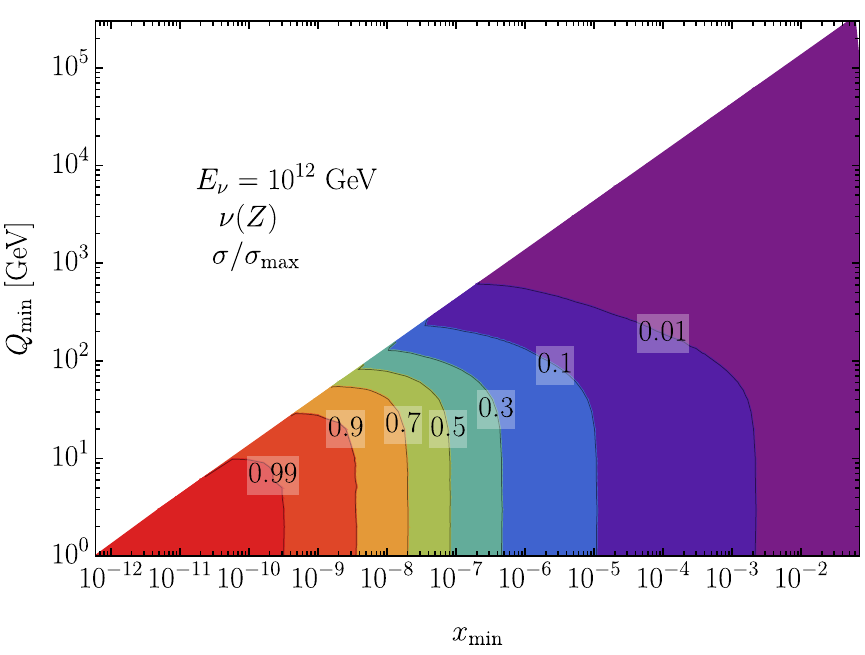}
\includegraphics[width=0.32\textwidth]{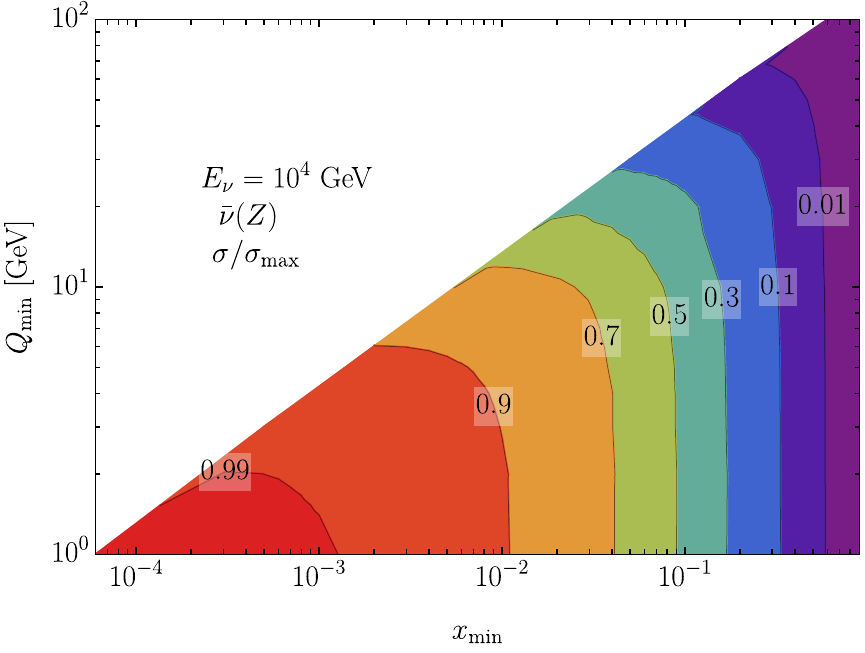}
\includegraphics[width=0.32\textwidth]{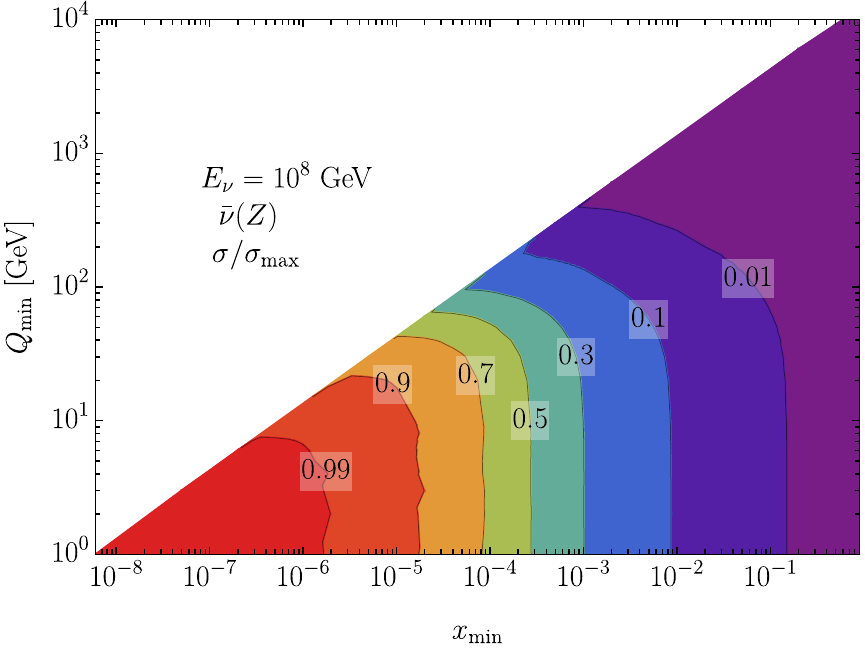}
\includegraphics[width=0.32\textwidth]{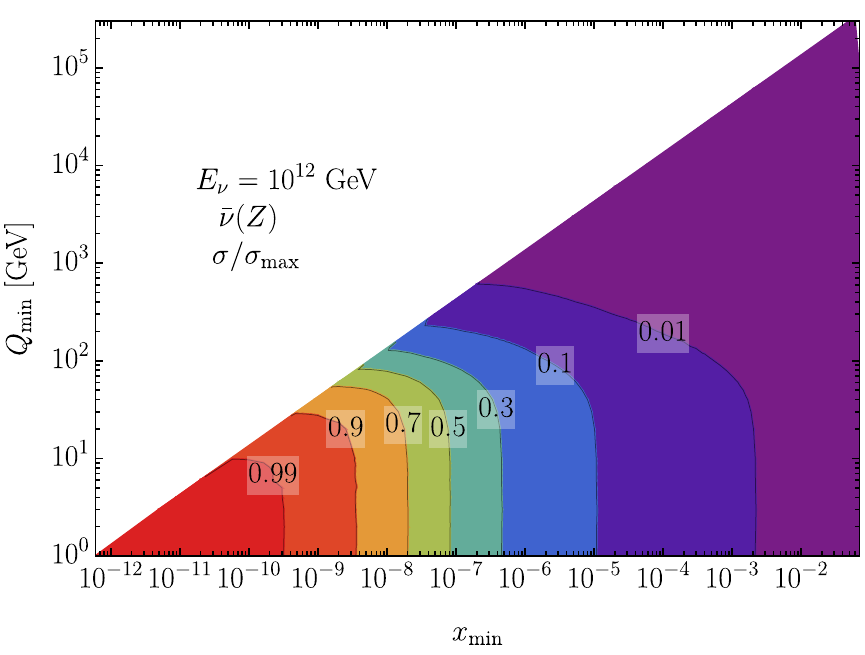}
    \caption{The ratio contours of neutrino-isoscalar DIS cross sections with variations of integration limits $(x_{\min},Q_{\min})$, with respect to the maximal one with 
    $x_{\min}=\frac{Q^2}{2m_NE_\nu}$ and $Q_{\min}=1~\GeV$.}
    \label{fig:xQmin}
\end{figure}

\begin{figure}
    \centering
\includegraphics[width=0.45\textwidth]{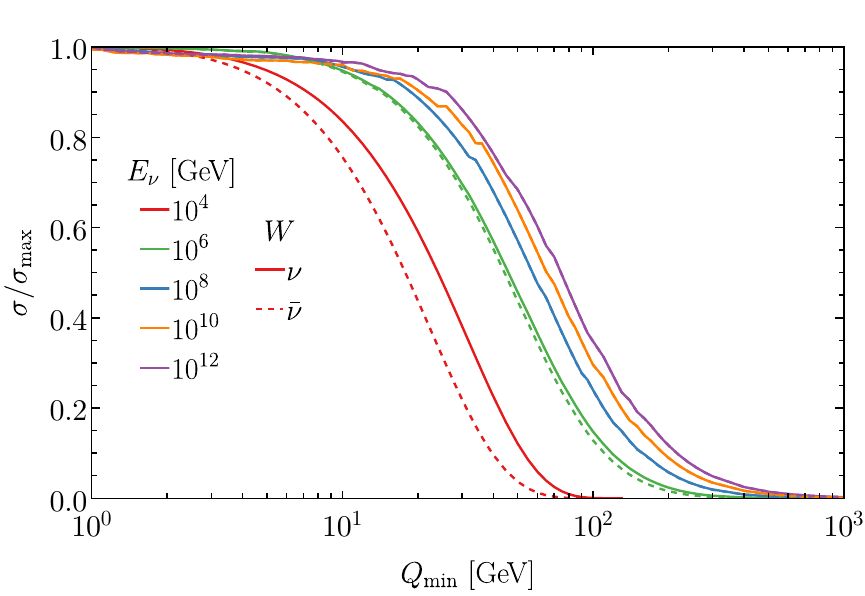}
\includegraphics[width=0.45\textwidth]{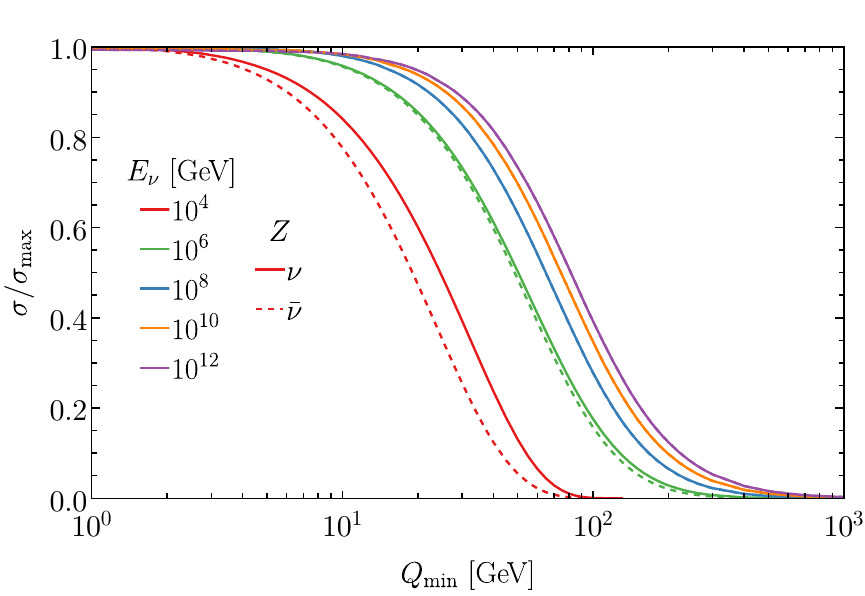}    
    \caption{Similar to Fig.~\ref{fig:scanQ}, but with an extended $Q_{\min}$ scan, as a projection of Fig.~\ref{fig:xQmin} to the $Q_{\min}$ direction with $x_{\min}=Q_{\min}^2/(2m_NE_\nu)$.}
    \label{fig:scanQ2}
\end{figure}

\begin{figure}
    \centering
\includegraphics[width=0.45\textwidth]{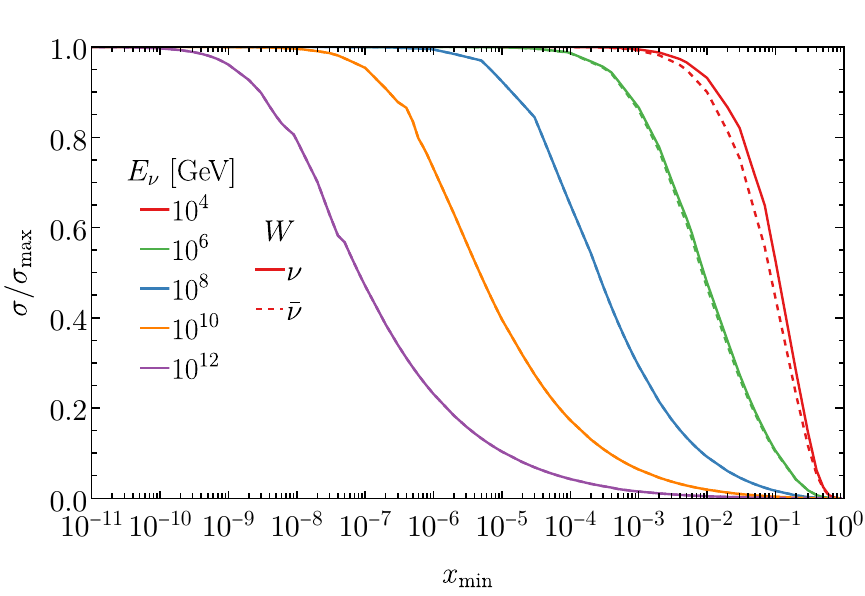}
\includegraphics[width=0.45\textwidth]{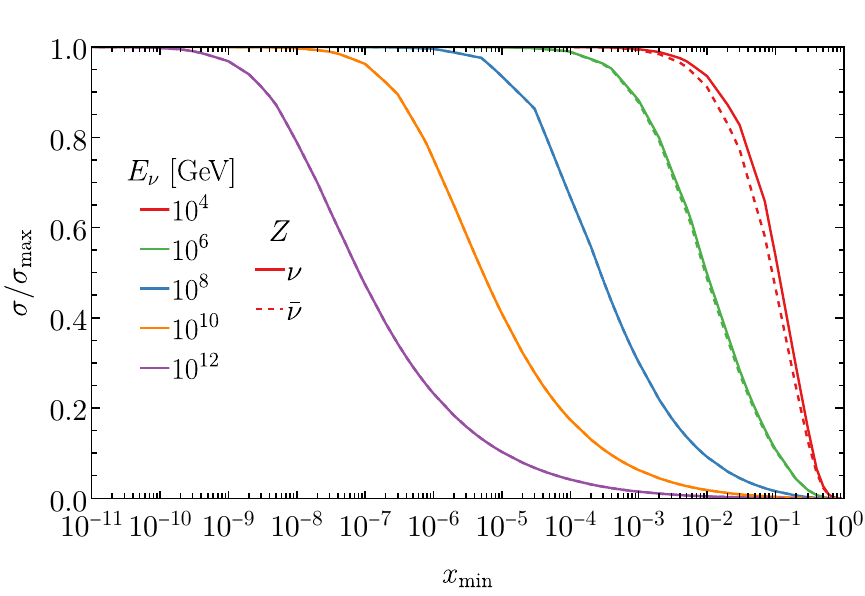}    
    \caption{Similar to Fig.~\ref{fig:scanx}, but with an extended $x_{\min}$ scan as a projection of Fig.~\ref{fig:xQmin} to the $x_{\min}$ direction.}
    \label{fig:scanx2}
\end{figure}

In Sec.~\ref{sec:xQ}, we have performed the two separate scans over $x_{\min}$ and $Q_{\min}$ in the low-$x$ and low-$Q$ region.
We found the extrapolation region only makes up to a percent level for the total cross sections.
These scans can be extended to higher $x$ and larger $Q$ values,
which can give us an idea of the integration contribution from various kinematic $(x,Q)$ regions.   
A more transparent and direct way can come from a joint two-dimensional scan of $x_{\min}$ and $Q_{\min}$.
Therefore, we perform the integration of Eq.~(\ref{eq:intXS}) in the trapezoid region,
\begin{equation}
    Q\in[Q_{\min},\sqrt{2m_NE_\nu}], 
    x\in[\max(x_{\min},Q^2/(2m_NE_\nu),1].
\end{equation}
With a few representative energies, we show the percentage of the total cross section as contour plots in Fig.~\ref{fig:xQmin}.
The projections to one-dimensional $x_{\min}$ and $Q_{\min}$ directions 
are plotted in Figs.~\ref{fig:scanQ2}-\ref{fig:scanx2}, similar to Figs.~\ref{fig:scanQ}-\ref{fig:scanx}.
We found that the integrated cross sections decreases 
drastically around $Q_{\min}\sim M_{W,Z}$ and $x\sim x_{W,Z}=M_{W,Z}^2/(2m_NE_\nu)$.
It indicates that a large part of the total cross section comes from the integration around the region $(x,Q)\sim(M_{W,Z}^2/(2m_NE_\nu),M_{W,Z})$, which we call the {\it important $(x,Q)$ kinematics}.
This can be understood naturally in terms of the integration in Eq.~(\ref{eq:diffXS_mass}).
On one side in the asymptotic limit $Q\gg M_{W,Z}$, the integrand dies away quickly due to the suppression from the prefactor, $1/(1+Q^2/M_{W,Z}^2)^2$.
That is, the large-$Q$ region does not contribute much to the total cross section in the integration.
Conversely, for $Q\ll M_{W,Z}$, the prefactor becomes a constant; in that case, the integration is mainly driven by the region with large structure functions. 
In comparison with $F_2$, $F_L$ and $F_3$ are generally small. 
Due to the DGLAP evolution effect, $F_2$ generally increases with the scale, $Q^2$, which can be seen in some
simplistic parametric forms, such as those of Refs.~\cite{Haidt:1999ps,Block:2006dz,Berger:2007vf}.
Consequently, the weight contribution to the cross-section integration grows with energy $Q$, which is explored in Ref.~\cite{Illarionov:2011wc} and leads to the Froissart bound~\cite{Froissart:1961ux,Martin:2009iq}.

\section{The Earth absorption}
\label{app:abspt}
The neutrinos measured at the IceCube can come from both the northern and southern skies. The neutrinos from the northern sky need to pass through the whole earth before arriving at the IceCube detectors located at the geographic South Pole.
As a result, the upward-going events, proportional to the neutrino arrival flux with the zenith angle $\theta\in[90^\circ,180^\circ]$, are subject to the Earth's absorption rate, which in turn depends on neutrino scattering cross sections.
The IceCube collaboration has taken the CSMS calculation~\cite{Cooper-Sarkar:2011jtt} in the corresponding experimental simulation and event analyses,
for the neutrino DIS cross sections based on up-going muon neutrinos~\cite{IceCube:2017roe} and the high-energy starting events (HESE) sample~\cite{IceCube:2020rnc}.

\begin{figure}
    \centering
    \includegraphics[width=0.41\textwidth]{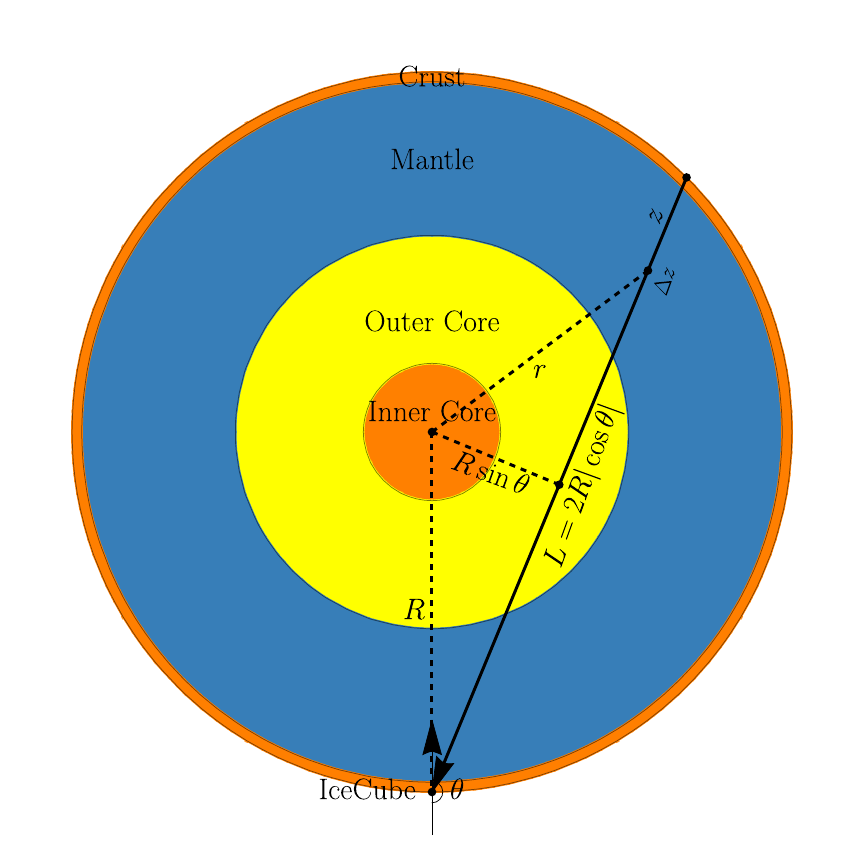}
    \includegraphics[width=0.58\textwidth]{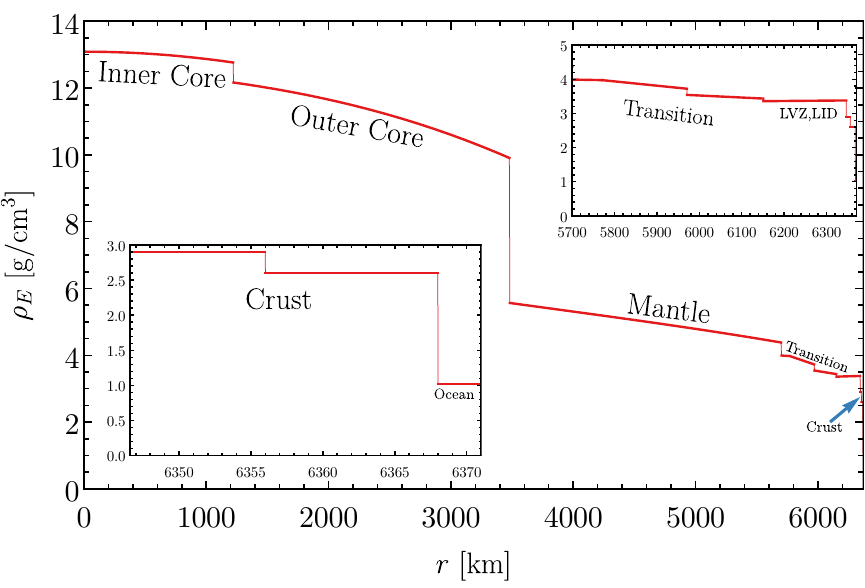}    
    \caption{(Left) A schematic path for neutrino traveling through the chord in the direction of zenith angle $\theta$. The Earth's internal structure is assumed to be spherical symmetric, with density (right) taken from the Preliminary Reference Earth Model (PREM)~\cite{Dziewonski:1981xy}.}
    \label{fig:earth}
\end{figure}

Recall that high-energy neutrinos detected at IceCube mainly come from atmospheric and astrophysical resources~\cite{IceCube:2015gsk,IceCube:2020wum}. Based on the simplest single-power-law flux model, 
the neutrino flux can be parameterized as
\begin{equation}
\Phi_{\nu}(E_\nu)=\phi\cdot\left(\frac{E_\nu}{100~\rm TeV}\right)^{-\gamma},
\end{equation}
where $\phi$ indicates the value at $E_\nu=100~\TeV$, and $\gamma$ is the power law spectral index, with both fitted from astrophysical data~\cite{IceCube:2015gsk,IceCube:2020wum}.
The event rate of upward-going neutrino-induced muons observed in the IceCube observatory can be written as
\begin{equation}\label{eq:event}
N_{\nu(\bar{\nu})}(E_\nu,\theta)\sim
\sigma_{\nu(\bar{\nu})}^{W}(E_\nu)
\Phi_{\nu}(E_\nu)
\mathcal{P}_{\rm trans}(E_\nu,\theta).
\end{equation}
Here $\sigma_{\nu(\bar{\nu})}^{W}(E_\nu)$ is the charged-current cross section of the (anti)neutrino scattered with the IceCube material, such as H$_2$O.
$\mathcal{P}_{\rm trans}(E_\nu,\theta)$ denotes the neutrino's transmission (or survival) probability when passing through the earth in the direction of zenith angle $\theta$.
For neutrino travels through the path of chord $z\in[0,L=2R|\cos\theta|]$ as shown in Fig.~\ref{fig:earth} (left), the transmission probability can be obtained in terms of
\begin{eqnarray}\label{eq:dP}
\mathcal{P}_{\rm trans}(E_\nu,\theta)
&=\prod_{\Delta z}P_{\alpha\alpha}(E_\nu,\Delta z)
\exp{-\Delta z/\lambda(r,E_\nu)}.
\end{eqnarray}
$P_{\alpha\alpha}(E_\nu,\Delta z)$ is the oscillation probability of neutrino flavor $\alpha=e,\mu,\tau$. Also,
$\lambda(r,E_\nu)=\frac{1}{n_N(r)\sigma_{\nu(\bar{\nu})}(E_\nu)}$ is the mean-free path, which depends on the earth's nucleon density $n_N(r)=\rho_E(r)/m_N$ at a distance to the earth center as
\begin{eqnarray}
r=\sqrt{(R\sin\theta)^2+(R|\cos\theta|-z)^2}=
\sqrt{R^2+z^2+2Rz\cos\theta},
\end{eqnarray}
as well as the absorption cross section $\sigma_{\nu(\bar{\nu})}(E_\nu)=\sigma_{\nu(\bar{\nu})}^{W}(E_\nu)+\sigma_{\nu(\bar{\nu})}^{Z}(E_\nu)$, including both NC and CC scatterings.
In principle, neutrino-electron scattering should be included as well. However, the corresponding cross section only contributes at most $2\sim3\%$ to the total rate~\cite{Gauld:2019pgt}, with an exception around the Glashow resonance region.\footnote{See App.~\ref{sec:Glashow} for its details.
Meanwhile, as pointed in Refs.~\cite{Zhou:2019vxt,Zhou:2019frk}, $W$-boson production (WBP) will also contribute to the Earth's absorption, albeit with a smaller size, due to the suppression of the nucleon's photon content as well as the $W$-boson threshold.}
As a consequence, we will neglect these minor effects in this estimation.

Moreover, we consider the earth's isotopic constituents as isoscalars and neglect the nuclear effect, which in principle can be included when knowing the earth's element abundances~\cite{Klein:2020nuk}.
Furthermore, the NC interactions, as well as the $\tau$ decays in $\nu_{\tau}$ CC scattering, will ``regenerate" neutrinos with lower energies, which is not considered here.
Integrating out the exponent in Eq.~(\ref{eq:dP}) ends up with~\cite{Gonzalez-Garcia:2007wfs}
\begin{equation}
\mathcal{P}_{\rm trans}(E_\nu,\theta)=
P_{\alpha\alpha}(E_\nu,L)
\exp{-X(\theta)\sigma(E_\nu)},
\end{equation}
where 
\begin{equation}
P_{\alpha\alpha}(E_\nu,L)=1-\sin^22\theta_{\alpha\alpha}\sin^2\frac{\Delta m^{2}_{\alpha\alpha}L}{4E_\nu}\simeq1
\end{equation}
when $E_\nu\gtrsim1~\TeV$~\cite{Zyla:2020zbs}, and
\begin{equation}
    X(\theta)=\int_{0}^{L=2R|\cos\theta|}\dd z\rho_E(\sqrt{R^2+z^2+2Rz\cos\theta})/m_N.
\end{equation}
Here, $\rho_E(r)$ is the earth matter density, which can be assumed to be spherically symmetric from the 
Preliminary Reference Earth Model (PREM)~\cite{Dziewonski:1981xy},
as shown in Fig.~\ref{fig:earth} (right).

\begin{figure}
    \centering  
    \includegraphics[width=0.49\textwidth]{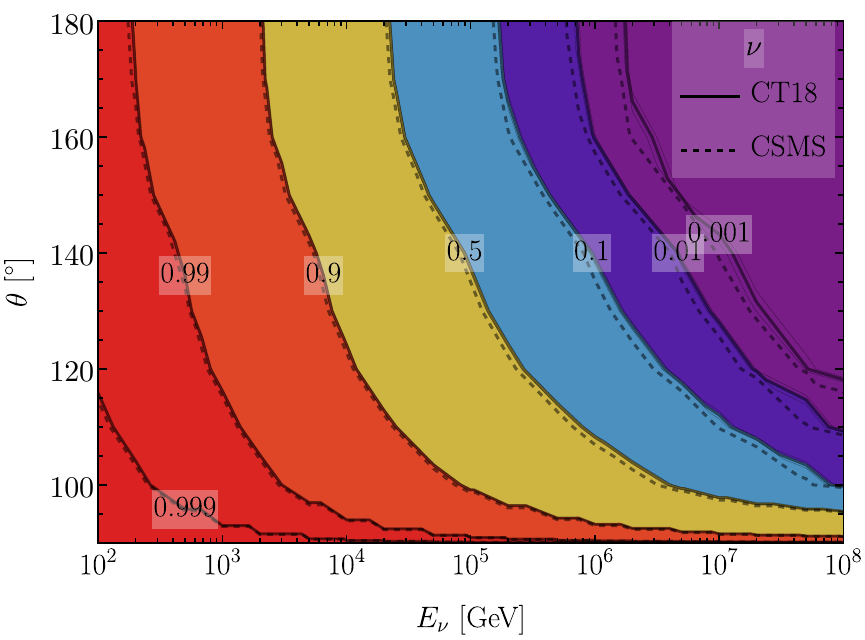}
    \includegraphics[width=0.49\textwidth]{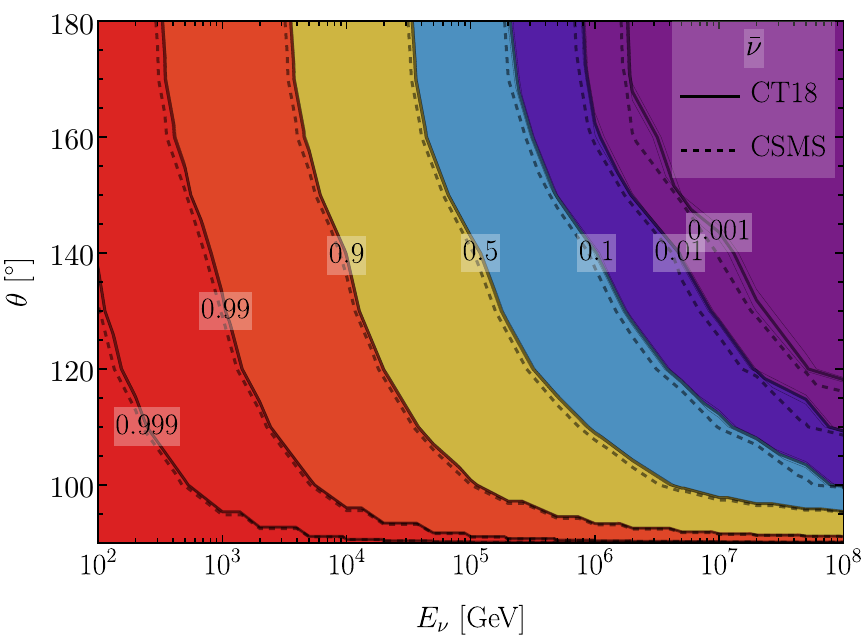}    
    \caption{The neutrino (left) and antineutrino (right) transmission probability $\mathcal{P}_{\rm trans}$ when passing the Earth, based on the CT18 and CSMS predictions for neutrino-isoscalar charged- and neutral-current DIS cross sections. The small thin bands correspond to the CT18 PDF variation.}
    \label{fig:trans}
\end{figure}
Knowing the neutrino cross sections from the CT18 and CSMS predictions, we can calculate the transmission probability, with results shown in Fig.~\ref{fig:trans}, with the CSMS one agreeing with the IceCube simulation quite well~\cite{IceCube:2017roe}.
However, we note that as shown in Fig.~\ref{fig:compare}, the CT18 prediction yields a smaller cross section than the CSMS one by about 10 percent, hence giving a larger transmission probability. Consequently, from Eq.~(\ref{eq:event}), for a given number of observed event $N_{\nu(\bar{\nu})}$, the CT18 predictions would give a smaller product of $\sigma_{\nu(\bar{\nu})}^{W}\Phi_{\nu}(E_\nu)$ as compared to the CSMS prediction.
This would have an entangled impact on the final measured cross section as well as other parameters, such as $\phi,~\gamma$ obtained by the IceCube analysis~\cite{IceCube:2017roe,Miarecki:2016kku}. 


Assuming a well-determined astrophysical neutrino flux, the IceCube Observatory may measure the neutrino cross section {\em without} being affected by the earth's absorption.
In other words, neutrino events {\em from the Southern Sky} could be used to determine the neutrino cross section in better systematics if enough statistics, due to the advantage of being free from the earth model of the nuclear isotopic abundance, density, and other parameters. However, this method faces a disadvantage due to a large background of cosmic-ray muon events in the ice Cherenkov detector as well as lower statistics. A more complete work remains to be done with experimental simulations.

\section{Neutrino-electron scatterings and the Glashow Resonance}
\label{sec:Glashow}

\begin{figure}
    \centering
    \includegraphics[width=0.32\textwidth]{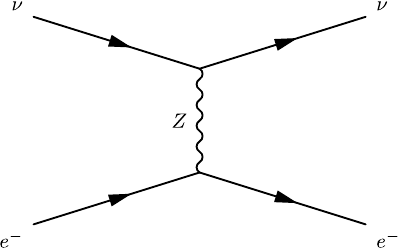}
    \includegraphics[width=0.32\textwidth]{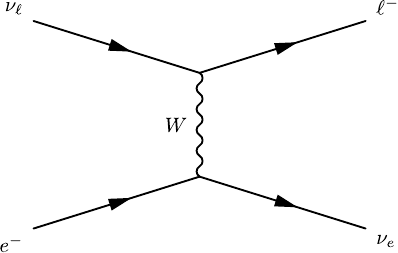}    
    \includegraphics[width=0.32\textwidth]{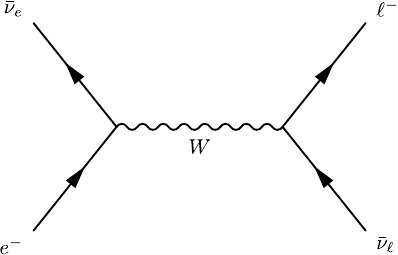}    
    \caption{Feynman diagrams for the neutrino-electron scattering.}
    \label{fig:feynve}
\end{figure}

\begin{figure}
    \centering
    \includegraphics{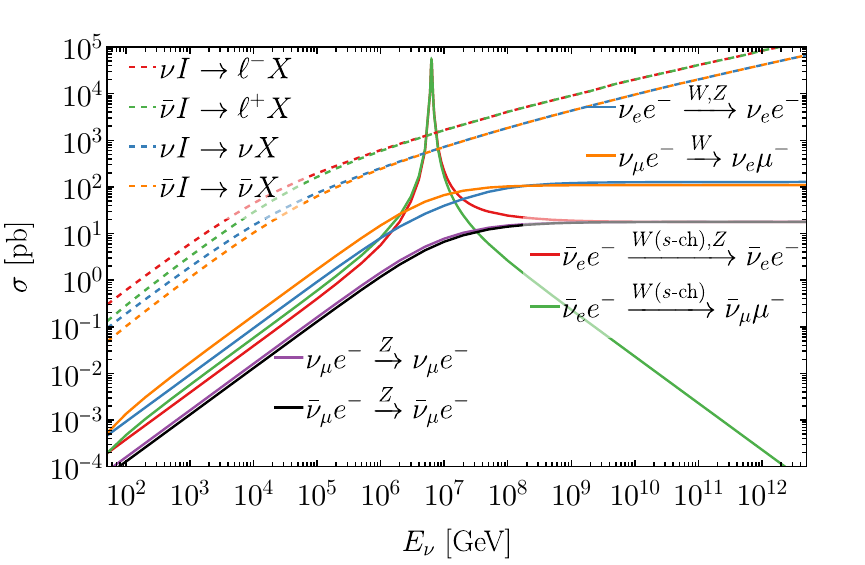}
    \caption{Comparison of the neutrino-electron and neutrino-isoscalar DIS scattering cross sections. The neutrino-isoscalar cross section is directly taken from the CT18 predictions without nuclear corrections.}
    \label{fig:Glashow}
\end{figure}

In addition to scattering from QCD matter as explored above, neutrinos can also interact with electrons, as representative diagrams shown in Fig.~\ref{fig:feynve}.
Different from neutrino-nucleon scattering, which involves nonperturbative parton structures, neutrino-electron scattering is fully perturbative in terms of the EW interaction.
We show the corresponding cross sections compared with the (anti)neutrino-nucleon DIS cross section without nuclear corrections in Fig.~\ref{fig:Glashow}.
The detailed description of the (anti)neutrino-nucleon cross section can be found in Sec.~\ref{sec:smallx}.
Generally speaking, the (anti)neutrino-electron cross sections are two or three magnitudes smaller than the neutrino-nucleon ones.

Similar to nucleon ones, neutrino-electron scatterings involve both the charged and neutral current interactions mediated by $W$ and $Z$ bosons. The neutral-current couplings 
are smaller than the charged-current ones, which explains the smaller cross sections for the processes $\nu_{\mu}e^-\xrightarrow{Z}\nu_\mu e^-$ and $\bar{\nu}_\mu e^-\xrightarrow{Z}\bar{\nu}_\mu e^-$ than 
$\nu_{e}e^-\xrightarrow{W,Z}\nu_e e^-$ and $\nu_\mu e^-\xrightarrow{W}\nu_e \mu^-$.
For sole NC scatterings, $\nu_\mu e^-\xrightarrow{Z}\nu_{\mu} e^-$ is more-or-less the same for $\bar{\nu}_{\mu} e^-\xrightarrow{Z} \bar{\nu}_{\mu} e^-$ due to the same coupling, while the minor difference is originated from the spin correlation.
In comparison, $\nu_e e^-\xrightarrow{W,Z}\nu_e e^-$ gives a smaller (larger) cross section than $\nu_\mu e^-\xrightarrow{W}\nu_e\mu^-$ when $E_\nu\lesssim10^{8}~\GeV(E_\nu\gtrsim10^{8}~\GeV)$, as a result of the destructive (constructive) interference between CC and NC interactions.
The tau-neutrino cross sections behave more-or-less the same as the muonic ones, with only a minor correction from the heavier tau mass whenever a final-state tau lepton shows up.

Distinct from scattering off nucleons, a new process emerges in the form of $s$-channel Glashow resonance~\cite{Glashow:1960zz} production, depicted in Fig.~\ref{fig:feynve} (right).
The corresponding cross section can be written as~\cite{Barger:2014iua}
\begin{equation}
\sigma_{W(s\textrm{-ch})}=24\pi\Gamma_W^2 
\mathcal{B}(W^-\to e^-\bar{\nu}_e)\mathcal{B}(W^-\to f\bar{f}')
\frac{s/M_W^2}{(s-M_W^2)^2+(M_W\Gamma_W)^2},
\end{equation}
where $\mathcal{B}(W^-\to f\bar{f}')$ is the corresponding decay branch fraction.
Here, $W$ bosons can subsequently decay into leptons as well as hadronic final states. The resonance peak appears at
\begin{equation}
s=2m_eE_\nu=M_W^2\implies E_\nu=\frac{M_W^2}{2m_e}\simeq 6.32\times10^{6}~\GeV.
\end{equation}
A sizable higher-order correction is found in Ref.~\cite{Gauld:2019pgt}, which is beyond the precision in our estimation.
 In comparison with other resonance decay channels, the $\bar{\nu}_e e^-\xrightarrow{W(s\textrm{-ch}),Z}\bar{\nu}_e e^-$ scattering also involves the neutral-current interaction, via the $t$-channel $Z$-mediated diagram also shown in Fig.~\ref{fig:feynve}, which dominates the high-energy tail.
 In contrast, the $\bar{\nu}_ee^-\xrightarrow{W(s\textrm{-ch})}\bar{\nu}_\mu\mu^-$ scattering involves only the CC interactions via $s$-channel diagram, resulting in a suppressed cross section at high neutrino energies.
In the IceCube measurement, one Glashow resonance event was reported around the neutrino energy $E_{\nu}=(6.05\pm0.72)\times10^{6}~\GeV$~\cite{IceCube:2021rpz}.
When neutrinos pass the earth, the neutrino-electron scattering will contribute to the earth's absorption. 
However, in comparison with the neutrino-nucleon (or -nucleus) cross sections shown in Fig.~\ref{fig:ICresults}, the neutrino-electron process is generally much smaller, being negligible for most scenarios. 

\bibliographystyle{utphys}
\bibliography{ref}

\providecommand{\href}[2]{#2}\begingroup\raggedright\begin{thebibliography}{100}

\bibitem{Zyla:2020zbs}
{\bfseries Particle Data Group} Collaboration, P.~Zyla {\em et~al.}, ``{Review
  of Particle Physics},'' \href{http://dx.doi.org/10.1093/ptep/ptaa104}{{\em
  PTEP} {\bfseries 2020} no.~8, (2020) 083C01}.

\bibitem{Ackermann:2022rqc}
M.~Ackermann {\em et~al.}, ``{High-energy and ultra-high-energy neutrinos: A
  Snowmass white paper},''
  \href{http://dx.doi.org/10.1016/j.jheap.2022.08.001}{{\em JHEAp} {\bfseries
  36} (2022) 55--110}, \href{http://arxiv.org/abs/2203.08096}{{\ttfamily
  arXiv:2203.08096 [hep-ph]}}.

\bibitem{IceCube:2003llu}
{\bfseries IceCube} Collaboration, J.~Ahrens {\em et~al.}, ``{Sensitivity of
  the IceCube detector to astrophysical sources of high energy muon
  neutrinos},''
  \href{http://dx.doi.org/10.1016/j.astropartphys.2003.09.003}{{\em Astropart.
  Phys.} {\bfseries 20} (2004) 507--532},
  \href{http://arxiv.org/abs/astro-ph/0305196}{{\ttfamily
  arXiv:astro-ph/0305196}}.

\bibitem{vanSanten:2017chb}
{\bfseries IceCube-Gen2} Collaboration, J.~van Santen, ``{IceCube-Gen2: the
  next-generation neutrino observatory for the South Pole},''
  \href{http://dx.doi.org/10.22323/1.301.0991}{{\em PoS} {\bfseries ICRC2017}
  (2018) 991}.

\bibitem{KM3Net:2016zxf}
{\bfseries KM3Net} Collaboration, S.~Adrian-Martinez {\em et~al.}, ``{Letter of
  intent for KM3NeT 2.0}''
  \href{http://dx.doi.org/10.1088/0954-3899/43/8/084001}{{\em J. Phys. G}
  {\bfseries 43} no.~8, (2016) 084001},
  \href{http://arxiv.org/abs/1601.07459}{{\ttfamily arXiv:1601.07459
  [astro-ph.IM]}}.

\bibitem{Avrorin:2011zzc}
A.~Avrorin {\em et~al.}, ``{The gigaton volume detector in Lake Baikal},''
  \href{http://dx.doi.org/10.1016/j.nima.2010.09.137}{{\em Nucl. Instrum. Meth.
  A} {\bfseries 639} (2011) 30--32}.

\bibitem{GRAND:2018iaj}
{\bfseries GRAND} Collaboration, J.~\'Alvarez-Mu\~niz {\em et~al.}, ``{The
  Giant Radio Array for Neutrino Detection (GRAND): Science and Design},''
  \href{http://dx.doi.org/10.1007/s11433-018-9385-7}{{\em Sci. China Phys.
  Mech. Astron.} {\bfseries 63} no.~1, (2020) 219501},
  \href{http://arxiv.org/abs/1810.09994}{{\ttfamily arXiv:1810.09994
  [astro-ph.HE]}}.

\bibitem{Olinto:2017xbi}
A.~V. Olinto {\em et~al.}, ``{POEMMA: Probe Of Extreme Multi-Messenger
  Astrophysics},'' \href{http://dx.doi.org/10.22323/1.301.0542}{{\em PoS}
  {\bfseries ICRC2017} (2018) 542},
  \href{http://arxiv.org/abs/1708.07599}{{\ttfamily arXiv:1708.07599
  [astro-ph.IM]}}.

\bibitem{P-ONE:2020ljt}
{\bfseries P-ONE} Collaboration, M.~Agostini {\em et~al.}, ``{The Pacific Ocean
  Neutrino Experiment},''
  \href{http://dx.doi.org/10.1038/s41550-020-1182-4}{{\em Nature Astron.}
  {\bfseries 4} no.~10, (2020) 913--915},
  \href{http://arxiv.org/abs/2005.09493}{{\ttfamily arXiv:2005.09493
  [astro-ph.HE]}}.

\bibitem{Ye:2022vbk}
Z.~P. Ye {\em et~al.}, ``{Proposal for a neutrino telescope in South China
  Sea},'' \href{http://arxiv.org/abs/2207.04519}{{\ttfamily arXiv:2207.04519
  [astro-ph.HE]}}.

\bibitem{Denton:2020jft}
P.~B. Denton and Y.~Kini, ``{Ultra-High-Energy Tau Neutrino Cross Sections with
  GRAND and POEMMA},''
  \href{http://dx.doi.org/10.1103/PhysRevD.102.123019}{{\em Phys. Rev. D}
  {\bfseries 102} (2020) 123019},
  \href{http://arxiv.org/abs/2007.10334}{{\ttfamily arXiv:2007.10334
  [astro-ph.HE]}}.

\bibitem{IceCube:2017roe}
{\bfseries IceCube} Collaboration, M.~G. Aartsen {\em et~al.}, ``{Measurement
  of the multi-TeV neutrino cross section with IceCube using Earth
  absorption},'' \href{http://dx.doi.org/10.1038/nature24459}{{\em Nature}
  {\bfseries 551} (2017) 596--600},
  \href{http://arxiv.org/abs/1711.08119}{{\ttfamily arXiv:1711.08119
  [hep-ex]}}.

\bibitem{IceCube:2020rnc}
{\bfseries IceCube} Collaboration, R.~Abbasi {\em et~al.}, ``{Measurement of
  the high-energy all-flavor neutrino-nucleon cross section with IceCube},''
  \href{http://arxiv.org/abs/2011.03560}{{\ttfamily arXiv:2011.03560
  [hep-ex]}}.

\bibitem{Donini:2018tsg}
A.~Donini, S.~Palomares-Ruiz, and J.~Salvado, ``{Neutrino tomography of
  Earth},'' \href{http://dx.doi.org/10.1038/s41567-018-0319-1}{{\em Nature
  Phys.} {\bfseries 15} no.~1, (2019) 37--40},
  \href{http://arxiv.org/abs/1803.05901}{{\ttfamily arXiv:1803.05901
  [hep-ph]}}.

\bibitem{IceCube:2018pgc}
{\bfseries IceCube} Collaboration, M.~G. Aartsen {\em et~al.}, ``{Measurements
  using the inelasticity distribution of multi-TeV neutrino interactions in
  IceCube},'' \href{http://dx.doi.org/10.1103/PhysRevD.99.032004}{{\em Phys.
  Rev. D} {\bfseries 99} no.~3, (2019) 032004},
  \href{http://arxiv.org/abs/1808.07629}{{\ttfamily arXiv:1808.07629
  [hep-ex]}}.

\bibitem{Valera:2022ylt}
V.~B. Valera, M.~Bustamante, and C.~Glaser, ``{The ultra-high-energy
  neutrino-nucleon cross section: measurement forecasts for an era of cosmic
  EeV-neutrino discovery},''
  \href{http://dx.doi.org/10.1007/JHEP06(2022)105}{{\em JHEP} {\bfseries 06}
  (2022) 105}, \href{http://arxiv.org/abs/2204.04237}{{\ttfamily
  arXiv:2204.04237 [hep-ph]}}.

\bibitem{Valera:2022wmu}
V.~B. Valera, M.~Bustamante, and C.~Glaser, ``{Near-future discovery of the
  diffuse flux of ultrahigh-energy cosmic neutrinos},''
  \href{http://dx.doi.org/10.1103/PhysRevD.107.043019}{{\em Phys. Rev. D}
  {\bfseries 107} no.~4, (2023) 043019},
  \href{http://arxiv.org/abs/2210.03756}{{\ttfamily arXiv:2210.03756
  [astro-ph.HE]}}.

\bibitem{Feng:2017uoz}
J.~L. Feng, I.~Galon, F.~Kling, and S.~Trojanowski, ``{ForwArd Search
  ExpeRiment at the LHC},''
  \href{http://dx.doi.org/10.1103/PhysRevD.97.035001}{{\em Phys. Rev. D}
  {\bfseries 97} no.~3, (2018) 035001},
  \href{http://arxiv.org/abs/1708.09389}{{\ttfamily arXiv:1708.09389
  [hep-ph]}}.

\bibitem{FASER:2018bac}
{\bfseries FASER} Collaboration, A.~Ariga {\em et~al.}, ``{Technical Proposal
  for FASER: ForwArd Search ExpeRiment at the LHC},''
  \href{http://arxiv.org/abs/1812.09139}{{\ttfamily arXiv:1812.09139
  [physics.ins-det]}}.

\bibitem{Anchordoqui:2021ghd}
L.~A. Anchordoqui {\em et~al.}, ``{The Forward Physics Facility: Sites,
  experiments, and physics potential},''
  \href{http://dx.doi.org/10.1016/j.physrep.2022.04.004}{{\em Phys. Rept.}
  {\bfseries 968} (2022) 1--50},
  \href{http://arxiv.org/abs/2109.10905}{{\ttfamily arXiv:2109.10905
  [hep-ph]}}.

\bibitem{Feng:2022inv}
J.~L. Feng {\em et~al.}, ``{The Forward Physics Facility at the High-Luminosity
  LHC},'' \href{http://arxiv.org/abs/2203.05090}{{\ttfamily arXiv:2203.05090
  [hep-ex]}}.

\bibitem{Collins:1989gx}
J.~C. Collins, D.~E. Soper, and G.~F. Sterman, ``{Factorization of Hard
  Processes in QCD},'' \href{http://dx.doi.org/10.1142/9789814503266_0001}{{\em
  Adv. Ser. Direct. High Energy Phys.} {\bfseries 5} (1989) 1--91},
  \href{http://arxiv.org/abs/hep-ph/0409313}{{\ttfamily arXiv:hep-ph/0409313}}.

\bibitem{H1:2015ubc}
{\bfseries H1, ZEUS} Collaboration, H.~Abramowicz {\em et~al.}, ``{Combination
  of measurements of inclusive deep inelastic ${e^{\pm }p}$ scattering cross
  sections and QCD analysis of HERA data},''
  \href{http://dx.doi.org/10.1140/epjc/s10052-015-3710-4}{{\em Eur. Phys. J. C}
  {\bfseries 75} no.~12, (2015) 580},
  \href{http://arxiv.org/abs/1506.06042}{{\ttfamily arXiv:1506.06042
  [hep-ex]}}.

\bibitem{Belyaev:2021cyr}
I.~Belyaev, G.~Carboni, N.~Harnew, C.~Matteuzzi, and F.~Teubert, ``{The history
  of LHCb},'' \href{http://dx.doi.org/10.1140/epjh/s13129-021-00002-z}{{\em
  Eur. Phys. J. H} {\bfseries 46} no.~1, (2021) 3},
  \href{http://arxiv.org/abs/2101.05331}{{\ttfamily arXiv:2101.05331
  [physics.hist-ph]}}.

\bibitem{Hou:2019efy}
T.-J. Hou {\em et~al.}, ``{New CTEQ global analysis of quantum chromodynamics
  with high-precision data from the LHC},''
  \href{http://dx.doi.org/10.1103/PhysRevD.103.014013}{{\em Phys. Rev. D}
  {\bfseries 103} no.~1, (2021) 014013},
  \href{http://arxiv.org/abs/1912.10053}{{\ttfamily arXiv:1912.10053
  [hep-ph]}}.

\bibitem{Gao:2021fle}
J.~Gao, T.~J. Hobbs, P.~M. Nadolsky, C.~Sun, and C.~P. Yuan, ``{General
  heavy-flavor mass scheme for charged-current DIS at NNLO and beyond},''
  \href{http://dx.doi.org/10.1103/PhysRevD.105.L011503}{{\em Phys. Rev. D}
  {\bfseries 105} no.~1, (2022) L011503},
  \href{http://arxiv.org/abs/2107.00460}{{\ttfamily arXiv:2107.00460
  [hep-ph]}}.

\bibitem{Salam:2008qg}
G.~P. Salam and J.~Rojo, ``{A Higher Order Perturbative Parton Evolution
  Toolkit (HOPPET)},'' \href{http://dx.doi.org/10.1016/j.cpc.2008.08.010}{{\em
  Comput. Phys. Commun.} {\bfseries 180} (2009) 120--156},
  \href{http://arxiv.org/abs/0804.3755}{{\ttfamily arXiv:0804.3755 [hep-ph]}}.

\bibitem{Vermaseren:2005qc}
J.~A.~M. Vermaseren, A.~Vogt, and S.~Moch, ``{The Third-order QCD corrections
  to deep-inelastic scattering by photon exchange},''
  \href{http://dx.doi.org/10.1016/j.nuclphysb.2005.06.020}{{\em Nucl. Phys. B}
  {\bfseries 724} (2005) 3--182},
  \href{http://arxiv.org/abs/hep-ph/0504242}{{\ttfamily arXiv:hep-ph/0504242}}.

\bibitem{Vogt:2006bt}
A.~Vogt, S.~Moch, and J.~Vermaseren, ``{Third-order QCD results on form factors
  and coefficient functions},''
  \href{http://dx.doi.org/10.1016/j.nuclphysbps.2006.09.101}{{\em Nucl. Phys. B
  Proc. Suppl.} {\bfseries 160} (2006) 44--50},
  \href{http://arxiv.org/abs/hep-ph/0608307}{{\ttfamily arXiv:hep-ph/0608307}}.

\bibitem{Moch:2007rq}
S.~Moch, M.~Rogal, and A.~Vogt, ``{Differences between charged-current
  coefficient functions},''
  \href{http://dx.doi.org/10.1016/j.nuclphysb.2007.09.022}{{\em Nucl. Phys. B}
  {\bfseries 790} (2008) 317--335},
  \href{http://arxiv.org/abs/0708.3731}{{\ttfamily arXiv:0708.3731 [hep-ph]}}.

\bibitem{Davies:2016ruz}
J.~Davies, A.~Vogt, S.~Moch, and J.~A.~M. Vermaseren, ``{Non-singlet
  coefficient functions for charged-current deep-inelastic scattering to the
  third order in QCD},'' \href{http://dx.doi.org/10.22323/1.265.0059}{{\em PoS}
  {\bfseries DIS2016} (2016) 059},
  \href{http://arxiv.org/abs/1606.08907}{{\ttfamily arXiv:1606.08907
  [hep-ph]}}.

\bibitem{Moch:2004xu}
S.~Moch, J.~A.~M. Vermaseren, and A.~Vogt, ``{The Longitudinal structure
  function at the third order},''
  \href{http://dx.doi.org/10.1016/j.physletb.2004.11.063}{{\em Phys. Lett. B}
  {\bfseries 606} (2005) 123--129},
  \href{http://arxiv.org/abs/hep-ph/0411112}{{\ttfamily arXiv:hep-ph/0411112}}.

\bibitem{Eskola:2021nhw}
K.~J. Eskola, P.~Paakkinen, H.~Paukkunen, and C.~A. Salgado, ``{EPPS21: a
  global QCD analysis of nuclear PDFs},''
  \href{http://dx.doi.org/10.1140/epjc/s10052-022-10359-0}{{\em Eur. Phys. J.
  C} {\bfseries 82} no.~5, (2022) 413},
  \href{http://arxiv.org/abs/2112.12462}{{\ttfamily arXiv:2112.12462
  [hep-ph]}}.

\bibitem{Kusina:2020lyz}
A.~Kusina {\em et~al.}, ``{Impact of LHC vector boson production in heavy ion
  collisions on strange PDFs},''
  \href{http://dx.doi.org/10.1140/epjc/s10052-020-08532-4}{{\em Eur. Phys. J.
  C} {\bfseries 80} no.~10, (2020) 968},
  \href{http://arxiv.org/abs/2007.09100}{{\ttfamily arXiv:2007.09100
  [hep-ph]}}.

\bibitem{Cooper-Sarkar:2011jtt}
A.~Cooper-Sarkar, P.~Mertsch, and S.~Sarkar, ``{The high energy neutrino
  cross-section in the Standard Model and its uncertainty},''
  \href{http://dx.doi.org/10.1007/JHEP08(2011)042}{{\em JHEP} {\bfseries 08}
  (2011) 042}, \href{http://arxiv.org/abs/1106.3723}{{\ttfamily arXiv:1106.3723
  [hep-ph]}}.

\bibitem{Gandhi:1998ri}
R.~Gandhi, C.~Quigg, M.~H. Reno, and I.~Sarcevic, ``{Neutrino interactions at
  ultrahigh-energies},''
  \href{http://dx.doi.org/10.1103/PhysRevD.58.093009}{{\em Phys. Rev. D}
  {\bfseries 58} (1998) 093009},
  \href{http://arxiv.org/abs/hep-ph/9807264}{{\ttfamily arXiv:hep-ph/9807264}}.

\bibitem{Connolly:2011vc}
A.~Connolly, R.~S. Thorne, and D.~Waters, ``{Calculation of High Energy
  Neutrino-Nucleon Cross Sections and Uncertainties Using the MSTW Parton
  Distribution Functions and Implications for Future Experiments},''
  \href{http://dx.doi.org/10.1103/PhysRevD.83.113009}{{\em Phys. Rev. D}
  {\bfseries 83} (2011) 113009},
  \href{http://arxiv.org/abs/1102.0691}{{\ttfamily arXiv:1102.0691 [hep-ph]}}.

\bibitem{Bertone:2018dse}
V.~Bertone, R.~Gauld, and J.~Rojo, ``{Neutrino Telescopes as QCD
  Microscopes},'' \href{http://dx.doi.org/10.1007/JHEP01(2019)217}{{\em JHEP}
  {\bfseries 01} (2019) 217}, \href{http://arxiv.org/abs/1808.02034}{{\ttfamily
  arXiv:1808.02034 [hep-ph]}}.

\bibitem{Garcia:2020jwr}
A.~Garcia, R.~Gauld, A.~Heijboer, and J.~Rojo, ``{Complete predictions for
  high-energy neutrino propagation in matter},''
  \href{http://dx.doi.org/10.1088/1475-7516/2020/09/025}{{\em JCAP} {\bfseries
  09} (2020) 025}, \href{http://arxiv.org/abs/2004.04756}{{\ttfamily
  arXiv:2004.04756 [hep-ph]}}.

\bibitem{Jeong:2023hwe}
Y.~S. Jeong and M.~H. Reno, ``{Neutrino cross sections: Interface of shallow-
  and deep-inelastic scattering for collider neutrinos},''
  \href{http://dx.doi.org/10.1103/PhysRevD.108.113010}{{\em Phys. Rev. D}
  {\bfseries 108} no.~11, (2023) 113010},
  \href{http://arxiv.org/abs/2307.09241}{{\ttfamily arXiv:2307.09241
  [hep-ph]}}.

\bibitem{Candido:2023utz}
A.~Candido, A.~Garcia, G.~Magni, T.~Rabemananjara, J.~Rojo, and R.~Stegeman,
  ``{Neutrino Structure Functions from GeV to EeV Energies},''
  \href{http://arxiv.org/abs/2302.08527}{{\ttfamily arXiv:2302.08527
  [hep-ph]}}.

\bibitem{Arguelles:2015wba}
C.~A. Arg\"uelles, F.~Halzen, L.~Wille, M.~Kroll, and M.~H. Reno,
  ``{High-energy behavior of photon, neutrino, and proton cross sections},''
  \href{http://dx.doi.org/10.1103/PhysRevD.92.074040}{{\em Phys. Rev. D}
  {\bfseries 92} no.~7, (2015) 074040},
  \href{http://arxiv.org/abs/1504.06639}{{\ttfamily arXiv:1504.06639
  [hep-ph]}}.

\bibitem{Formaggio:2012cpf}
J.~A. Formaggio and G.~P. Zeller, ``{From eV to EeV: Neutrino Cross Sections
  Across Energy Scales},''
  \href{http://dx.doi.org/10.1103/RevModPhys.84.1307}{{\em Rev. Mod. Phys.}
  {\bfseries 84} (2012) 1307--1341},
  \href{http://arxiv.org/abs/1305.7513}{{\ttfamily arXiv:1305.7513 [hep-ex]}}.

\bibitem{Illarionov:2011wc}
A.~Y. Illarionov, B.~A. Kniehl, and A.~V. Kotikov, ``{Ultrahigh-energy
  neutrino-nucleon deep-inelastic scattering and the Froissart bound},''
  \href{http://dx.doi.org/10.1103/PhysRevLett.106.231802}{{\em Phys. Rev.
  Lett.} {\bfseries 106} (2011) 231802},
  \href{http://arxiv.org/abs/1105.2829}{{\ttfamily arXiv:1105.2829 [hep-ph]}}.

\bibitem{MINERvA:2016oql}
{\bfseries MINERvA} Collaboration, J.~Mousseau {\em et~al.}, ``{Measurement of
  Partonic Nuclear Effects in Deep-Inelastic Neutrino Scattering using
  MINERvA},'' \href{http://dx.doi.org/10.1103/PhysRevD.93.071101}{{\em Phys.
  Rev. D} {\bfseries 93} no.~7, (2016) 071101},
  \href{http://arxiv.org/abs/1601.06313}{{\ttfamily arXiv:1601.06313
  [hep-ex]}}.

\bibitem{Buckley:2014ana}
A.~Buckley, J.~Ferrando, S.~Lloyd, K.~Nordstr\"om, B.~Page, M.~R\"ufenacht,
  M.~Sch\"onherr, and G.~Watt, ``{LHAPDF6: parton density access in the LHC
  precision era},''
  \href{http://dx.doi.org/10.1140/epjc/s10052-015-3318-8}{{\em Eur. Phys. J. C}
  {\bfseries 75} (2015) 132}, \href{http://arxiv.org/abs/1412.7420}{{\ttfamily
  arXiv:1412.7420 [hep-ph]}}.

\bibitem{Bertone:2013vaa}
V.~Bertone, S.~Carrazza, and J.~Rojo, ``{APFEL: A PDF Evolution Library with
  QED corrections},'' \href{http://dx.doi.org/10.1016/j.cpc.2014.03.007}{{\em
  Comput. Phys. Commun.} {\bfseries 185} (2014) 1647--1668},
  \href{http://arxiv.org/abs/1310.1394}{{\ttfamily arXiv:1310.1394 [hep-ph]}}.

\bibitem{NNPDF:2017mvq}
{\bfseries NNPDF} Collaboration, R.~D. Ball {\em et~al.}, ``{Parton
  distributions from high-precision collider data},''
  \href{http://dx.doi.org/10.1140/epjc/s10052-017-5199-5}{{\em Eur. Phys. J. C}
  {\bfseries 77} no.~10, (2017) 663},
  \href{http://arxiv.org/abs/1706.00428}{{\ttfamily arXiv:1706.00428
  [hep-ph]}}.

\bibitem{NNPDF:2021njg}
{\bfseries NNPDF} Collaboration, R.~D. Ball {\em et~al.}, ``{The path to proton
  structure at 1\% accuracy},''
  \href{http://dx.doi.org/10.1140/epjc/s10052-022-10328-7}{{\em Eur. Phys. J.
  C} {\bfseries 82} no.~5, (2022) 428},
  \href{http://arxiv.org/abs/2109.02653}{{\ttfamily arXiv:2109.02653
  [hep-ph]}}.

\bibitem{Bailey:2020ooq}
S.~Bailey, T.~Cridge, L.~A. Harland-Lang, A.~D. Martin, and R.~S. Thorne,
  ``{Parton distributions from LHC, HERA, Tevatron and fixed target data:
  MSHT20 PDFs},'' \href{http://dx.doi.org/10.1140/epjc/s10052-021-09057-0}{{\em
  Eur. Phys. J. C} {\bfseries 81} no.~4, (2021) 341},
  \href{http://arxiv.org/abs/2012.04684}{{\ttfamily arXiv:2012.04684
  [hep-ph]}}.

\bibitem{Gao:2013bia}
J.~Gao and P.~Nadolsky, ``{A meta-analysis of parton distribution functions},''
  \href{http://dx.doi.org/10.1007/JHEP07(2014)035}{{\em JHEP} {\bfseries 07}
  (2014) 035}, \href{http://arxiv.org/abs/1401.0013}{{\ttfamily arXiv:1401.0013
  [hep-ph]}}.

\bibitem{Yan:2022pzl}
M.~Yan, T.-J. Hou, P.~Nadolsky, and C.~P. Yuan, ``{A CT18 global PDF fit at the
  leading order in QCD},'' \href{http://arxiv.org/abs/2205.00137}{{\ttfamily
  arXiv:2205.00137 [hep-ph]}}.

\bibitem{Xie:2021equ}
{\bfseries CTEQ-TEA} Collaboration, K.~Xie, T.~J. Hobbs, T.-J. Hou, C.~Schmidt,
  M.~Yan, and C.~P. Yuan, ``{Photon PDF within the CT18 global analysis},''
  \href{http://dx.doi.org/10.1103/PhysRevD.105.054006}{{\em Phys. Rev. D}
  {\bfseries 105} no.~5, (2022) 054006},
  \href{http://arxiv.org/abs/2106.10299}{{\ttfamily arXiv:2106.10299
  [hep-ph]}}.

\bibitem{Guzzi:2011ew}
M.~Guzzi, P.~M. Nadolsky, H.-L. Lai, and C.~P. Yuan, ``{General-Mass Treatment
  for Deep Inelastic Scattering at Two-Loop Accuracy},''
  \href{http://dx.doi.org/10.1103/PhysRevD.86.053005}{{\em Phys. Rev. D}
  {\bfseries 86} (2012) 053005},
  \href{http://arxiv.org/abs/1108.5112}{{\ttfamily arXiv:1108.5112 [hep-ph]}}.

\bibitem{Forte:2010ta}
S.~Forte, E.~Laenen, P.~Nason, and J.~Rojo, ``{Heavy quarks in deep-inelastic
  scattering},'' \href{http://dx.doi.org/10.1016/j.nuclphysb.2010.03.014}{{\em
  Nucl. Phys. B} {\bfseries 834} (2010) 116--162},
  \href{http://arxiv.org/abs/1001.2312}{{\ttfamily arXiv:1001.2312 [hep-ph]}}.

\bibitem{Thorne:2012az}
R.~S. Thorne, ``{Effect of changes of variable flavor number scheme on parton
  distribution functions and predicted cross sections},''
  \href{http://dx.doi.org/10.1103/PhysRevD.86.074017}{{\em Phys. Rev. D}
  {\bfseries 86} (2012) 074017},
  \href{http://arxiv.org/abs/1201.6180}{{\ttfamily arXiv:1201.6180 [hep-ph]}}.

\bibitem{Blumlein:2014fqa}
J.~Bl\"umlein, A.~Hasselhuhn, and T.~Pfoh, ``{The $O(\alpha_s^2)$ heavy quark
  corrections to charged current deep-inelastic scattering at large
  virtualities},''
  \href{http://dx.doi.org/10.1016/j.nuclphysb.2014.01.023}{{\em Nucl. Phys. B}
  {\bfseries 881} (2014) 1--41},
  \href{http://arxiv.org/abs/1401.4352}{{\ttfamily arXiv:1401.4352 [hep-ph]}}.

\bibitem{Blumlein:2016xcy}
J.~Bl\"umlein, G.~Falcioni, and A.~De~Freitas, ``{The Complete $O(\alpha_s^2)$
  Non-Singlet Heavy Flavor Corrections to the Structure Functions
  $g_{1,2}^{ep}(x,Q^2)$, $F_{1,2,L}^{ep}(x,Q^2)$,
  $F_{1,2,3}^{\nu(\bar{\nu})}(x,Q^2)$ and the Associated Sum Rules},''
  \href{http://dx.doi.org/10.1016/j.nuclphysb.2016.06.018}{{\em Nucl. Phys. B}
  {\bfseries 910} (2016) 568--617},
  \href{http://arxiv.org/abs/1605.05541}{{\ttfamily arXiv:1605.05541
  [hep-ph]}}.

\bibitem{Behring:2015roa}
A.~Behring, J.~Bl\"umlein, A.~De~Freitas, A.~Hasselhuhn, A.~von Manteuffel, and
  C.~Schneider, ``{O($\alpha_s^3$) heavy flavor contributions to the charged
  current structure function $xF_3(x,Q^2)$ at large momentum transfer},''
  \href{http://dx.doi.org/10.1103/PhysRevD.92.114005}{{\em Phys. Rev. D}
  {\bfseries 92} no.~11, (2015) 114005},
  \href{http://arxiv.org/abs/1508.01449}{{\ttfamily arXiv:1508.01449
  [hep-ph]}}.

\bibitem{Behring:2016hpa}
A.~Behring, J.~Bl\"umlein, G.~Falcioni, A.~De~Freitas, A.~von Manteuffel, and
  C.~Schneider, ``{Asymptotic 3-loop heavy flavor corrections to the charged
  current structure functions $F_L^{W^+-W^-}(x,Q^2)$ and
  $F_2^{W^+-W^-}(x,Q^2)$},''
  \href{http://dx.doi.org/10.1103/PhysRevD.94.114006}{{\em Phys. Rev. D}
  {\bfseries 94} no.~11, (2016) 114006},
  \href{http://arxiv.org/abs/1609.06255}{{\ttfamily arXiv:1609.06255
  [hep-ph]}}.

\bibitem{Berger:2016inr}
E.~L. Berger, J.~Gao, C.~S. Li, Z.~L. Liu, and H.~X. Zhu, ``{Charm-Quark
  Production in Deep-Inelastic Neutrino Scattering at Next-to-Next-to-Leading
  Order in QCD},'' \href{http://dx.doi.org/10.1103/PhysRevLett.116.212002}{{\em
  Phys. Rev. Lett.} {\bfseries 116} no.~21, (2016) 212002},
  \href{http://arxiv.org/abs/1601.05430}{{\ttfamily arXiv:1601.05430
  [hep-ph]}}.

\bibitem{Gao:2017kkx}
J.~Gao, ``{Massive charged-current coefficient functions in deep-inelastic
  scattering at NNLO and impact on strange-quark distributions},''
  \href{http://dx.doi.org/10.1007/JHEP02(2018)026}{{\em JHEP} {\bfseries 02}
  (2018) 026}, \href{http://arxiv.org/abs/1710.04258}{{\ttfamily
  arXiv:1710.04258 [hep-ph]}}.

\bibitem{Ball:2017otu}
R.~D. Ball, V.~Bertone, M.~Bonvini, S.~Marzani, J.~Rojo, and L.~Rottoli,
  ``{Parton distributions with small-x resummation: evidence for BFKL dynamics
  in HERA data},'' \href{http://dx.doi.org/10.1140/epjc/s10052-018-5774-4}{{\em
  Eur. Phys. J. C} {\bfseries 78} no.~4, (2018) 321},
  \href{http://arxiv.org/abs/1710.05935}{{\ttfamily arXiv:1710.05935
  [hep-ph]}}.

\bibitem{xFitterDevelopersTeam:2018hym}
{\bfseries xFitter Developers' Team} Collaboration, H.~Abdolmaleki {\em
  et~al.}, ``{Impact of low-$x$ resummation on QCD analysis of HERA data},''
  \href{http://dx.doi.org/10.1140/epjc/s10052-018-6090-8}{{\em Eur. Phys. J. C}
  {\bfseries 78} no.~8, (2018) 621},
  \href{http://arxiv.org/abs/1802.00064}{{\ttfamily arXiv:1802.00064
  [hep-ph]}}.

\bibitem{Balitsky:1978ic}
I.~I. Balitsky and L.~N. Lipatov, ``{The Pomeranchuk Singularity in Quantum
  Chromodynamics},'' {\em Sov. J. Nucl. Phys.} {\bfseries 28} (1978) 822--829.

\bibitem{Balitsky:1979ns}
I.~I. Balitsky and L.~N. Lipatov, ``{Calculation of Meson Meson Interaction
  Cross-Section in Quantum Chromodynamics (In Russian)},'' {\em JETP Lett.}
  {\bfseries 30} (1979) 355.

\bibitem{Fadin:1975cb}
V.~S. Fadin, E.~A. Kuraev, and L.~N. Lipatov, ``{On the Pomeranchuk Singularity
  in Asymptotically Free Theories},''
  \href{http://dx.doi.org/10.1016/0370-2693(75)90524-9}{{\em Phys. Lett. B}
  {\bfseries 60} (1975) 50--52}.

\bibitem{Lipatov:1976zz}
L.~N. Lipatov, ``{Reggeization of the Vector Meson and the Vacuum Singularity
  in Nonabelian Gauge Theories},'' {\em Sov. J. Nucl. Phys.} {\bfseries 23}
  (1976) 338--345.

\bibitem{Kuraev:1976ge}
E.~A. Kuraev, L.~N. Lipatov, and V.~S. Fadin, ``{Multi - Reggeon Processes in
  the Yang-Mills Theory},'' {\em Sov. Phys. JETP} {\bfseries 44} (1976)
  443--450.

\bibitem{Kuraev:1977fs}
E.~A. Kuraev, L.~N. Lipatov, and V.~S. Fadin, ``{The Pomeranchuk Singularity in
  Nonabelian Gauge Theories},'' {\em Sov. Phys. JETP} {\bfseries 45} (1977)
  199--204.

\bibitem{H1:2013ktq}
{\bfseries H1} Collaboration, V.~Andreev {\em et~al.}, ``{Measurement of
  inclusive $e p$ cross sections at high $Q^2$ at $\sqrt s =$ 225 and 252 GeV
  and of the longitudinal proton structure function $F_L$ at HERA},''
  \href{http://dx.doi.org/10.1140/epjc/s10052-014-2814-6}{{\em Eur. Phys. J. C}
  {\bfseries 74} no.~4, (2014) 2814},
  \href{http://arxiv.org/abs/1312.4821}{{\ttfamily arXiv:1312.4821 [hep-ex]}}.

\bibitem{Bonvini:2016wki}
M.~Bonvini, S.~Marzani, and T.~Peraro, ``{Small-$x$ resummation from HELL},''
  \href{http://dx.doi.org/10.1140/epjc/s10052-016-4445-6}{{\em Eur. Phys. J. C}
  {\bfseries 76} no.~11, (2016) 597},
  \href{http://arxiv.org/abs/1607.02153}{{\ttfamily arXiv:1607.02153
  [hep-ph]}}.

\bibitem{Bonvini:2017ogt}
M.~Bonvini, S.~Marzani, and C.~Muselli, ``{Towards parton distribution
  functions with small-$x$ resummation: HELL 2.0}''
  \href{http://dx.doi.org/10.1007/JHEP12(2017)117}{{\em JHEP} {\bfseries 12}
  (2017) 117}, \href{http://arxiv.org/abs/1708.07510}{{\ttfamily
  arXiv:1708.07510 [hep-ph]}}.

\bibitem{Guzzi:2021fre}
M.~Guzzi {\em et~al.}, ``{NNLO constraints on proton PDFs from the SeaQuest and
  STAR experiments and other developments in the CTEQ-TEA global analysis},''
  \href{http://dx.doi.org/10.21468/SciPostPhysProc.8.005}{{\em SciPost Phys.
  Proc.} {\bfseries 8} (2022) 005},
  \href{http://arxiv.org/abs/2108.06596}{{\ttfamily arXiv:2108.06596
  [hep-ph]}}.

\bibitem{Golec-Biernat:1998zce}
K.~J. Golec-Biernat and M.~Wusthoff, ``{Saturation effects in deep inelastic
  scattering at low $Q^2$ and its implications on diffraction},''
  \href{http://dx.doi.org/10.1103/PhysRevD.59.014017}{{\em Phys. Rev. D}
  {\bfseries 59} (1998) 014017},
  \href{http://arxiv.org/abs/hep-ph/9807513}{{\ttfamily arXiv:hep-ph/9807513}}.

\bibitem{PDF4LHCWorkingGroup:2022cjn}
{\bfseries PDF4LHC Working Group} Collaboration, R.~D. Ball {\em et~al.},
  ``{The PDF4LHC21 combination of global PDF fits for the LHC Run III},''
  \href{http://dx.doi.org/10.1088/1361-6471/ac7216}{{\em J. Phys. G} {\bfseries
  49} no.~8, (2022) 080501}, \href{http://arxiv.org/abs/2203.05506}{{\ttfamily
  arXiv:2203.05506 [hep-ph]}}.

\bibitem{Amoroso:2022eow}
S.~Amoroso {\em et~al.}, ``{Snowmass 2021 Whitepaper: Proton Structure at the
  Precision Frontier},''
  \href{http://dx.doi.org/10.5506/APhysPolB.53.12-A1}{{\em Acta Phys. Polon. B}
  {\bfseries 53} no.~12, (2022) 12--A1},
  \href{http://arxiv.org/abs/2203.13923}{{\ttfamily arXiv:2203.13923
  [hep-ph]}}.

\bibitem{Courtoy:2022ocu}
A.~Courtoy, J.~Huston, P.~Nadolsky, K.~Xie, M.~Yan, and C.~P. Yuan, ``{Parton
  distributions need representative sampling},''
  \href{http://dx.doi.org/10.1103/PhysRevD.107.034008}{{\em Phys. Rev. D}
  {\bfseries 107} no.~3, (2023) 034008},
  \href{http://arxiv.org/abs/2205.10444}{{\ttfamily arXiv:2205.10444
  [hep-ph]}}.

\bibitem{Kulagin:2004ie}
S.~A. Kulagin and R.~Petti, ``{Global study of nuclear structure functions},''
  \href{http://dx.doi.org/10.1016/j.nuclphysa.2005.10.011}{{\em Nucl. Phys. A}
  {\bfseries 765} (2006) 126--187},
  \href{http://arxiv.org/abs/hep-ph/0412425}{{\ttfamily arXiv:hep-ph/0412425}}.

\bibitem{Kulagin:2014vsa}
S.~A. Kulagin and R.~Petti, ``{Nuclear parton distributions and the Drell-Yan
  process},'' \href{http://dx.doi.org/10.1103/PhysRevC.90.045204}{{\em Phys.
  Rev. C} {\bfseries 90} no.~4, (2014) 045204},
  \href{http://arxiv.org/abs/1405.2529}{{\ttfamily arXiv:1405.2529 [hep-ph]}}.

\bibitem{Alekhin:2017fpf}
S.~I. Alekhin, S.~A. Kulagin, and R.~Petti, ``{Nuclear Effects in the Deuteron
  and Constraints on the d/u Ratio},''
  \href{http://dx.doi.org/10.1103/PhysRevD.96.054005}{{\em Phys. Rev. D}
  {\bfseries 96} no.~5, (2017) 054005},
  \href{http://arxiv.org/abs/1704.00204}{{\ttfamily arXiv:1704.00204
  [nucl-th]}}.

\bibitem{Paakkinen:2017jpo}
P.~Paakkinen, ``{Nuclear parton distribution functions},'' {\em Frascati Phys.
  Ser.} (2017) 33--40, \href{http://arxiv.org/abs/1802.05927}{{\ttfamily
  arXiv:1802.05927 [hep-ph]}}.

\bibitem{Ethier:2020way}
J.~J. Ethier and E.~R. Nocera, ``{Parton Distributions in Nucleons and
  Nuclei},'' \href{http://dx.doi.org/10.1146/annurev-nucl-011720-042725}{{\em
  Ann. Rev. Nucl. Part. Sci.} {\bfseries 70} (2020) 43--76},
  \href{http://arxiv.org/abs/2001.07722}{{\ttfamily arXiv:2001.07722
  [hep-ph]}}.

\bibitem{Muzakka:2022wey}
K.~F. Muzakka {\em et~al.}, ``{Compatibility of neutrino DIS data and its
  impact on nuclear parton distribution functions},''
  \href{http://dx.doi.org/10.1103/PhysRevD.106.074004}{{\em Phys. Rev. D}
  {\bfseries 106} no.~7, (2022) 074004},
  \href{http://arxiv.org/abs/2204.13157}{{\ttfamily arXiv:2204.13157
  [hep-ph]}}.

\bibitem{Ruiz:2023ozv}
R.~Ruiz {\em et~al.}, ``{Target mass corrections in lepton-nucleus DIS: theory
  and applications to nuclear PDFs},''
  \href{http://arxiv.org/abs/2301.07715}{{\ttfamily arXiv:2301.07715
  [hep-ph]}}.

\bibitem{TheATLAScollaboration:2015lnm}
``{Measurement of $W\rightarrow\mu\nu$ production in $p$+Pb collision at
  $\sqrt{s_{\text{NN}}}=5.02$ TeV with ATLAS detector at the LHC},''.
  \url{http://cds.cern.ch/record/2055677}.

\bibitem{ATLAS:2015mwq}
{\bfseries ATLAS} Collaboration, G.~Aad {\em et~al.}, ``{$Z$ boson production
  in $p+$Pb collisions at $\sqrt{s_{NN}}=5.02$ TeV measured with the ATLAS
  detector},'' \href{http://dx.doi.org/10.1103/PhysRevC.92.044915}{{\em Phys.
  Rev. C} {\bfseries 92} no.~4, (2015) 044915},
  \href{http://arxiv.org/abs/1507.06232}{{\ttfamily arXiv:1507.06232
  [hep-ex]}}.

\bibitem{CMS:2015ehw}
{\bfseries CMS} Collaboration, V.~Khachatryan {\em et~al.}, ``{Study of W boson
  production in pPb collisions at $\sqrt{s_{\mathrm{NN}}} =$ 5.02 TeV},''
  \href{http://dx.doi.org/10.1016/j.physletb.2015.09.057}{{\em Phys. Lett. B}
  {\bfseries 750} (2015) 565--586},
  \href{http://arxiv.org/abs/1503.05825}{{\ttfamily arXiv:1503.05825
  [nucl-ex]}}.

\bibitem{CMS:2015zlj}
{\bfseries CMS} Collaboration, V.~Khachatryan {\em et~al.}, ``{Study of Z boson
  production in pPb collisions at $\sqrt {s_{NN}} = 5.02$ TeV},''
  \href{http://dx.doi.org/10.1016/j.physletb.2016.05.044}{{\em Phys. Lett. B}
  {\bfseries 759} (2016) 36--57},
  \href{http://arxiv.org/abs/1512.06461}{{\ttfamily arXiv:1512.06461
  [hep-ex]}}.

\bibitem{CMS:2019leu}
{\bfseries CMS} Collaboration, A.~M. Sirunyan {\em et~al.}, ``{Observation of
  nuclear modifications in W$^\pm$ boson production in pPb collisions at
  $\sqrt{s_\mathrm{NN}} =$ 8.16 TeV},''
  \href{http://dx.doi.org/10.1016/j.physletb.2019.135048}{{\em Phys. Lett. B}
  {\bfseries 800} (2020) 135048},
  \href{http://arxiv.org/abs/1905.01486}{{\ttfamily arXiv:1905.01486
  [hep-ex]}}.

\bibitem{ALICE:2016rzo}
{\bfseries ALICE} Collaboration, J.~Adam {\em et~al.}, ``{W and Z boson
  production in p-Pb collisions at $\sqrt{s_{\rm NN}}$ = 5.02 TeV},''
  \href{http://dx.doi.org/10.1007/JHEP02(2017)077}{{\em JHEP} {\bfseries 02}
  (2017) 077}, \href{http://arxiv.org/abs/1611.03002}{{\ttfamily
  arXiv:1611.03002 [nucl-ex]}}.

\bibitem{Senosi:2015omk}
{\bfseries ALICE} Collaboration, K.~Senosi, ``{Measurement of W-boson
  production in p-Pb collisions at the LHC with ALICE},''
  \href{http://dx.doi.org/10.22323/1.238.0042}{{\em PoS} {\bfseries Bormio2015}
  (2015) 042}, \href{http://arxiv.org/abs/1511.06398}{{\ttfamily
  arXiv:1511.06398 [hep-ex]}}.

\bibitem{LHCb:2014jgh}
{\bfseries LHCb} Collaboration, R.~Aaij {\em et~al.}, ``{Observation of $Z$
  production in proton-lead collisions at LHCb},''
  \href{http://dx.doi.org/10.1007/JHEP09(2014)030}{{\em JHEP} {\bfseries 09}
  (2014) 030}, \href{http://arxiv.org/abs/1406.2885}{{\ttfamily arXiv:1406.2885
  [hep-ex]}}.

\bibitem{ALICE:2020jff}
{\bfseries ALICE} Collaboration, S.~Acharya {\em et~al.}, ``{Z-boson production
  in p-Pb collisions at $\sqrt{s_{\mathrm{NN}}}=8.16$ TeV and Pb-Pb collisions
  at $\sqrt{s_{\mathrm{NN}}}=5.02$ TeV},''
  \href{http://dx.doi.org/10.1007/JHEP09(2020)076}{{\em JHEP} {\bfseries 09}
  (2020) 076}, \href{http://arxiv.org/abs/2005.11126}{{\ttfamily
  arXiv:2005.11126 [nucl-ex]}}.

\bibitem{IceCube-Gen2:2020qha}
{\bfseries IceCube-Gen2} Collaboration, M.~G. Aartsen {\em et~al.},
  ``{IceCube-Gen2: the window to the extreme Universe},''
  \href{http://dx.doi.org/10.1088/1361-6471/abbd48}{{\em J. Phys. G} {\bfseries
  48} no.~6, (2021) 060501}, \href{http://arxiv.org/abs/2008.04323}{{\ttfamily
  arXiv:2008.04323 [astro-ph.HE]}}.

\bibitem{FASER:2020gpr}
{\bfseries FASER} Collaboration, H.~Abreu {\em et~al.}, ``{Technical Proposal:
  FASERnu},'' \href{http://arxiv.org/abs/2001.03073}{{\ttfamily
  arXiv:2001.03073 [physics.ins-det]}}.

\bibitem{Lai:1996mg}
H.~L. Lai, J.~Huston, S.~Kuhlmann, F.~I. Olness, J.~F. Owens, D.~E. Soper,
  W.~K. Tung, and H.~Weerts, ``{Improved parton distributions from global
  analysis of recent deep inelastic scattering and inclusive jet data},''
  \href{http://dx.doi.org/10.1103/PhysRevD.55.1280}{{\em Phys. Rev. D}
  {\bfseries 55} (1997) 1280--1296},
  \href{http://arxiv.org/abs/hep-ph/9606399}{{\ttfamily arXiv:hep-ph/9606399}}.

\bibitem{Martin:2009iq}
A.~D. Martin, W.~J. Stirling, R.~S. Thorne, and G.~Watt, ``{Parton
  distributions for the LHC},''
  \href{http://dx.doi.org/10.1140/epjc/s10052-009-1072-5}{{\em Eur. Phys. J. C}
  {\bfseries 63} (2009) 189--285},
  \href{http://arxiv.org/abs/0901.0002}{{\ttfamily arXiv:0901.0002 [hep-ph]}}.

\bibitem{Thorne:1997ga}
R.~S. Thorne and R.~G. Roberts, ``{An Ordered analysis of heavy flavor
  production in deep inelastic scattering},''
  \href{http://dx.doi.org/10.1103/PhysRevD.57.6871}{{\em Phys. Rev. D}
  {\bfseries 57} (1998) 6871--6898},
  \href{http://arxiv.org/abs/hep-ph/9709442}{{\ttfamily arXiv:hep-ph/9709442}}.

\bibitem{Thorne:2006qt}
R.~S. Thorne, ``{A Variable-flavor number scheme for NNLO},''
  \href{http://dx.doi.org/10.1103/PhysRevD.73.054019}{{\em Phys. Rev. D}
  {\bfseries 73} (2006) 054019},
  \href{http://arxiv.org/abs/hep-ph/0601245}{{\ttfamily arXiv:hep-ph/0601245}}.

\bibitem{Cooper-Sarkar:2010yul}
{\bfseries H1, ZEUS} Collaboration, A.~Cooper-Sarkar, ``{Proton Structure from
  HERA to LHC},'' in {\em {40th International Symposium on Multiparticle
  Dynamics}}.
\newblock 12, 2010.
\newblock \href{http://arxiv.org/abs/1012.1438}{{\ttfamily arXiv:1012.1438
  [hep-ph]}}.

\bibitem{Yang:1998zb}
U.-K. Yang and A.~Bodek, ``{Parton distributions, $d / u$, and higher twist
  effects at high x},''
  \href{http://dx.doi.org/10.1103/PhysRevLett.82.2467}{{\em Phys. Rev. Lett.}
  {\bfseries 82} (1999) 2467--2470},
  \href{http://arxiv.org/abs/hep-ph/9809480}{{\ttfamily arXiv:hep-ph/9809480}}.

\bibitem{Bodek:2002vp}
A.~Bodek and U.~K. Yang, ``{Modeling deep inelastic cross-sections in the few
  GeV region},'' \href{http://dx.doi.org/10.1016/S0920-5632(02)01755-3}{{\em
  Nucl. Phys. B Proc. Suppl.} {\bfseries 112} (2002) 70--76},
  \href{http://arxiv.org/abs/hep-ex/0203009}{{\ttfamily arXiv:hep-ex/0203009}}.

\bibitem{Bodek:2003wd}
A.~Bodek and U.~K. Yang, ``{Modeling neutrino and electron scattering inelastic
  cross- sections in the few GeV region with effective LO PDFs TV Leading
  Order},'' in {\em {2nd International Workshop on Neutrino-Nucleus
  Interactions in the Few GeV Region}}.
\newblock 8, 2003.
\newblock \href{http://arxiv.org/abs/hep-ex/0308007}{{\ttfamily
  arXiv:hep-ex/0308007}}.

\bibitem{Bodek:2004pc}
A.~Bodek, I.~Park, and U.-k. Yang, ``{Improved low $Q^2$ model for neutrino and
  electron nucleon cross sections in few GeV region},''
  \href{http://dx.doi.org/10.1016/j.nuclphysbps.2004.11.208}{{\em Nucl. Phys. B
  Proc. Suppl.} {\bfseries 139} (2005) 113--118},
  \href{http://arxiv.org/abs/hep-ph/0411202}{{\ttfamily arXiv:hep-ph/0411202}}.

\bibitem{Bodek:2010km}
A.~Bodek and U.-k. Yang, ``{Axial and Vector Structure Functions for Electron-
  and Neutrino- Nucleon Scattering Cross Sections at all $Q^2$ using Effective
  Leading order Parton Distribution Functions},''
  \href{http://arxiv.org/abs/1011.6592}{{\ttfamily arXiv:1011.6592 [hep-ph]}}.

\bibitem{Bodek:2021bde}
A.~Bodek, U.~K. Yang, and Y.~Xu, ``{Inelastic Axial and Vector Structure
  Functions for Lepton-Nucleon Scattering 2021 Update},''
  \href{http://arxiv.org/abs/2108.09240}{{\ttfamily arXiv:2108.09240
  [hep-ph]}}.

\bibitem{Collins:2021vke}
J.~Collins, T.~C. Rogers, and N.~Sato, ``{Positivity and renormalization of
  parton densities},''
  \href{http://dx.doi.org/10.1103/PhysRevD.105.076010}{{\em Phys. Rev. D}
  {\bfseries 105} no.~7, (2022) 076010},
  \href{http://arxiv.org/abs/2111.01170}{{\ttfamily arXiv:2111.01170
  [hep-ph]}}.

\bibitem{Pumplin:2001ct}
J.~Pumplin, D.~Stump, R.~Brock, D.~Casey, J.~Huston, J.~Kalk, H.~L. Lai, and
  W.~K. Tung, ``{Uncertainties of predictions from parton distribution
  functions. 2. The Hessian method},''
  \href{http://dx.doi.org/10.1103/PhysRevD.65.014013}{{\em Phys. Rev. D}
  {\bfseries 65} (2001) 014013},
  \href{http://arxiv.org/abs/hep-ph/0101032}{{\ttfamily arXiv:hep-ph/0101032}}.

\bibitem{Pumplin:2002vw}
J.~Pumplin, D.~R. Stump, J.~Huston, H.~L. Lai, P.~M. Nadolsky, and W.~K. Tung,
  ``{New generation of parton distributions with uncertainties from global QCD
  analysis},'' \href{http://dx.doi.org/10.1088/1126-6708/2002/07/012}{{\em
  JHEP} {\bfseries 07} (2002) 012},
  \href{http://arxiv.org/abs/hep-ph/0201195}{{\ttfamily arXiv:hep-ph/0201195}}.

\bibitem{IceCube:2013gge}
{\bfseries IceCube} Collaboration, M.~G. Aartsen {\em et~al.}, ``{Search for a
  diffuse flux of astrophysical muon neutrinos with the IceCube 59-string
  configuration},'' \href{http://dx.doi.org/10.1103/PhysRevD.89.062007}{{\em
  Phys. Rev. D} {\bfseries 89} no.~6, (2014) 062007},
  \href{http://arxiv.org/abs/1311.7048}{{\ttfamily arXiv:1311.7048
  [astro-ph.HE]}}.

\bibitem{IceCube:2021rpz}
{\bfseries IceCube} Collaboration, M.~G. Aartsen {\em et~al.}, ``{Detection of
  a particle shower at the Glashow resonance with IceCube},''
  \href{http://dx.doi.org/10.1038/s41586-021-03256-1}{{\em Nature} {\bfseries
  591} no.~7849, (2021) 220--224},
  \href{http://arxiv.org/abs/2110.15051}{{\ttfamily arXiv:2110.15051
  [hep-ex]}}. [Erratum: Nature 592, E11 (2021)].

\bibitem{Miarecki:2016kku}
S.~C. Miarecki, {\em {Earth versus Neutrinos: Measuring the total
  muon-neutrino-to-nucleon cross section at ultra-high energies through
  differential Earth absorption of muon neutrinos from cosmic rays using the
  IceCube Detector}}.
\newblock PhD thesis, UC, Berkeley, 1, 2016.
\newblock \url{{https://escholarship.org/uc/item/7q09d51t}}.

\bibitem{Gaisser:2014eaa}
T.~K. Gaisser, ``{Atmospheric Lepton Fluxes},''
  \href{http://dx.doi.org/10.1051/epjconf/20159905002}{{\em EPJ Web Conf.}
  {\bfseries 99} (2015) 05002},
  \href{http://arxiv.org/abs/1412.6424}{{\ttfamily arXiv:1412.6424
  [astro-ph.HE]}}.

\bibitem{OPERA:2010cos}
{\bfseries OPERA} Collaboration, N.~Agafonova {\em et~al.}, ``{Measurement of
  the atmospheric muon charge ratio with the OPERA detector},''
  \href{http://dx.doi.org/10.1140/epjc/s10052-010-1284-8}{{\em Eur. Phys. J. C}
  {\bfseries 67} (2010) 25--37},
  \href{http://arxiv.org/abs/1003.1907}{{\ttfamily arXiv:1003.1907 [hep-ex]}}.

\bibitem{Bustamante:2017xuy}
M.~Bustamante and A.~Connolly, ``{Extracting the Energy-Dependent
  Neutrino-Nucleon Cross Section above 10 TeV Using IceCube Showers},''
  \href{http://dx.doi.org/10.1103/PhysRevLett.122.041101}{{\em Phys. Rev.
  Lett.} {\bfseries 122} no.~4, (2019) 041101},
  \href{http://arxiv.org/abs/1711.11043}{{\ttfamily arXiv:1711.11043
  [astro-ph.HE]}}.

\bibitem{IceCube:2021keu}
{\bfseries IceCube} Collaboration, R.~Abbasi {\em et~al.}, ``{Measuring the
  Neutrino Cross Section Using 8 years of Upgoing Muon Neutrinos Observed with
  IceCube},'' \href{http://dx.doi.org/10.22323/1.395.1158}{{\em PoS} {\bfseries
  ICRC2021} (2021) 1158}, \href{http://arxiv.org/abs/2108.04965}{{\ttfamily
  arXiv:2108.04965 [astro-ph.HE]}}.

\bibitem{NuTeV:2005wsg}
{\bfseries NuTeV} Collaboration, M.~Tzanov {\em et~al.}, ``{Precise measurement
  of neutrino and anti-neutrino differential cross sections},''
  \href{http://dx.doi.org/10.1103/PhysRevD.74.012008}{{\em Phys. Rev. D}
  {\bfseries 74} (2006) 012008},
  \href{http://arxiv.org/abs/hep-ex/0509010}{{\ttfamily arXiv:hep-ex/0509010}}.

\bibitem{Seligman:1997fe}
W.~G. Seligman, \href{http://dx.doi.org/10.2172/1421736}{{\em {A
  Next-to-Leading Order QCD Analysis of Neutrino - Iron Structure Functions at
  the Tevatron}}}.
\newblock PhD thesis, Nevis Labs, Columbia U., 1997.

\bibitem{NOMAD:2007krq}
{\bfseries NOMAD} Collaboration, Q.~Wu {\em et~al.}, ``{A Precise measurement
  of the muon neutrino-nucleon inclusive charged current cross-section off an
  isoscalar target in the energy range 2.5 \ensuremath{<} E(nu) \ensuremath{<}
  40 GeV by NOMAD},''
  \href{http://dx.doi.org/10.1016/j.physletb.2007.12.027}{{\em Phys. Lett. B}
  {\bfseries 660} (2008) 19--25},
  \href{http://arxiv.org/abs/0711.1183}{{\ttfamily arXiv:0711.1183 [hep-ex]}}.

\bibitem{FASER:2019dxq}
{\bfseries FASER} Collaboration, H.~Abreu {\em et~al.}, ``{Detecting and
  Studying High-Energy Collider Neutrinos with FASER at the LHC},''
  \href{http://dx.doi.org/10.1140/epjc/s10052-020-7631-5}{{\em Eur. Phys. J. C}
  {\bfseries 80} no.~1, (2020) 61},
  \href{http://arxiv.org/abs/1908.02310}{{\ttfamily arXiv:1908.02310
  [hep-ex]}}.

\bibitem{FASER:2021mtu}
{\bfseries FASER} Collaboration, H.~Abreu {\em et~al.}, ``{First neutrino
  interaction candidates at the LHC},''
  \href{http://dx.doi.org/10.1103/PhysRevD.104.L091101}{{\em Phys. Rev. D}
  {\bfseries 104} no.~9, (2021) L091101},
  \href{http://arxiv.org/abs/2105.06197}{{\ttfamily arXiv:2105.06197
  [hep-ex]}}.

\bibitem{FASER:2023zcr}
{\bfseries FASER} Collaboration, H.~Abreu {\em et~al.}, ``{First Direct
  Observation of Collider Neutrinos with FASER at the LHC},''
  \href{http://arxiv.org/abs/2303.14185}{{\ttfamily arXiv:2303.14185
  [hep-ex]}}.

\bibitem{McGowan:2022nag}
J.~McGowan, T.~Cridge, L.~A. Harland-Lang, and R.~S. Thorne, ``{Approximate
  N$^{3}$LO Parton Distribution Functions with Theoretical Uncertainties:
  MSHT20aN$^3$LO PDFs},'' \href{http://arxiv.org/abs/2207.04739}{{\ttfamily
  arXiv:2207.04739 [hep-ph]}}.

\bibitem{LHCb:2015swx}
{\bfseries LHCb} Collaboration, R.~Aaij {\em et~al.}, ``{Measurements of prompt
  charm production cross-sections in $pp$ collisions at $ \sqrt{s}=13 $ TeV},''
  \href{http://dx.doi.org/10.1007/JHEP03(2016)159}{{\em JHEP} {\bfseries 03}
  (2016) 159}, \href{http://arxiv.org/abs/1510.01707}{{\ttfamily
  arXiv:1510.01707 [hep-ex]}}. [Erratum: JHEP 09, 013 (2016), Erratum: JHEP 05,
  074 (2017)].

\bibitem{LHCb:2017vec}
{\bfseries LHCb} Collaboration, R.~Aaij {\em et~al.}, ``{Measurement of the
  $B^{\pm}$ production cross-section in pp collisions at $\sqrt{s} =$ 7 and 13
  TeV},'' \href{http://dx.doi.org/10.1007/JHEP12(2017)026}{{\em JHEP}
  {\bfseries 12} (2017) 026}, \href{http://arxiv.org/abs/1710.04921}{{\ttfamily
  arXiv:1710.04921 [hep-ex]}}.

\bibitem{Kovchegov:2019atj}
Y.~V. Kovchegov, {\em {Chapter 9: From Parton Saturation to Proton Spin: The
  Impact of BFKL Equation and Reggeon Evolution}},
  \href{http://dx.doi.org/10.1142/9789811231124_0009}{pp.~203--238}.
\newblock 2021.
\newblock \href{http://arxiv.org/abs/1911.02651}{{\ttfamily arXiv:1911.02651
  [hep-ph]}}.

\bibitem{Haidt:1999ps}
D.~Haidt, ``{Log-log behaviour of $F_2$ at low x in the $Q^2$ range from 0.1
  GeV$^2$ to 35 GeV$^2$},''
  \href{http://dx.doi.org/10.1016/S0920-5632(99)00670-2}{{\em Nucl. Phys. B
  Proc. Suppl.} {\bfseries 79} (1999) 186--188}.

\bibitem{Block:2006dz}
M.~M. Block, E.~L. Berger, and C.-I. Tan, ``{Small x Behavior of Parton
  Distributions from the Observed Froissart Energy Dependence of the Deep
  Inelastic Scattering Cross Section},''
  \href{http://dx.doi.org/10.1103/PhysRevLett.97.252003}{{\em Phys. Rev. Lett.}
  {\bfseries 97} (2006) 252003},
  \href{http://arxiv.org/abs/hep-ph/0610296}{{\ttfamily arXiv:hep-ph/0610296}}.

\bibitem{Berger:2007vf}
E.~L. Berger, M.~M. Block, and C.-I. Tan, ``{Analytic Expression for the Joint
  x and $Q^2$ Dependences of the Structure Functions of Deep Inelastic
  Scattering},'' \href{http://dx.doi.org/10.1103/PhysRevLett.98.242001}{{\em
  Phys. Rev. Lett.} {\bfseries 98} (2007) 242001},
  \href{http://arxiv.org/abs/hep-ph/0703003}{{\ttfamily arXiv:hep-ph/0703003}}.

\bibitem{Froissart:1961ux}
M.~Froissart, ``{Asymptotic behavior and subtractions in the Mandelstam
  representation},'' \href{http://dx.doi.org/10.1103/PhysRev.123.1053}{{\em
  Phys. Rev.} {\bfseries 123} (1961) 1053--1057}.

\bibitem{Dziewonski:1981xy}
A.~M. Dziewonski and D.~L. Anderson, ``{Preliminary reference earth model},''
  \href{http://dx.doi.org/10.1016/0031-9201(81)90046-7}{{\em Phys. Earth
  Planet. Interiors} {\bfseries 25} (1981) 297--356}.

\bibitem{IceCube:2015gsk}
{\bfseries IceCube} Collaboration, M.~G. Aartsen {\em et~al.}, ``{A combined
  maximum-likelihood analysis of the high-energy astrophysical neutrino flux
  measured with IceCube},''
  \href{http://dx.doi.org/10.1088/0004-637X/809/1/98}{{\em Astrophys. J.}
  {\bfseries 809} no.~1, (2015) 98},
  \href{http://arxiv.org/abs/1507.03991}{{\ttfamily arXiv:1507.03991
  [astro-ph.HE]}}.

\bibitem{IceCube:2020wum}
{\bfseries IceCube} Collaboration, R.~Abbasi {\em et~al.}, ``{The IceCube
  high-energy starting event sample: Description and flux characterization with
  7.5 years of data},''
  \href{http://dx.doi.org/10.1103/PhysRevD.104.022002}{{\em Phys. Rev. D}
  {\bfseries 104} (2021) 022002},
  \href{http://arxiv.org/abs/2011.03545}{{\ttfamily arXiv:2011.03545
  [astro-ph.HE]}}.

\bibitem{Gauld:2019pgt}
R.~Gauld, ``{Precise predictions for multi-TeV and PeV energy neutrino
  scattering rates},''
  \href{http://dx.doi.org/10.1103/PhysRevD.100.091301}{{\em Phys. Rev. D}
  {\bfseries 100} no.~9, (2019) 091301},
  \href{http://arxiv.org/abs/1905.03792}{{\ttfamily arXiv:1905.03792
  [hep-ph]}}.

\bibitem{Zhou:2019vxt}
B.~Zhou and J.~F. Beacom, ``{Neutrino-nucleus cross sections for W-boson and
  trident production},''
  \href{http://dx.doi.org/10.1103/PhysRevD.101.036011}{{\em Phys. Rev. D}
  {\bfseries 101} no.~3, (2020) 036011},
  \href{http://arxiv.org/abs/1910.08090}{{\ttfamily arXiv:1910.08090
  [hep-ph]}}.

\bibitem{Zhou:2019frk}
B.~Zhou and J.~F. Beacom, ``{W-boson and trident production in
  TeV\textendash{}PeV neutrino observatories},''
  \href{http://dx.doi.org/10.1103/PhysRevD.101.036010}{{\em Phys. Rev. D}
  {\bfseries 101} no.~3, (2020) 036010},
  \href{http://arxiv.org/abs/1910.10720}{{\ttfamily arXiv:1910.10720
  [hep-ph]}}.

\bibitem{Klein:2020nuk}
S.~R. Klein, S.~A. Robertson, and R.~Vogt, ``{Nuclear effects in high-energy
  neutrino interactions},''
  \href{http://dx.doi.org/10.1103/PhysRevC.102.015808}{{\em Phys. Rev. C}
  {\bfseries 102} no.~1, (2020) 015808},
  \href{http://arxiv.org/abs/2001.03677}{{\ttfamily arXiv:2001.03677
  [hep-ph]}}.

\bibitem{Gonzalez-Garcia:2007wfs}
M.~C. Gonzalez-Garcia, F.~Halzen, M.~Maltoni, and H.~K.~M. Tanaka,
  ``{Radiography of earth's core and mantle with atmospheric neutrinos},''
  \href{http://dx.doi.org/10.1103/PhysRevLett.100.061802}{{\em Phys. Rev.
  Lett.} {\bfseries 100} (2008) 061802},
  \href{http://arxiv.org/abs/0711.0745}{{\ttfamily arXiv:0711.0745 [hep-ph]}}.

\bibitem{Glashow:1960zz}
S.~L. Glashow, ``{Resonant Scattering of Antineutrinos},''
  \href{http://dx.doi.org/10.1103/PhysRev.118.316}{{\em Phys. Rev.} {\bfseries
  118} (1960) 316--317}.

\bibitem{Barger:2014iua}
V.~Barger, L.~Fu, J.~G. Learned, D.~Marfatia, S.~Pakvasa, and T.~J. Weiler,
  ``{Glashow resonance as a window into cosmic neutrino sources},''
  \href{http://dx.doi.org/10.1103/PhysRevD.90.121301}{{\em Phys. Rev. D}
  {\bfseries 90} (2014) 121301},
  \href{http://arxiv.org/abs/1407.3255}{{\ttfamily arXiv:1407.3255
  [astro-ph.HE]}}.

\end{thebibliography}\endgroup
\end{document}